\documentstyle[epsfig]{article}
\def\abstract#1{\vskip 7mm 
        \begin{center}{\large Abstract}\par \smallskip
                \begin{minipage}[c]{14cm}
                        \small #1
                \end{minipage}
        \end{center}
}
\def\title#1{\begin{center}{\Large\bf #1}\end{center}}
\def\author#1{\vskip 5mm \begin{center}{#1}\end{center}}
\def\address#1{\begin{center}{\it #1}\end{center}}

\def\eq{\begin{equation}} 
\def\en{\end{equation}} 
\newcommand{\beqn}{\begin{eqnarray}} 
\newcommand{\eeqn}{\end{eqnarray}} 
\newcommand\Sum{\displaystyle\sum} 
\newcommand\Frac{\displaystyle\frac} 
\def\part{\partial} 
 
\def\part{\partial}

\newcommand{\sect}{\setcounter{equation}{0}\section}

\makeatletter
\@ifundefined{lesssim}{}{}
\@ifundefined{gtrsim}{}{}
\def\vereq#1#2{\lower3pt\vbox{\baselineskip1.5pt \lineskip1.5pt
\ialign{$\m@th#1\hfill##\hfil$\crcr#2\crcr\sim\crcr}}}
\makeatother

\begin{document}
{{
\title{\vspace{5.6cm}
{\huge Natural beauty of the standard model}\\
\vspace{0.5cm}
-A derivation of the electro-weak unified and quantum-gravity theory
without assuming a Higgs particle-\\
\vspace{1.4cm}
(PhD. Thesis, submitted to the University of Tokyo
\footnote{This thesis was, however, rejected by the judging committee, 
without even mentioning sections 4-7. They forced me to omit 
sections 3-7 at all. Thus, only the revised version of Section 2 
was approved by the judging committee.})\\
\vspace{14mm}
Miyuki Nishikawa}
\address{
  Department of Physics,
  Graduate School of Science, \\
  University of Tokyo, Hongo, Bunkyo, Tokyo 113-0033, Japan}
\vspace{1.4cm}
\begin{center}{\Large Abstract}\end{center} 
{\normalsize\hspace{3mm} We study the asymptotic behavior of a singular potential that arises under several frequently occurring analytic behaviors of the
eigen functions (of the Schr\"{o}dinger eigenvalue problem) without 
introducing cut-offs. Instead, in our analyses we focus on power 
behaviors of eigen functions. We find that the asymptotic behavior of 
the singular potential crucially depends on the analytic property of 
the eigen functions near the singular point. \\
\\ \indent
 Then, we discuss the self-consistency condition for the spherical 
symmetric Klein-Gordon equation, and discuss a natural possibility
 that gravity and weak coupling constants $g_G$ and $g_W$ may be defined 
after $g_{EM}$. From this point of view, gravity and the weak force are 
subsidiary, derived from electricity. Particularly, $SU(2)_L\times U(1)$ 
unification is derived from the $L^2$ normalizability condition, without assuming a phase transition. A possible origin of the Higgs mechanism is proposed. Each particle pair of the standard model is associated with the corresponding asymptotic expansion of an eigen function.\\
\\ \indent
 Next we consider the meaning of internal and external degrees
of freedom for a 2 body problem, and find a complex $U(1)$ phase of spins, which can not reduce to the local motion of an external observer. These degrees of freedom are inherent to the Poincar\'{e} group, and can be expressed in terms of asymmetric spinor representations. Then we try to derive all gauge fields via this nonintegrable complex $U(1)$ phase. As a spin-off, supersymmetry is regarded as a kind of Mach's principle for spinning frames-or the Ptolemaic (geocentric) theory to confuse a rotating frame with an inertial frame.\\
\\ \indent
 Furthermore, we review classical experimental backgrounds for 
general relativity and try to explain them within the range of 
special relativity, and discuss possible solutions for 
paradoxes in quantum gravity.\\
\\ \indent
 Taking angular momentums into account to improve above discussions, we can explain the smallness of neutrino mass without assuming the see-saw mechanism.
A natural geometric interpretation of the quark flavor mixing angle is added in the Conclusion.
}}
\newpage{\pagestyle{empty}
\vspace{7cm}\begin{center}
{\bf Dream of Meeting Wein}\\
\vspace{4.9mm} Since when we were born, \\
each of us has been spinning \\
an invisible red thread. \\
On meeting and passing by someone on streets, \\
the thread becomes intertwining and entangled. \\
The locus of people all over the world constitutes a huge net. \\
Drawing in the thread of mine, \\
so quietly as not to break the net, \\
and unraveling knots to know whereabouts, \\
selecting one out of many lines, \\
and following it, \\
I wonder when the dream comes true.
\vspace{5cm}\end{center}
}\newpage
\tableofcontents
\listoffigures
\newpage
\section{Introduction}
 The current standard model of particles coincides so beautifully with experiments\cite{Wein}. The aim of this thesis is to investigate the necessity or naturalness of this model, with assumptions as minimal as possible. However, our starting point is rather special in some aspects as follows:
\begin{description}
	\item[(1)] 
In contrast to the recent trend of supersymmetry and superstring theories, we shall pay our maximal respects to the present standard model. Particularly, the difficulty in quantum gravity consists in its smallness beyond our ability of measurement and non-renormalizability. However, in the standard model of particles, all particles except photons and neutrinos obtain masses through the Higgs mechanism. That is why we naturally consider the Higgs mechanism, as the origin of mass generation, has an intimate relation to quantum gravity. The possibility that the standard model makes the ultimate theory by itself is very attractive.  
	\item[(2)] 
A Higgs particle is the last `particle' that appears in the standard model but has never been observed yet. In fact, it is an exceptional field in the sense that all elementary particles need its nonzero vacuum expectation value only to obtain nonzero masses of themselves. Furthermore, the theoretical description of this process called Higgs mechanism\cite{Higgs} is mathematically tricky. The origin of the Higgs field, and whether it is an existing elementary particle at all, or some composite particles, are unclear. We even call it `an auxiliary Higgs field', which stands well for its artificial nature. \\
 For this reason, we shall not assume the existence of a Higgs particle in this thesis. A Higgs field can be an artificial tool, like an auxiliary line to 
solve an elementary geometric problem, a subsidiary field derived after spontaneous symmetry breaking of $SU(2)_L\times U(1)_Y\to U(1)_{EM}$.
        \item[(3)] 
There are a few mathematically subtle problems in the construction of quantum field theory. For instance, 
\begin{description}
	\item[(i)] 
In usual dimensional counting, momentum has the dimension one. But 
a function $f(x)$, when differentiated $n$ times,  
does not always behave like one with its power smaller by $n$. 
This inevitable uncertainty may be essential in general theory of  
renormalization, including quantum gravity. 
	\item[(ii)] 
In usual, we assume that a scattered particle asymptotically behaves like a free particle in the long distance limit after interactions. But strictly speaking, a free state wave function has a constant absolute value and thus its norm diverges if integrated over an infinite spatial region. We might as well consider that all physical particles are constrained in the universe by gravity.
	\item[(iii)] 
We should be most careful and sincere in applying the particle physics to cosmology. For example, the Penrose' singularity theorem and the existence of a black hole in general relativity rather show a limit on the applicability of the theory. Therefore, a naive combination of the standard model and general relativity, or extrapolation to the two extreme scales leads to physically misleading results. 
\end{description}
        \item[(4)] 
Neutrino mass has been a hot topic since Super-Kamiokande observed the possible oscillation phenomena in 1998. However, a massive neutrino is absent in the original standard model, and a naive insertion of a mass term or new particles breaks the subtle mechanism of anomaly cancellation\cite{Wein}\cite{Doba}, including the quantization of charges. In addition, a massive neutrino inevitably generates a nonzero anomalous magnetic moment of itself\cite{Nu}, which has never been observed to be of positive value\cite{Part}. Historically Kobayashi and Maskawa\cite{KobaMas} predicted the third generation of quarks from the observed ${\cal CP}$ violating decay $K^0_L\to 2\pi$. The existence of the third generation well accounts for this phenomenon by a naturally induced flavor-changing $3\times 3$ matrix of quarks, while for a lepton sector, no corresponding flavor-changing decay has been observed. $W^{\pm}$ bosons couple only to left-handed particles. These phenomena are better explained by the right-left asymmetry of the weak interaction, traced back to the masslessness of neutrinos. Such difference between quarks and leptons might well be reduced to the largest difference between the corresponding forces, i.e., weak and strong interactions. 
\end{description}
Motivated by the above consideration and with help of many previous works of pioneers to construct a consistent theory of quantum gravity with minimal assumptions  briefly reviewed in appendix \ref{QGMA}, we study the natural possibility to unify all the existing 4 kinds of interactions into a common phase of a wave function. The key concept is the singularity of a phase as the origin of gauge fields\cite{Dirac}. If this is the case, all interactions are subsidiary force derived from the original $U(1)$ gauge potential. In addition, gravity and weak coupling constants $g_G$ and $g_W$ can be defined after $g_{EM}$.

 Section 2 is a note on singularities, especially a general class of singularities made from usual Taylor expansions by finite times of operations defined there, with its application to the short distance limit potential problem of the one-dimensional Schr\"{o}dinger equation.

 However this is only a preparation for a physically important generalization of the results to higher dimensions and the long distance limit of the Klein-Gordon equation in section 3. This section is our main contribution. We make use of the previous results to build the self-consistency conditions of powers for the general 2 body problems of elementary particles.

 Neglecting spins and spherical asymmetry, we shall derive several general, though qualitative theorems on the relations of the (scalar) bosons and the potentials each of them feels. Particularly, $SU(2)_L \times U(1)$-unification is derived without assuming a phase transition. Furthermore, this can account for the smallness of neutrino mass without assuming the see-saw or other non-standard mechanisms. In the next subsection, a possible origin of the Higgs mechanism is proposed. If this is the case, $W^{\pm}$-bosons are altered forms of photons and then we can define $Z^0$-bosons as subsidiary particles obtained by projecting the $SU(2)_L$ Lie algebra generated by $W^{\pm}$-bosons onto the orthogonal complement of photons.

 In the next subsection, we shall deal with gravity within the range of the standard model, respecting the result of Weinberg\cite{WeinG} that derives the Einstein equation without assuming a curved space-time, not giving gravitons special treatments. The inferred shape of gravity potential involves both the Newtonian potential and its Schwarzschild correction. According to this, it is natural to consider that there is no black hole, for no singularity appears in the Klein-Gordon equation in a flat space-time. In the next subsection, we apply previous results to more general 2 body problems and associate every particle pair of the standard model with their corresponding asymptotic expansion of an eigen function, thus unifying all 4 kinds of interactions. 

 Then we turn around to section 4, where we consider the meaning of internal and external degrees of freedom, and find that two degrees of freedom naturally arises from the Poincar\'{e} transformation rule for the two bodies. This can be 
the origin of a nonintegrable complex $U(1)$ phase $\theta$. Furthermore, a geometric derivation of gauge fields is explored. In section 5, we shall express the gravitational, electric, magnetic and gluon fields in a unified way from this phase $\theta$. 

 The section 6 is rather supplementary, where we review classical experimental backgrounds for general relativity and try to explain them within the range of 
special relativity, and discuss several paradoxes in quantum gravity.

 Finally, in section 7 we include angular momentums and thus give a more strict discussion of the results of the section 3. We try to answer the mystery of gravity, and of small but nonzero neutrino masses.

Nonetheless, Prof. Koshiba's work is still stimulating our ambitious attempts for the ultimate theory after his reception of the Nobel Prize in 2002. Now in 2009 we feel a bit more close to the theory, for Prof. Kobayashi, Maskawa, and Nambu also received the Prize. A natural geometric interpretation of the quark flavor mixing angle is added in the Conclusion.
\newpage
\sect{On singular potential of the Schr\"{o}dinger equation}\label{Sing}
\subsection{Introduction to this section} 
 Quantum theory is afflicted with inevitable ambiguity beyond 
measurements. This work is motivated by an attempt to shed light on such  
ambiguity. For example, the analyticity of an eigen function can not be  
determined by finite times of measurements. But if the eigen function 
becomes very sharp at a point, average momentum is dominated by the point, 
and the theory may not be sensitive to such analyticity. Moreover, if 
a continuous eigen function takes both plus and minus values, it  
must also take 0, which is the qualitative fact not sensitive to the  
exact shape of the eigen function. In this section we first distinguish 
these two kinds of singularities in subsection 2. Then we proceed 
to construct several frequently occurring analytic behaviors of 
possible singularities in subsection 3. Extension to a general spatial 
dimension $N$ is discussed in subsection 4. Physical explanation of the results and possible applications are discussed in subsection 5.
 Notice that we never use the term `wave function' here, but use 
`eigen function'. This is because the renormalization technique 
means to cut-off high frequency modes of eigen functions. Here, 
instead, we respect the $C^2$-class analyticity of eigen functions. 
This focus on $C^2$-class eigen functions is minor to traditional 
approach to respect the $L^2$-condition {\footnote{The natural 
$L^2$-condition is discussed in an appendix and in 
the section \ref{AHU}.}, 
and hence the application of the present analyses may be limited. 
However, we believe that $C^2$-class analyticity is independent 
of the $L^2$-condition 
{\footnote{An $C^2$-class eigen function defined on a finite closed 
region always satisfies $L^2$-condition, but not, if defined on a 
finite opened region.}}, and nonetheless important. 
Thus in this section we shall study the asymptotic behavior 
of a singular potential with a focus on $C^2$-class analyticity of 
eigen functions. However, we do not restrict our discussion only on the 
$C^2$-class eigen functions.}.
\subsection{Possibility of singularity and domain of definition } 
 For simplicity, let us first consider a one-dimensional  
Schr\"{o}dinger equation,  
\beqn y''=Vy-Ey. \label{ko1 } \eeqn 

In this section, we will say that a function
$y(x)$ has a potential $V(x)$ iff $y$ is a $C^2$-class
function defined in $(0, a)$ and further if there exist a function
$V(x)$ and a constant $E$ which satisfy (\ref{ko1 }) in $(0, a)$.

In fact, any $C^2$-class function $y$ satisfies (\ref{ko1 }) if  
we take 
\beqn V=y''/y, \;\; E=0. \label{ko2 } \eeqn 
Here the replacement of the constant $E\to E'$ is equivalent to  
$V\to V -E'+E$, so from now on we take $E=0$. There are 2 possible  
cases for (\ref{ko2 }) to have a singularity: 
 
\vspace{.2 in}
(I) There exists a $x$ such that $y(x)=0$, $0<x<a$,  

\vspace{.2 in} 
(II) $y''$ does not converge (for $x\to +0$ or $x\to a-0$). 

\vspace{.2 in}
\noindent Both cases include a removable singularity in which case 
a pole and a zero point cancel out. 
\subsection{Construction of several frequently occurring powers of 
possible singularities} 
 Now let us move the possible singularity to $x=0$ by the 
redefinition of the origin and consider the behavior of $V$ as  
$x\to +0$. Let $y(z)$ be the natural analytic continuation of $y(x)$ 
(from the real axis) to the complex plane. We first consider
 
\vspace{.2 in} 
\noindent (CASE 1) $y(z)$ has no essential singularity at $z=0$. 
 
\vspace{.1 in} 
\noindent (a) If $y(z)$ can be Laurent expanded around $z=0$ as  
\beqn y=\sum_{n=k }^\infty a_nz^n, \;\; a_k\neq 0 \label{ko4 }  
\eeqn 
and the expansion is uniformly convergent on any compact subset in the annular domain $D=\{ z|0<|z|<r\}$\cite{Ahlf}\footnote{From now on, we assume this 
condition of uniformly convergence on compact sets without further mentioning.}, then we find that the potential $V$ behaves as
\beqn \frac{y'' }{y }=\frac{\sum_{n=k }^\infty  
a_nn(n-1)z^{n-2 }}{\sum_{n=k }^\infty a_nz^n} \to\left  
\{ \begin{array}{l}\frac{a_d }{a_k }d(d-1)z^{d-2-k }\;\; (0\leq  
k)  \\ 
k(k-1)z^{-2 }\;\; (k<0) \end{array} \right. , \label{ko5 } \eeqn 

with $d$ the lowest power such that $a_d\neq 0$ and $1<d$ (if  
there is no such $d$, $a_d=0$). 
 
\vspace{.1 in} 
\noindent (b) When we replace the power of the finite number of 
terms in the a (a) expansion with an arbitrary real number, we obtain

{\footnote{From now on, the expansion coefficients are all real  
except if mentioned, and the branch is chosen so that the 
function takes unique real value at $z\to +0$. More precisely, a  
branching point with the power of an irrational number is an  
essential singularity, but the difference is not important  
here. All the terms are arranged in the order of ascending powers. For example, let us assume $y(x)= 7x^{\frac{73}{95}}+5x^{\sqrt 5}+3x^{\pi}+
x^{2\sqrt 5}+3x^{\frac{77}{5}}$. Then, there is the natural unique analytic 
continuation of $y(r)$ from the real axis to the complex region around $z=0$ 
except the cut {\bf Re} $z\leq 0$.}}  
\beqn \frac{y'' }{y }\to\left\{ \begin{array}{l}\frac{a_d }{a_k } 
d(d-1)z^{d-2-k 
}\;\;\left (\begin{array}{l}\mbox{if}\;  y = a_0 +a_1z +a_dz^d\cdots  
\;\; \mbox{or} \\ 
y = a_1z +z_dz^d\cdots \end{array} \right )\\ 
k(k-1)z^{-2 }\;\; (k<0)\;\; \mbox{(otherwise)} \end{array} \right. , 
\label{ko6 } 
\eeqn 
where $a_d$ is the coefficient of the lowest power except for  
$0, 1$. 
Thus far, the powers $\nu$ where the potential can  behave like  
$V\to x^\nu$ as $x\to +0$ are 
\beqn 
\mbox{for (I),   }\qquad & & \nu =-2 \; ;\; -1\leq \nu , \\
\mbox{for (II),   }\qquad & & -2\leq \nu <-1,\;\; -1<\nu <0. \eeqn 
\noindent (c) If $y(z)$ admits the expansion 
\beqn y=\sum_{n=l }^ka_n(\log z)^n, \;\; a_l\neq 0, \label{ko7 }  
\eeqn 
then we find 
\beqn \frac{y'' }{y }=\frac{\sum_{n=l }^kna_n\{ (n-1)(\log z)^{n- 
2 }-(\log  
z)^{n-1 }\} }{z^2\sum_{n=l }^ka_n(\log z)^n }\to\frac{-k }{z^2 
\log z } \label{ko8 } \;\; ,  
\eeqn 
where $\log z$ diverges as $z\to 0$, but for an arbitrary integer  
$n$, $z(\log z)^n$ tends to $0$. So we can regard $\log z$ as `an 
infinitesimal negative power' $z^{-\epsilon }\; (\epsilon >0)$. Then we  
can generalize the type (b) expansion by the replacement of 
finite number of terms 
\beqn a_nz^n\to z^n\sum_{m=l_n }^{k_n }a_{mn }(\log z)^m\;\;  
(m\in {\bf R}). \label{ko9 }  
\eeqn 
This has the effect of 
\beqn \left \{ \begin{array}{l}z^{d-2-k }\to z^{d-2-k }(\log z) 
^m\;\; 
       (m\in {\bf R})\\ z^{-2 }\to z^{-2 }/\log z \end{array} 
\right. \label{ko10 } \eeqn 
in (\ref{ko6 }), i.e., 
{\footnote{For a $C^2$-class function $y$, $-1-\epsilon$ is  
impossible. And for (II), the region of $\nu$ is invariant.}}  
\beqn \mbox{for (I), }\qquad \nu =-2 (+\epsilon )\; ;\;\; -1\leq \nu . 
\label{ko11 }\eeqn 
Let us call this type of expansion the type (c). For the type (c) expansions, 
we can define the index of power $k_y, \mu_y, \nu_y (z\to +0)$ 
as follows: \beqn y\to z^{k_y }, \;\;\frac{y' }{y }\to 
z^{\mu_y }, \;\;\frac {y'' }{y }\to  
z^{\nu_y }.  \label{ko12 } \eeqn 
Type (c) property is invariant under finite times of  
summations, subtractions, and differentiations. 
  
\vspace{.1 in} 
\noindent (d) When we apply finite times of summations,  
subtractions, multiplications, divisions (by $\neq 0$),  
differentiations, and compositions (with the shape of $f(g(z)), \; 0 
\leq k_g, \; g( +0)= +0$ where $f, g$ are the type (c) expansions),  
$k_y, \mu_y, \nu_y$ can also be defined. As an arbitrary type (d)  
expansion $f(z)$ has a countable number of terms and a  
nonzero `radius of convergence 
{\footnote{The meaning of this term is different from the usual one 
because $z=0$ can be a singularity point.}}  
' $r$ where the expansion converges for $0<|z|<r$, it can be written as  
\beqn f(z)=\sum_{n=0 }^\infty f_n \;\; . \label{ko13 } \eeqn 
As the `principal part' which satisfies $k_{f_n}<0$ consists  
of finite number of terms, a type (d) expansion diverges or  
converges monotonically as $z\to +0$, so enables  the expansion of  
(\ref{ko13 }) in the order of ascending powers. As the expansion  
is almost the same as that of the type (c) (the only differences are the  
multiplications by $(\log z)^n$ for an infinite number of terms  
and the appearance of the terms like $\log (z\log z)$), the  
region of $\nu_y$ remains unchanged. We just use the 
symbolic notation $\epsilon^2=\epsilon$ for an infinitesimal power 
like $1/\log (-\log z)$. \\

Next we consider
 
\noindent (CASE 2) $y(z)$ has an isolated essential  
singularity at $z=0$. Then Weierstrass' theorem of complex analysis 
shows that a sequence of points can converge to any  
value depending on its approach to an essential singularity (with  
infinite order) \cite{Ahlf}. 
But now that we study only the  
case along the real axis $z\to +0$, the limit is sometimes well  
defined. Let us study the following cases.  
 
\vspace{.1 in} 
\noindent (e) When the following expansion is possible (the type (e)): 
$y=\pm e^{f(z) }$, where $f$ is a type (d) expansion, then we can  
define the finite values $\mu_y, \nu_y$ by  

\beqn \left \{ \begin{array}{l}\frac{y' }{y }=f'\to z^{\mu_y },  
\;\; \mu_y=k_f +\mu_f , 
\nonumber \\ 
\frac{y'' }{y }=f'^2 +f''\to z^{\nu_y }, \;\;\nu_y\geq {\mbox{min}}(2k_f  
+2\mu_f, \;  
k_f +\nu_f) .\end{array} \right. \label{ko14 } \eeqn 
Let us consider the region of $\nu_y$. For $k_f\geq  
0$ it is the same as for the type (d). For  
\beqn y=e^{az^k } \;\; (a, \; k\in {\bf R}, \;\; k\leq 0) \label{ko15 }  
\eeqn 
satisfies 
\beqn 
\frac{y'' }{y }=a^2k^2z^{2k-2 } +ak(k-1)z^{k-2 }\to\left\{  
\begin{array}{l} Cz^{-2\pm\epsilon }\;\;(k=-\epsilon )
\mbox{ (or higher order) } \nonumber \\ 
a^2k^2z^{2k-2 }\;\; (k<0) \end{array} \right. \label{ko16 }  
\eeqn 
where $C$ is a constant, combination with the type (c) case leads 
to the region of $\nu_y$ as:  
 
\beqn 
\mbox{For (I),   }\qquad & & \nu_y\leq -2 +\epsilon \; ; \;\; -1\leq \nu_y , \\
\mbox{For (II),   }\qquad & & \mbox{an arbitrary negative number except for }-1, \eeqn  
\noindent Let us then consider if we can fill the remaining `window'  
of the region of $\nu_y$ for (I),  
 
$-2 +\epsilon <\nu_y<-1$. 
 
\vspace{.1 in} 
\noindent (f) When we can write $y=f_0 +\sum_{n=1 }^{m }(\pm )e^ 
{f_n }$, where $f_0$ and $f_n$ is of the type (d), $k_{f_n }<0$, 
and ($\pm$) takes each of the signatures $+-$, then 
we can assume that each term in $\sum$ is ordered in the  
increasing absolute values for $z\to +0$. Because 
\beqn e^{az^k }\to\left \{\begin{array}{l}z^0\;\;(k\geq 0, \;  
a\neq 0)\nonumber \\ 
0\;\;\; (k<0, \; a<0)\nonumber \\ 

\infty\;\; (k<0, \; a>0) \end{array}\right. \label{ko18 } \eeqn  
and $y\to 0$ for (I), we obtain 
 
\beqn y=\left (\sum_{n=0 }^\infty\sum_{m=l_n }^{m_n }a_{nm }z^n 
(\log z)^m\right ) +\sum_{n=1 }^l(\pm ) 
e^{\sum_{i=k_n }^\infty\sum_{j=l_{ni } }^{k_{ni } }a_{nij }z^i(\log z)^j } 
. \label{ko19 } \eeqn  
\indent If the second term sum at the R.H.S. is not 0, we can write 
\beqn k_l\leq\cdots \leq k_1\leq 0, \;\; a_{nk_nk_{ni } }<0. \label{ko20 }  
\eeqn  
The following example illustrates the meaning of above operation 
to arrange all terms in an ascending order.
We can symbolically arrange several terms in an ascending order of $z$ 
in the limit $z\to +0$ as follows: 
\eq e^{\frac{1}{z^2}},\quad e^{-\frac{\log z}{z}},\quad 
e^{\frac{1}{z}},\quad z^{-2},\quad 1\simeq
 e^{\pm\frac{1}{\log z}},\quad z,\quad e^{-\log^2 z},\quad 
e^{-\frac{1}{z}},\quad e^{-\frac{1}{z^2}}.\en

On the other hand, for the case (I) ($y$ is of $C^2$-class), 
the first term sum of (\ref{ko19 }) can be written  
as  
\beqn (\;\; )=a_{10 }z +\sum_{n=2 }^\infty\sum_{m=l_n }^{m_n } 
\cdots , \;\; m_2=0.   
\label{ko21 } \eeqn  
 
As 
\beqn y''\to \left \{  
{\begin{array}{l}a_{nm}(z^n(\log z)^m)''\to n(n-1)a_{nm}z^{n-m\epsilon -2 }  
\;\;\left (\begin{array}{l}  
\mbox{The term such that } \\  
\mbox{$n-m\epsilon$ is the smallest} \end{array}\right )  
\;\; (^\exists a_{nm} \neq 0) \\ 
\left [\{ {a_{nk_nk_{ni } }z^{k_n }(\log z)^{k_{ni} }\}'}^2
+\{ a_{nk_nk_{ni } }z^{k_n }(\log z)^{k_{ni} }\}'' \right ] \\ 
\hspace{49mm} 
\times e^{\sum_{i=k_n }^\infty\sum_{j=l_{ni }}^{k_{ni } } 
a_{nij }z^i(\log z)^j}\;\;\; ( ^\forall a_{nm }=0) \end{array}} 
\right.\hspace{-2cm}\label{ko22 } \eeqn  
for  $z\to +0$, we find 
\beqn \frac{y'' }{y }\to \left \{ \begin{array}{l} 
n(n-1)\frac{a_{nm}}{a_{10 }}z^{n-m\epsilon -3 }\;\;  
(a_{10 }\neq 0\; \mbox{and}\;  ^\exists a_{nm}\neq 0)\\ 
Cz^{2k_n-2k_{ni }\epsilon -2 }
e^{a_{nk_nk_{ni } }z^{k_n } (\log z)^{k_{ni } }}\to 0\;\;  
(a_{10 }\neq 0\; \mbox{and}\; ^\forall a_{nm }=0 ) \\ 
n(n-1)z^{-2 }\;\; (a_{10 }= 0\; \mbox{and}\;  ^\exists a_{nm }\neq 0) \\ 
C'z^{2k_n-2k_{ni }\epsilon-2 }\;\;(a_{10 }= 0\; 
\mbox{and}\; ^\forall a_{nm }=0 )  
\end{array} \right. , \label{ko23 } \eeqn 
where $C, (0<)C'$ are constants. 
The possible values of $\nu_y$ for (I) remain unchanged:
\eq\nu_y\leq -2 +\epsilon \; ; \;\; -1\leq \nu_y\; . 
\en
\vspace{.1 in} 
\noindent (g) If the expansion is obtained from the type (f)
expansions by finite times of summations, subtractions,
multiplications, divisions (by $\neq 0$), differentiations, and
compositions (with the shape of $f(g(z)), \; 0\leq k_g, \; g( +0)=
+0$ where $f, g$ are the type (f) expansions), then the expansion
becomes very complicated compared to an ordinary Laurent expansion.
However, such an expansion has a countable 
number of terms and a nonzero `radius of convergence' $r$ 
{\footnote{Of course, the meaning is different from the usual  
one. }} 
where $y$ is analytic for $0<|z|<r$. This can also be ordered  
partially in the ascending powers and we can write the first  
term explicitly, and so monotonically diverges or converges but  
never oscillates as  $z\to +0$. Its general shape is the whole  
sum  
\beqn (1)_i +(2)_j +\cdots +(m)_k \; , \label{ko25 } \eeqn  
where 
\beqn (1)_i \;\; & := & \mbox{\huge (}\sum_{n\in \{ n\}_i 
}^\infty\sum_{m_1,\cdots , m_{d_i }=-\infty }^{m_{i1 }, \cdots , m_{id_i 
} }a_{inm_1\cdots m_{d_i } }z^n(-\log z)^{m_1 }(-\log (-z/\log 
z))^{m_2}\nonumber \\ &    &  \hspace{4.2 cm}\cdots (-\log (-z/(-\log  
(-z/\log\cdots z))))^{m_{d_i } } \mbox{\huge )}_i, \nonumber \\ 
(2)_{\pm j } & := & \sum_{i\in \{ i\}_j }(\pm )e^{\pm (1)_i },  
\nonumber \\ 
(3)_{\pm k } & := & \sum_{j\in \{ j\}_k }(\pm )e^{\pm (2)_j },  
\nonumber \\ 
\vdots \label{ko24 } \eeqn 
Here the ($\pm$) in front of $e$ takes each of the  
signatures depending on each $i$ (or $j, k, \cdots $), while the  
$\pm$ in the exponent of $e$ and in front of $j, k, \cdots $  
takes the signature such that the coefficient of the first term  
in $\sum$ is of the same signature as $j$ after choosing the  
signatures. Each term is ordered in partially ascending powers 
with regard to any sums. The sum with index $n$ is performed  
according to the monotonically non-decreasing sequence of real  
numbers $\{ n_i\}\; (-\infty <n_i)$ depending on $i$. In the  
same manner, the sum with index  $i, j, \cdots $ is performed  
according to the finite, monotonically non-decreasing sequence $\{  
i_j\} , \;\{ j_k\}\cdots $ of natural numbers. $m_{i_1 },  
\cdots , m_{i_{d_i } }$ take finite values, but they increase  
in correspondence with $n$ and grows $\to\infty$ as $n\to  
\infty$, and depend on $i$. $d_i$ is the maximal `depth' of the  
composition of $\log$s, or the number of $\log$s, depending on $i$  
and of finite value. 
{\footnote{The power is smaller when $m_1 +m_2 +\cdots +m_{i_1 } 
$ is greater for the same $n$, and when it is also the same and  
$m_1$ is smaller, and when it is also the same and $m_2$ is 
smaller, ..., and so on.} } 
 
As the sum of the shape of $(m)_i$ can always be represented  
as the exponent of the infinite sum of the same shape,  
\beqn (m)_i & = & (\pm )e^{(m)_0 }, \;\; (m)_0:=\log \left (\mbox{sum  
of the finite number of ${e^{(m-1)_i } }$s}\right ) \nonumber \\ 
& = & (m-1)_1 +\log \left ({1\pm e^{(m-1)'_2 } +\cdots } 
\right ), \label{ko26 } \eeqn 
the type (g) expansion can in fact be written in only `one term'  
$\exp (m)_{i +1 }$. 
 
Now, for the part of $i\leq 0$ in $(m)_i$, satisfying $0\leq  
k_{(m)_i }$, $\exp (m)_i$ can be written within the shape of $(m) 
_i$ as the composition of $e^z$ and $(m)_i$, Then we can write for  
(I) 
\beqn y=bz +\sum_{n=2 }^\infty a_nz^n\sim +\sum_{i<0 } 
(\pm )e^{-b_iz^i\sim\cdots }+\sum_{j<0 }(\pm )e^{-e^{c_jz^j\sim\cdots }. 
.. }  \nonumber \\ 
+\sum_{k<0 }(\pm )e^{-e^{e^{d_kz^k\sim\cdots }\cdots } 
\cdots }\cdots , \label{ko27 } \eeqn 
where $b_i, c_j, d_k, \cdots >0$, $\sim$ represents the  
abbreviation of $\log z\sim$, and $\cdots$ the higher order  
terms. The power of $y''/y$ depends on whether $b=0$  
or not, and what is the first of $b_i, c_j, d_k, \cdots $ such  
that the corresponding term is not 0: 
\beqn \frac{y'' }{y }\to\left \{ \begin{array}{l}
n(n-1)\frac{a_n}{b}\;z^{n-m\epsilon -3 }\;\; (b\neq 0\;  
\mbox{and}\;  ^\exists a_n\neq 0, \; n-m\epsilon \geq 2) \\ 
(\pm )\;0\;\; (b\neq 0\; \mbox{and}\; ^\forall a_n=0\; \mbox{and}\;  
 ^\exists b_i\;\mbox{or} \; c_j\; \mbox{or}\; d_k\cdots >0) \\ 
+n(n-1)z^{-2 }\;\; (b=0\; \mbox{and} \;  ^\exists a_n\neq 0) \\ 
+Cz^{2i\pm 2\epsilon -2 }\;\; (b= ^\forall a_n=0\; \mbox{and} \;  ^\exists  
b_i>0) \\ 
+\infty\;\; (b= ^\forall a_n= ^\forall b_i=0\;  
\mbox{and} \;  ^\exists c_j\; \mbox{or} \; d_k\;  
\mbox{or} \cdots >0)\end{array} \right. ,\label{ko29 } \eeqn 
where $C$ is a positive constant and $^\forall b_i=0$ means 
that there is no term in $\sum_{i<0 }$. 
Notice that the first 2 lines of (\ref{ko29 }) can take both signs, 
in contrast that other 3 lines can take only positive sign. 

We therefore find 
\eq\mbox{for (I), }\qquad\nu_y\leq -2 +\epsilon , \;\; -1\leq \nu_y, \en 
where $\epsilon$ represents the power like $\log z\sim$. 
 
\vspace{.1 in} 
It is unclear to us whether there are other cases, but we just
mention that the type (g) expansion is self-contained in the sense that it is
closed under usual operations.
 
\vspace{.2 in} Finally, we consider

\noindent  (CASE 3) $y(z)$ has a non-isolated essential  
singularity at $z=0$. 
 
\vspace{.1 in} 
\noindent (h) Even if we allow complex coefficients in (g),
the discussion above is almost valid. The exception occurs
when $a$ is complex $e^{az }$ shows oscillatory behavior, and  
so $y$ is not monotonic as $z\to 0$ and generally has an accumulation 
point of poles or essential singularities, keeping us away from  
defining $k_y$, $\mu_y$, or $\nu_y$. For example,  
\beqn y=z^5\sin (z^{-1 }) \eeqn  
satisfies the condition of (I) and the term with the  
smallest power in $y$ cancels that of $y''$, yet higher order  
oscillation remains.  
 
\vspace{.1 in} 
There may be cases other than discussed above, but in such cases
$\nu_y$ would not be physical, even if it could be defined.
\subsection{Extension to higher dimensions} 
We can extend the results to dimension $N$ as follows. If we 
assume that the eigen function $y$ is a N-dimensional spherical 
symmetric function $R(r)$ and take the constant term $a=0$,  
(\ref{ko29 }) is clearly replaced by  
\beqn \frac{\Delta R(r) }{R(r)} & = & 
 \frac{R''}{R}+\frac{N-1}{r}\frac{R'}{R} \nonumber \\ 
 & \to & \left \{ \begin{array}{l} +(N-1)r^{-2}\;\; 
\mbox{(or higher order)}\;\; (b\neq 0) \\ 
+n(n+N-2)r^{-2 }\;\; (^\exists a_n\neq 0\; \mbox{is the first term} ) \\ 
+(-ib_i)^2r^{2i\pm 2\epsilon -2 }\;\;\mbox{(or higher order)}\;\; 
(b= ^\forall a_n=0\; \mbox{and} 
\;  ^\exists  b_i>0) \\ 
+\infty\;\; (b= ^\forall a_n= ^\forall b_i=0\; \mbox{and} \;  ^\exists  
c_j\; \mbox{or} \; d_k\;  
\mbox{or} \cdots >0)\end{array} \right. .\label{ko30 } \eeqn 
Notice that we neglected the angular dependence for simplicity.
There is no reason to assume that $R(r)$ is $C^2$-class\cite{Schiff}. 
For the realistic $N=3$ case, the weaker $L^2$ condition to allow logarithmic 
divergence is equivalent to $R(r)$ with the first term $r^{(-3/2)}$ 
or higher order, with the only difference that $-\frac32\leq n$ instead of 
$2\leq n$ in (\ref{ko30 }).
Above results show that for a physical dimension $N=1, 2, 3$, the sign of 
a potential $V\to r^\nu$ with a $C^2$-class $R(r)$ must be positive for 
$\nu\leq -2+\epsilon\;\; (r\to 0)$, but can be negative for other cases.
In fact, a negative potential for $\nu <-2$ can appear, when we take 
$b_i$ pure imaginary in (\ref{ko27 }) and (\ref{ko30 }). 
But this corresponds to the oscillatory behavior of a type (h) expansion 
and may not be distinguished from usual analytic points by finite 
times of measurements.
\subsection{Physical explanation of the results and possible applications} 
The previous results are not mathematically perfect, but show that very  
wide types of functions such that closed in usual operations, 
only by satisfying the second order differential equation, 
can restrict the behavior of the potential. 
Or physically, if there exists a wave function that  
can be applied to every point of the world, the point of nonzero  
charge should also be included in the domain,  
which determines the shape of a force.
 
Notice that the difficulties caused by point-like  
particles may be absent here. If we assume that the existence of  
an eigen function is more fundamental than that of a potential,  
there can be the region where the potential is not defined (where  
the eigen function is 0). Even if analyticity of a
matter field is not a quantity distinguished by finite times  
of measurement, this inevitable ambiguity may be  
the origin of gauge uncertainty\cite{Simon}. 

 Notice also that the type (g) expansion is valid under the 
special rule that we must not decompose an exponent $e^{f(z)}$ for 
$k_f<0$ until the end of the calculation 
(that is, we always maintain the term like 
$e^{-1/z}$ as a single term. As long as $0\leq k_f$, 
the Taylor expansion of $e^{f(z)}$ causes no problem). 
Then a type (g) expansion diverges or converges monotonically 
as $z\to +0$, and hence enables the expansion of  
(\ref{ko27 }) in the order of ascending powers of $z$.
Each expansion has several infinite series of different order.  
Having nonzero `radius of convergence', it can be calculated as  
a usual function. Instead, near $z=0$, if we do not obey the  
rule and try to calculate by extracting all the terms below a certain  
order, the result, even if finite, may depend on the arrangement  
of terms. (It is known in mathematics that an infinite series that 
does not converge absolutely does not always converge to a unique value.) 
This implies an interesting non-commutative property.  
 
Possible physical applications of the results are as follows. 
The first application is to general relativity, where the results show directly  
that in quantum mechanics, an eigen function and a potential  
obey different transformation rules under a nonlinear coordinate  
transformation. This is because not the value of an eigen function $R(r)$ 
itself but the ratio of $R(r)$ to its second derivative $R''(r)$ is important for a potential singularity. The integral constants, i.e., coefficients of the constant and linear terms vanish after differentiation. 
In usual quantum mechanics, the special property of an eigen  
function, that its 0-points are of order one and near them it  
behaves like $\sin x$, saves the potential from divergence. 
We can learn from our results that this paradox leads to an information loss problem and quantum mechanics are valid only in the flat spacetime. 

Another application is to the general theory of renormalization.  
The above consideration explains why some theories nonrenormalizable  
in the usual sense are partially computable. 
The first example of such cases is quantum gravity, where the one  
loop quantum corrections to the Newton potential are determined  
by assuming the Einstein-Hilbert action and the perturbation  
around the flat metric and calculating the effective action\cite{Dono} 
\cite{Doba}\cite{Naka }. 
The result is  
\beqn V(r)=-\frac{Gm_1m_2}{r}\left ({1-\frac{G(m_1+m_2)}{rc^2} 
-\frac{127Gh}{30\pi^2r^2c^3}}\right ),\label{Dono} \eeqn 
where $G, h, c, m_1, m_2$ are respectively the Newton constant,  
Planck constant, the speed of light in a vacuum, and the masses of  
the particles. This naturally contains both two kinds of corrections at the distance. The second term is the Schwarzschild relativistic correction, and the third is the first order quantum correction. The first term in (\ref{Dono})  
is an attractive force and others are repulsive. They correspond to the type (e)  
singularity of the eigen function. Of course, whether or not the assumption  
is valid for very high energy is another story\cite{Wein}, though. 
  
The second example is perturbative QCD, where it is shown that in a confined theory  the poles and branch points of the true Green functions are generated by the  Physical states of hadrons in the unitarity relation, and no singularities related to the underlying quark and gluon degrees of freedom should appear \cite{Oehm }. Detailed discussions about these topics - Borel summation, renormalons, Landau singularities - are in \cite{Capr }, so I only mention here that the Callan-Symanzik equation 
\beqn \mu\frac{d}{d\mu}g_\mu =\beta (g_\mu )\eeqn 
just means that the cutoff scale $\mu (g)$ as the function of a coupling constant,  when differentiated once,  
does not always behave like one with its power smaller by $1$. 
(It seems peculiar that the cutoff scale depends on a coupling constant,  
but the idea of multi-valued coupling constant is interesting.) 
 
Other applications may include the spherical-symmetric part of the  
effective field equation of the Higgs potential, 
where we can extend the potential to the more general functional of  
a scalar field $\phi$ without breaking gauge symmetry.
\subsection{Conclusion for this section} 
 Singularities of a potential $V$ are most likely good indicators 
of new physics at high energies. Thus, it is important to study possible 
singularities and clarify if they are consistent and admissible. 
The theme of this section is to study the asymptotic behavior of a 
singular potential that arises under several frequently occurring 
analytic behaviors of the eigen functions (of the Schr\"{o}dinger 
eigenvalue problem). We never introduce cut-offs. Instead, we 
study the effect of assuming a $C^2$-class eigen function. 
We find that the asymptotic behavior of the singular potential 
crucially depends on the analytic property of the eigen function near the
singular point. The results show that for a one-dimensional eigen function 
$y(x)$, the powers $\nu$ of the asymptotic behavior $V(x)\to x^\nu$ 
as $x\to +0$ satisfy $\nu\leq -2 +\epsilon$ or $-1\leq \nu$, 
where $\epsilon$ represents an infinitesimal positive constant. 
Particularly, $\nu< -2$ can occur when we admit an 
essential singularity of $y$, where the positive sign of $V(x)\;\; (x\to +0)$ 
corresponds to both exponential decay and divergence of $y(x)$ like 
$e^{\pm 1/x}$, and the negative sign to a pathological oscillatory 
behavior like $e^{\pm i/x}$. This section has something to do with the 
theory of computability \cite{Comp,Pen,Simon}. See the next section 
\cite{Mu3} for useful applications. 
\sect{An alternative to Higgs and unification}\label{AHU} 
\subsection{Introduction to this section} 
In usual dimensional counting, a momentum has dimension one. But 
a function $f(x)$, when differentiated $n$ times,  
does not always behave like one with its power smaller by $n$. 
For example, this can occur in the neighborhood of $x=0$ 
if the function $f(x)$ has an essential singularity at $x=0, 
f(0)\to 0 (x\to 0)$. Thus the dimension of a momentum is such an 
$\it operator$ that cannot be fixed unless the operand of the 
differential operator is explicit. This inevitable uncertainty 
discussed in section \ref{Sing} may be essential in general theory of 
renormalization, including quantum gravity\cite{Mu}. 

In this section, the settings to deal with this problem are as follows. 
Let us neglect spins at first and consider a 2 body problem of a pair of 
scalar particles $X, Y$ in the vacuum\footnote{
The vacuum in axiomatic quantum field theory is defined as 
the Poincar\'{e} invariant state where not any kind of a particle 
nor an anti particle is present and satisfies the spectrum condition\cite{AF}
 that every 4-momentum $p^\mu$ as an operator of parallel transformations 
is positive definite $p^\mu p_\mu\geq M^2,\;\; M>0$ and has a positive 
energy $E:=p^0>0$. In particular, the spacetime is uniform, isotropic, and flat. See also Appendix \ref{Spe}.}. 
We assume the spherical-symmetric part of the Klein-Gordon equation
\cite{Schiff} 
\begin{eqnarray}
\Delta R(r) = \left [\frac1{r^2}\frac{d}{dr}(r^2\frac{d}{dr})-
\frac{l(l+1)}{r^2}\right ]R(r) 
= -\frac{(E-e\phi (r))^2-M^2c^4}{{{\hbar}}^2c^2}R(r) &=:& V(r)R(r)
\label{KG}\\
\mbox{with a time-independent spherical-symmetric }U(1)
\mbox{ potential }\qquad A^\mu &:=&(\phi (r), 0, 0, 0), \label{Amu}
\end{eqnarray} 
where $E, M$ are respectively 
the energy and mass of the 2 body system. Notice that here 
$M$ means the reduced mass $\frac{M_XM_Y}{M_X+M_Y}$ in terms of 
the masses $M_X, M_Y$ of the scalar particles $X, Y$. 
$-e$ is simply a constant. It can be the same as the electric 
charge of $X$ if we regard $\phi (r)$ as the electric potential 
generated by $Y$ and $R(r)$ as the spherical symmetric part 
of the system, if both particles $X$ and $Y$ have an electrical 
charge. However, it can be 
only formally the charge of an electron, when we regard $X$ as 
a chargeless photon with its spin neglected. In this approximate 
(or toy model) case, the photon $X$ can naturally interact only 
with the gravitational potential $\phi_G(r)$. 

At first we neglect angular momentums -both spins and orbital 
angular momentums $l$-, which are discussed later in the section \ref{Mom}. 
Let us require the following self-consistent conditions for a realistic gauge field (\ref{Amu}): 
\begin{enumerate}
\item There exists at least one $C^2$-class eigen function $R(r)$ 
satisfying $(1)$ with $l=0$. 
\item The free field part of a Lagrangian is finite for the $R(r)$. 
\item This $R(r)$ in turn creates another potential $\phi '(r)\propto 
R$ and the corresponding force $\propto -$grad $R$.
\end{enumerate}
We can construct some possible powers of the potentials by assuming $R(r)$ 
to be a `generalized Taylor expansion' defined in section \ref{Sing}.
\subsection{Extension to the long distance limit and higher dimensions}
\label{3.2} 
We can extend previous results to $r\to \infty$ case in a spatial dimension $N$ as follows.  
Let us consider a spherical-symmetric Klein-Gordon equation 
with $l=0$ and a time-independent $U(1)$ gauge potential 
$A^\mu :=(\phi (r), 0, 0, 0)$ (only the first time component is 
nonzero and the rest $N-1$ components are $0$), 
\beqn -\Delta R(r)
& = & \frac{(E-e\phi )^2-M^2c^4}{{\hbar}^2c^2}R(r)\nonumber \\
& =: & -V(r)R(r). \eeqn
For simplicity, we assume that 
the eigen function $R(r)$ is a N-dimensional spherical symmetric function $R(r)$. We  change the variable to $z:=\frac{1}{r}$ and assume that $R(z)$ is a 
$C^2$ class function of the type (g) in the previous section \ref{Sing} (expanded as below)  
\beqn R=a+bz  & + & \sum_{n=2 }^\infty  
a_nz^n\sim\cdots +\sum_{i<0 }(\pm )e^{-b_iz^i\sim\cdots } 
\cdots  \nonumber \\ 
 & + & \sum_{j<0 }(\pm )e^{-e^{c_jz^j\sim\cdots }. 
.. }\cdots +\sum_{k<0 }(\pm )e^{-e^{e^{d_kz^k\sim\cdots }\cdots } 
\cdots }\cdots . \label{ko27' } \eeqn 
The previous result (\ref{ko30 }) for $a=0$ and $N\neq 1$ 
\beqn \frac{\Delta R(r) }{R(r)} & = & 
 \frac{R''}{R}+\frac{N-1}{r}\frac{R'}{R} \nonumber \\ 
 & \to & \left \{ \begin{array}{l} +(N-1)r^{-2}\;\; (b\neq 0) \\ 
+n(n+N-2)r^{-2 }\;\; (b=0\; \mbox{and} \;  ^\exists a_n\neq 0) \\ 
+(-ib_i)^2r^{2i\pm 2\epsilon -2 }\;\; (b= ^\forall a_n=0\; \mbox{and} \;  ^\exists  b_i>0) \\ 
+\infty\;\; (b= ^\forall a_n= ^\forall b_i=0\; \mbox{and} \;  ^\exists  
c_j\; \mbox{or} \; d_k\;  
\mbox{or} \cdots >0)\end{array} \right. .\label{ko30'} \eeqn 
  for the short distance limit is clearly replaced by 
\beqn \frac{\Delta R(r) }{R(r)} & = & 
 \frac{1}{R(z)}\left \{{\frac{dz}{dr}\frac{d}{dz} 
\left ({\frac{dz}{dr}\frac{dR(z)}{dz}}\right ) 
+(N-1)z\frac{dz}{dr}\frac{dR(z)}{dz}}\right \} \nonumber \\ 
 & = & z^4\frac{R''(z)}{R(z)}-z^3(N-3)\frac{R'(z)}{R(z)} \nonumber \\ 
 && \hspace{-14mm}\to\left \{ \begin{array}{l} (3-N)\frac{b}{a}z^{3}\;\; (a\neq 0\; 
  \mbox{and} \; b\neq 0\; \mbox{and} \;  N\neq 3) \\ 
(n-N+2)n\frac{a_n}{a}z^{n+2}\;\; \mbox{(or higher order)}\;\;
(a\neq 0\;  \mbox{and} \; b=0\; \mbox{and} \;  
^\exists a_n\neq 0 \; \mbox{and} \;  N\neq 3) \\ 
(n-1)n\frac{a_n}{a}z^{n+2}\;\; (a\neq 0\; \mbox{and} \; ^\exists a_n\neq 0\;  
\mbox{and} \;  N=3) \\ 
(\pm )\; 0\;\; (a\neq 0\; \mbox{and} \; b=^\forall a_n=0 \; \mbox{and} \; 
 ^\exists b_i\; \mbox{or} \; c_j\; \mbox{or} \; d_k\; \mbox{or} \cdots >0) \\ 
(3-N)z^{2}\;\; (a=0\; \mbox{and} \; b\neq 0\; \mbox{and} \; N\neq 3) \\ 
(n-1)n\frac{a_n}{b}z^{n+1}\;\; (a=0\; \mbox{and} \; b\neq 0\; \mbox{and} \;  
^\exists a_n\neq 0\; \mbox{and} \; N=3) \\ 
(\pm )\; 0\;\; (a\;  \mbox{or} \; b\neq 0\; \mbox{and} \; ^\forall a_n=0\;  
\mbox{and} \; ^\exists b_i\; \mbox{or} \; c_j\; \mbox{or} \; d_k\; \mbox{or} \cdots  
>0\; \mbox{and} \;  N=3) \\ 
(n-N+2)nz^{2}\;\; \mbox{(or higher order)}\;\; 
(a=b=0\; \mbox{and} \; ^\exists a_n\neq 0) \\ 
+(-ib_i)^2z^{2i\pm 2\epsilon +2 }\;\; (a=b= ^\forall a_n=0\; \mbox{and} \;  ^\exists  
b_i>0) \\ 
+\infty\;\; (a=b= ^\forall a_n= ^\forall b_i=0\; \mbox{and} \;  ^\exists  
c_j\; \mbox{or} \; d_k\;  
\mbox{or} \cdots >0)\end{array} \right. .\label{ko31 } \eeqn 
{\footnote{The line 2 includes the case $a\neq 0$ and $b=0$ and 
$^\exists a_n\neq 0$ and $N=n+2\neq 3$, when \\
$\frac{\Delta R(r) }{R(r)}\to (m-N+2)m\frac{a_m}{a}z^{m+2}$ or the like, where 
$a_m$ is the term next to $a_nz^n$. \\ 
The line 8 includes the case $a=b=0$ and $^\exists a_n\neq 0$ and $n=N-2$, when \\
$\frac{\Delta R(r) }{R(r)}\to (m-N+2)m\frac{a_m}{a_n}z^{m-n+2}$ or the like, where $a_m$ is the term next to $a_nz^n$.\\
In addition, line 2, 4, 7, 8 include the Yukawa potential case, when \\
$\phi\sim r^le^{-\frac{b_i}{z^i}}$ in (\ref{C3}), the only finite 
solutions are that of the footnote 6, i.e., \\
$\frac{\Delta R(z) }{R(z)}\to \frac{d}{a}{(b_ii)}^2z^le^{-\frac{b_i}{z^i}}\;\; 
\left (R(z)=\left \{{\begin{array}{l}
a+bz^n+dz^ke^{-\frac{b_i}{z^i}}+\cdots\;\; (b=0\; \mbox{if}\; N\neq n+2)\; \mbox{or}\\
az^n+dz^{k+n}e^{-\frac{b_i}{z^i}}+\cdots\;\; (N=n+2)
\end{array} }\right.\right )$, where 
$a, d\neq 0$ and $b_i, i>0$ and $k=2i-l$. \\ 
It is curious that there are some 
`degenerate' eigen functions for the same asymptotic potential, 
even if not normalizable for $N<4$.\\
Finally, this and line 9 allow for a single term with pure imaginary $b_i$, 
except for which an imaginary coefficient of the exponent 
leads to a non-physical oscillatory or imaginary potential. This is indeed 
the case for the Coulomb scattering of a photon.
}}
Noting that $2\leq n$ and $i<0$, we conclude the potential 
 $V(r)$ as $r\to\infty$ must be positive for ($N\leq 3$ and 
$\nu =-2$) or $-2< \nu$, where $\nu$ is the power of the potential 
 $V\to r^\nu$ as $r\to\infty$; can take both signs for 
other cases.
There is no reason to assume that $R(z)$ is $C^2$ class, but more
 natural normalizability condition that $R(r)$ is a $L^2$ function leads
 to small modification $N<2n$ instead of $2\leq n$ in (\ref{ko27' }) 
and so, $a=0$ if $0<N$ and R.H.S. of (\ref{ko31 }) is replaced by
\footnote{It is impossible for the R.H.S. to be
\beqn \to \left \{ \begin{array}{l} 
(n-1)n\frac{a_n}{b}z^{n+1}\;\; (a=0\; \mbox{and} \; b\neq 0\; \mbox{and} \;  
^\exists a_n\neq 0\; \mbox{and} \; 1<n\; \mbox{and} \; N=3) \\ 
(\pm )\; 0\;\; (a=0\;  \mbox{and} \; b\neq 0\; \mbox{and} \; ^\forall a_n=0\;  
\mbox{and} \; ^\exists b_i\; \mbox{or} \; c_j\; \mbox{or} \; d_k\; 

\mbox{or} \cdots >0\; \mbox{and} \;  N=3) \\ 
\end{array} \right. .\nonumber \eeqn 
because $b$ appears. In addition, `higher order' does not appear in case of 
$n\leq 2$ nor $N\leq 4$.\\
For the realistic $N=3$ case, the weaker $L^2$ condition to allow logarithmic 
divergence is equivalent to $R(z)$ of $C^1$ class, with only difference 
that $1<n$ instead of $2\leq n$ in (\ref{ko30 }) and (\ref{ko30 }).}
\beqn \to  \left \{ \begin{array}{l} 
(n-N+2)nz^{2}\;\; \mbox{(or higher order)}\;\; 
\left ({\begin{array}{l}
(a=b=0\; \mbox{and} \; ^\exists a_n\neq 0\; \mbox{and} \; N<2n)\; \mbox{or}\\
(a=0\;\mbox{and} \; ^\exists a_n\neq 0\;\mbox{and} \; \frac{N}{2}<n<1)
\end{array} }\right )\\ 
+(-ib_i)^2z^{2i\pm 2\epsilon +2 }\;\; (a=b= ^\forall a_n=0\; \mbox{and} \;  ^\exists  
b_i>0) \\ 
+\infty\;\; (a=b= ^\forall a_n= ^\forall b_i=0\; \mbox{and} \;  ^\exists  
c_j\; \mbox{or} \; d_k\;  
\mbox{or} \cdots >0)\end{array} \right. .\label{ko32 } \eeqn 
 In this case, the potential must be 
positive for ($\nu =-2$ and $N<n+2$) or $-2<\nu$.
 Notice that (\ref{ko30 }) for more
general cases of $N, a$ can be obtained from (\ref{ko31 }) by the trivial 
replacement $N\to 4-N$ and $z\to r$ with its power smaller by $4$. 
Then, we conclude the potential $V(r)$ as $r\to +0$ must be 
positive for ($1\leq N$ and $\nu =-2$) or $\nu < -2$, where $\nu$ is the power of the potential $V\to r^\nu$ as $r\to +0$; can take both signs for 
other cases. If we assume $R(r)$ is $L^2$ instead of $C^2$, 
$-2n<N$ instead of $2\leq n$, and so by renaming $a_1:=b$ the results are
\beqn & \frac{\Delta R(r) }{R(r)} & \nonumber \\
 & \to & \left \{ \begin{array}{l} 
(n+N-2)n\frac{a_n}{a}r^{n-2}\;\; \mbox{(or higher order)}\;\;
(a\neq 0\; \mbox{and} \; ^\exists a_n\neq 0\; \mbox{and} \;  
0<n) \\ 
(\pm )\; 0\;\; (a\;\mbox{or}\; a_{2-N}\neq 0\;  \mbox{and} \; ^\forall a_n=0\; 
\; \mbox{for} \;n\neq 2-N\; \mbox{and} \; ^\exists b_i\; \mbox{or} \; c_j\; 
\mbox{or} \; d_k\; \mbox{or} \cdots  >0) \\ 
(n+N-2)nr^{-2}\;\; \mbox{(or higher order)}\;\; 
\left ({\begin{array}{l}
(a=0\; \mbox{and} \; ^\exists a_n\neq 0)\; \mbox{or}\\
(^\exists a_n\neq 0\;\mbox{and} \; n<0)
\end{array} }\right ) \\ 
+(-ib_i)^2r^{2i\pm 2\epsilon -2 }\;\; (a=^\forall a_n=0\; \mbox{and} \; 
^\exists  b_i>0) \\
+\infty\;\; (a= ^\forall a_n= ^\forall b_i=0\; \mbox{and} \;  ^\exists  
c_j\; \mbox{or} \; d_k\;  
\mbox{or} \cdots >0)\end{array} \right. .\label{ko34 } \eeqn 
\setcounter{footnote}{1}
\footnote{The line 1,3 include the special case $N=2-n$, when 
$\frac{\Delta R(r) }{R(r)}\to$
$(m+N-2)m\frac{a_m}{a}r^{m-2}$, $(m+N-2)m\frac{a_m}{a_n}r^{m-n-2}$ 
or the like, respectively, where $a_m$ is the term next to $a_nr^n$
such that $m\neq 0$.
In addition, `higher order' does not appear in case of $n\leq -2$ 
nor $4\leq N$.}
Above results show that for a physical dimension $N=1, 2, 3$, 
the sign of a potential $V$ must be positive for
$\nu\leq -2+\epsilon\;\; (r\to 0)$ and
$-2-\epsilon\leq\nu\;\; (r\to\infty)$, but can be negative for other cases.
\subsection{Theorems for the long distance limit} 
Now we define the following conditions for later convenience. \\
The first are normalization conditions naturally required for the 
boson or fermion free field. We require 
the free field Lagrangian to be finite. The $L^2$ condition 
is equivalent to a finite mass term, a little different from our 
present condition to require finite free field terms. The meaning 
and necessity of our condition is discussed in Appendix \ref{L2}. 

 Let us take $F_{\mu\nu}:=\part_\mu A_\nu -\part_\nu A_\mu$ and 
denote the power as $n$ where the field behaves like $r^n$ for 
the corresponding cases $r\to +0$ or $r\to\infty$ on consideration. 
Clearly, 
 {\bf Normalization Conditions for Free fields} are \\
for {\bf Massive Boson fields in the Long distance limit (NC-MBL)},\\
$n\leq -\frac{N+1}{2}$, where ${\cal L}_{\mbox{free}}=
\{ (\part_\mu -ieA_\mu )\phi '\}^\dagger\{ (\part_\mu -ieA_\mu )\phi '\}
-M^2\phi '^\dagger\phi '-F_{\mu\nu}F^{\mu\nu}/4\;\; (M\neq 0)$;\\
for {\bf Massless Boson fields in the Long distance limit (NC-0BL)},\\
$n\leq \frac{1-N}{2}$, where ${\cal L}_{\mbox{free}}=
\{ (\part_\mu -ieA_\mu )\phi '\}^\dagger\{ (\part_\mu -ieA_\mu )\phi '\}
-F_{\mu\nu}F^{\mu\nu}/4$; \\
for {\bf Massive Fermion fields in the Long distance limit (NC-MFL)},\\
$n\leq -\frac{N+1}{2}$, where ${\cal L}_{\mbox{free}}=
\bar\psi\gamma^\mu (\part_\mu -ieA_\mu )\psi -M\bar\psi\psi 
-F_{\mu\nu}F^{\mu\nu}/4\;\; (M\neq 0)$;\\
for {\bf Massless Fermion fields in the Long distance limit (NC-0FL)},\\
$n\leq -\frac{N}{2}$, where ${\cal L}_{\mbox{free}}=
\bar\psi\gamma^\mu (\part_\mu -ieA_\mu )\psi -F_{\mu\nu}F^{\mu\nu}/4$;\\
all of them with the {\bf Exceptional rule for massless particles (NC-0Ex)} 
that \\
{\it an arbitrary constant can be added, with the next term $n\leq -1$ 
for bosons; }$n\leq -3$ {\it for fermions. }\\
For {\bf Boson fields in the Short distance limit (NC-BS)},\\
$\frac{1-N}{2}\leq n$, where ${\cal L}_{\mbox{free}}=
\{ (\part_\mu -ieA_\mu )\phi '\}^\dagger\{ (\part_\mu -ieA_\mu )\phi '\}
-M^2\phi '^\dagger\phi '-F_{\mu\nu}F^{\mu\nu}/4$;\\
for {\bf Fermion fields in the Short distance limit (NC-FS)},\\
$-\frac{N}{2}\leq n$, where ${\cal L}_{\mbox{free}}=
\bar\psi\gamma^\mu (\part_\mu -ieA_\mu )\psi -M\bar\psi\psi 
-F_{\mu\nu}F^{\mu\nu}/4$.\\
For all above cases, $=$ means the critical case of logarithmic divergence. 

The second condition we define for later convenience is the \\
{\bf Positive Potential Condition (PPC)} that \\ 
the potential $V(r)$ defined in (\ref{KG}) (as like in section \ref{Sing}) is positive.

 This is indeed satisfied for the non-relativistic approximation of 
a Klein-Gordon equation (\ref{KG}), if 
\beqn -V & = & \frac{(E-e\phi )^2-M^2c^4}{{\hbar}^2c^2} \label{C1} \\
& \approx & \frac{2M}{\hbar^2}(E-Mc^2-e\phi )<0,\label{C2}\eeqn
where $|E-Mc^2|, |e\phi |\ll Mc^2$. In addition, (\ref{C1}) shows that 
PPC is never satisfied for massless bosons. Indeed, {\bf PPC} means that 
the particle is in the bound state. `$-e\phi$' in (\ref{KG}) means 
the electric charge $-e$ times potential $\phi$ and then, the 
force is defined as $e\nabla\phi$. 

 Notice that in our definition (\ref{Amu}) of a gauge potential $A^\mu$, 
the self-consistent condition $3$ in previous introduction leads to 
another gauge potential and the corresponding field 
\beqn A'^\mu &:=&(\phi '(r),\; 0,\; 0,\; 0),\qquad 
\label{Amu'}\\
\mbox{As }\qquad \frac{\delta F_{\mu\nu}F^{\mu\nu}}{\delta A_\mu} 
&=&4\part_\nu (\part^\mu A^\nu -\part^\nu A^\mu )=-4\Delta A^\mu
\qquad\mbox{for (\ref{Amu})}, \\
\mbox{we can write }&&\{ (\part_\mu -ieA_\mu )\phi '\}^\dagger\{ (\part_\mu -ieA_\mu )\phi '\}-M^2\phi '^\dagger\phi '-F_{\mu\nu}F^{\mu\nu}/4\nonumber \\ 
&&=-F_{\mu\nu}F^{\mu\nu}/4-{F'^\dagger}_{\mu\nu}F'^{\mu\nu}/4-M^2\phi '^\dagger\phi ',\quad \\
\mbox{where }\qquad F'_{\mu\nu}&:=&(\part_\mu -ieA_\mu )A_\nu '
-(\part_\nu -ieA_\nu )A_\mu '.\eeqn 
\indent Remember that for the long distance limit $r:=|x-y|\to\infty$, every correlation function of the two points in a spatial position is decomposed into the multiplication of two linear functions\cite{Wein}:  \\
$<A(x^\mu )B(y^\nu )>\to <A(x^\mu )><B(y^\nu )>$. \\ 
Further if a conserved charge $Q$ is a well-defined operator, all the asymptotic fields transform linearly under $Q$\cite{Kugo}. That is why we focus on the asymptotic behaviors of fields under the linear Klein-Gordon equation (\ref{KG}). 

 Then, from the previous results we can verify the following theorems for the long distance limit. It is unclear to me whether or not previous results are valid for the Dirac equation, but for a moment we keep away the validity as a later discussion and just describe the specific results given by application for fermions by writing in a [ ]. Short distance behaviors and angular momentums are also neglected.\\
{\bf Theorem 1} \\ 
For the higher or smaller spatial dimension $N\neq 3$, a massless boson [or fermion] 
that behaves like $\sim \frac{1}r\;\; (r\to\infty )$ can not feel a dominant 
$\frac{1}{r^2}$-like long range force. \\
{\bf Proof} \\ 
For $N<3$, this is proven by taking $E=M=0$ in (\ref{C1}) and comparing with 
the line 5 of (\ref{ko31 }). The former violates PPC, which contradicts the 
latter condition that $V(+\infty)$ must be positive for $N<3$. For $3<N$, 
this is proven just because a nonzero $b$ in the line 5 breaks {\bf NC-0BL} 
and for $3<N$, that is a weaker condition than {\bf NC-MBL}, {\bf NC-0FL}, 
{\bf NC-0FL}.\\

 A static spherical symmetric electric field like $\sim \frac{1}{r^2}$ is 
of course experimentally observed and therefore, a photon behaves like 
$A^\mu =(\phi (r), 0, 0, 0),\;\;\phi(r)\sim\frac{1}{r}$. An interesting 
corollary of the above theorem is that, if $N<3$, a photon can not feel gravity and there is no gravitational lens! Some people might suspect that we can not 
always take $E=0$ because it means a virtual photon, but 
at least in situations a photon is bounded by the potential and another 
photon is not bounded, we can always take $E=0$ and thus we dare say \\
{\bf Theorem 2} \\ 
For $N>3$, if a charged static spherical symmetric black hole can exist, 
the electric field decreases more rapidly than $\frac{1}{r^{N-1}}$. \\
{\bf Proof} \\ 
Even a black hole has its gravitational potential $\sim\frac{1}{r}$ 
in the distance, for its density is finite and spherical symmetric, 
and obeys Gauss' law and $\frac{1}{r^2}$-law experimentally, 
and gravity is always attractive.  
A black hole is of course such a matter that even a light can 
not escape, therefore a bound state exists for a photon. 
Taking $E=M=0$, the only possible cases are line 5, 8 of (\ref{ko31 }). 
Strictly speaking, there might be exceptions for 
the theorem, where the black hole potential is deviated from 
the $\frac{1}{r}$-law because of the presence of another long range 
force that a photon can feel. In the standard model of particles, 
this is not the case, for no other long range force 
(gluons nor a photon) couple with a photon. Noting that 
if the asymptotic $\frac{1}{r^2}$-law of gravity holds, from (\ref{C1}), 
\beqn 
V(r) & = & \frac{M^2c^4-E^2+2eE\phi -(e\phi )^2}{{\hbar}^2c^2} \nonumber \\
& \to &  \left \{ \begin{array}{l} 
(1)\;\;M^2c^4-E^2\;\; (E^2\neq M^2c^4) \\ 
(2)\;\;2eE\phi\sim \frac{1}{r}\;\; (E^2=M^2c^4\neq 0) \\ 
(3)\;\;(e\phi )^2\sim \frac{1}{r^2}\;\; (E^2=M^2c^4=0) 
\end{array} \right. .\label{C3} \eeqn 
Thus the only ways for the massless photon to allow such a black hole are 
the lines 8, 9 of (\ref{ko31 })
{\footnote{
For $N<3$ (and also for $N=3$ if we do not allow logarithmic divergence), 
this can also be derived from {\bf NC-0BF} without assuming 
the $\frac{1}{r^2}$-law of gravity. For [fermions and] massive bosons, 
more severe than logarithmic divergence appears.}
}
i.e., 
\begin{description}
	\item[(line 8.)] 
This is (3) of (\ref{C3}), where the asymptotic $\frac{1}{r^2}$-law 
of gravity exactly holds and the photon remains massless, but 
the electric field behaves as if away from a polarized matter with 
no electric charge as a whole. The former requires that no 
density is present at distance. Gauss' law is geometric and 
valid in presence of gravity, therefore the latter requires 
real existence of the charge to cancel that of the black hole. 
Up to now all the particles with electric charge are massive, 
therefore the cancellation must be due to the electric charge 
density distributed in a finite region. From (\ref{C3}) 
and {\bf NC-0BL} and (\ref{ko31 }), such a `medium range' force 
can be felt dominant only by such fields that behave like 
($b=0$ and $\frac{N-1}{2}\leq n$). (For other particles, 
Notice that in quantum mechanics, even a particle in an empty 
metal sphere can `feel the outer world'. )
Or
	\item[(line 9.)] 
The electric field vanish exponentially, and $V(r)$ survive slowly 
than $\frac{1}{r^2}$. For each case of (\ref{C3}), 
(1) $i=-1$ (2) $i=-\frac12$ (3) no $i$ allowed.
Thus only possible cases are, either the photon `becomes massive' 
(i.e., $m\neq 0$) or otherwise $E\neq 0$. It is an interesting possibility 
that a massive static photon, that is not bounded because violating PPC, 
can create a $e^{\sqrt r}$-like electric field 
in the former case and even a massless photon can create the
Yukawa-type electric field in the latter case. 
Normalization condition is automatically satisfied for these 
exponentially vanishing solutions. Such a Yukawa-type electric field 
can be felt iff by a $b=0$ massless boson [or fermion] satisfying the
normalization condition. 
\\
 There is yet another possibility that asymptotic $\frac{1}{r^2}$-law 
of gravity changes to survive more slowly, because of 
the long tail of nonzero density the black hole is accompanied with. 

In this case, $i$ can take some negative value $\neq -1$ 
iff ($E^2=M^2c^4$ and $-1< i<0$) or $i<-1$, when the photon creates a neither 
long range nor Yukawa-type but rapidly vanishing electric field. \\
In addition, from (\ref{ko31 }), this is the only case for gluons 
to make $\phi\sim r$ potential at distance, when $i=-2$ regardless of 
$M, E$.
\end{description}
{\bf Theorem 3} \\ 
For $N=3$, a massless boson [or fermion] vanishes more rapidly than  
$\sim \frac{1}{r}\;\; (r\to\infty )$, if it feels a 
$\frac{1}{r^2}$-like dominant long range force. \\
{\bf Proof} \\ 
This is proven by taking $M=0$ in (\ref{C3}) and comparing with the 
line 8, 9 of (\ref{ko31 }), for they are the only cases for 
$\phi\sim\frac{1}{r}\;\; (r\to\infty)$ to exist. \\
\\
Thus, Theorem 2 and its proof hold also for $N=3$, only by adding 
the last of {(\bf line 8)} $<$and {(\bf line 9)}$>$ the following sentences:
`except for the line 6 $<$ and 7$>$ of (\ref{ko31 }), in which only 
logarithmic divergence appears for a massless boson that feels the photon 
which behaves like $b\neq 0$. But it causes a self contradiction 
to identify the massless boson as a photon feeling itself' . \\

 An interesting corollary of Theorem 1, 3 is \\
{\bf Corollary 1} \\ 
A massless gauge boson that feels a dominant 
$\frac{1}{r^2}$-like long range force can not create a long range force. \\ 
\\ 
This is a bit strange, for a photon can not feel $\frac{1}{r}$-like 
potential of gravitational lens. Maybe $\frac{1}{r^2}$-law 
of an electric field is only approximation in presence of gravity, 
or $\frac{1}{r^2}$-rule of gravity is only approximation 
in presence of an electric field, or the coexistence of a graviton 
and a photon leads to a contradiction in present theory and 
gravity should be derived from other forces. 
But this corollary well accounts for the properties of the 
standard model, for if a gluon or a glueball had an 
electric charge, it must be `massive', and if a photon had a color, 
the photon must be `massive', provided an isolated gluon or a glueball 
could be observed,  
In addition, a weak boson has an electric charge (this is followed 
by the experimental fact that an electron is suddenly created and 
comes out in $\beta$-decay), and then it must be `massive', 
regardless of the Higgs mechanism. Thus we come to \\
{\bf Corollary 2} \\ 
If a photon or graviton is massless and the self interactions of 
$W^\pm$ bosons are not so strong as to create a long range force 
which vanish more slowly than $\frac{1}{r^2}$, then the 
$W^\pm$ bosons can not create a long range force. In the same way, 
if a graviton is massless, the glue-balls (if exist) and pions 
with weak enough self interaction can not create a long range force, 
and even if a graviton is massive and a photon is massless, the 
electrically charged glueballs (if exist) and pions with weak 
enough self interaction can not.\\
If the standard model particles are to be unified some day in 
such a manner that a photon is massless and a graviton has an 
electric charge, then the graviton must be `massive'. 
In the same way, if a graviton is massless, then the photon 
must be `massive'. Conversely, 
if a graviton is massless and a photon or gluon is massive, 
then the photon or gluon must be `massive' (this is a tautology). \\

 By the way, from the argument of footnote 6, we can verify also \\
{\bf Theorem 4} \\ 
If a boson [or fermion] feels a dominant Yukawa-type potential, 
then the eigen function of the particle is not $L^2$. Particularly, 
such a boson [or fermion] must be massless for $N<5$; 
can be massive for $5\leq N$. 
\\
{\bf proof} \\ 
This is because if we take $E=Mc^2$ and the Yukawa-type potential 
$\phi\sim z^le^{-\frac{b_i}{z^i}}$ in (\ref{C3}), the only finite 
solutions are that of the footnote 6, i.e., \\
$\frac{\Delta R(z) }{R(z)}\to 
\frac{d}{a}{(b_ii)}^2z^le^{-\frac{b_i}{z^i}}\;\; 
\left (R(z)=\left \{{\begin{array}{l}
a+bz^n+dz^ke^{-\frac{b_i}{z^i}}+\cdots\;\; (b=0\; \mbox{if}\; N\neq n+2)\; \mbox{or}\\
az^n+dz^{k+n}e^{-\frac{b_i}{z^i}}+\cdots\;\; (N=n+2)
\end{array} }\right.\right )$, \\ 
where $a, d\neq 0$ and $b_i, i>0$ and $k=2i-l$, 
R(z)s vanish no more rapidly than $r^{2-N}$, to be compared 
with {\bf NC-MBL}, {\bf NC-0BL}, [{\bf NC-MFL}, {\bf NC-0FL}], 
and {\bf NC-0Ex}. Notice that for $N=3$, a logarithmic divergence inevitably 
appears even for a massless boson in the second case of \{ , and therefore 
this case is impossible for $3<N$. \\
It is surely a severe condition for realistic physics and thus \\
{\bf Corollary 3} \\ 
A short range force must always be dominated by a longer range force, 
or otherwise $N<5$ and not felt by a massive boson [and fermion], 
or otherwise $5\leq N$, \\
where the term `longer' used to notice that any force that survives 
more slowly than any Yukawa potential is allowed. \\

 From now on, we take $N=3$ and concentrate on the self-consistent 
conditions for the standard model. Then, if we do not take account of 
gravity as the dominant force, \\
{\bf Corollary 4} \\ 
A boson with a charge of a short range force must have another 
charge of a longer range force, or otherwise must be massless. \\
\\ 
Thus, the unification of $SU(2)_L\times U(1)$ can be proved without 
assuming experimental results. The corollary well accounts 
for the property of $W^\pm$ [and quarks and charged leptons] 
which are massive and have both $U(1)$ and $SU(2)_L$ charge. 
An equivalent proposition that 
`a particle with no charge of any long range force must not have 
a charge of any short range force, or otherwise must be massless' 
is satisfied for a $Z^0$ and a photon and a $\pi^0$ 
[and almost for neutrinos]. Only if we can neglect gravity$\cdots$.
\subsection{Origin of the Higgs mechanism}
Now, remember that a massless free spin $1$ boson is always 
identified with a photon and creates a $\frac{1}{r^2}$-like 
long range force which is identified with an electric field, 
and gravity couples equivalently to all matter 
\cite{WeinG}. Suppose that a massive boson has a charge. If we 
do not take account of gravity as the dominant force, and 
if it is the charge of a short range force, then from 
{\bf Corollary 2} the boson has a charge of a longer range force. 
Thus, we can naturally assume that a `boson with a charge of 
a short range force' has a charge of a long range force, 
say, electric charge, and call it $W^+$, for a photon of 
course exists and creates a long range force. 
Then, from the $CPT$-theorem, its anti-particle $W^-$ also 
exists. And this is the source of the short range force, 
therefore must be identified with the `photon that feels 
a $\frac{1}{r^2}$-force and can create a Yukawa-type electric field' 
i.e., the line 9 of (\ref{ko31 }), for the Gauss' law 
is valid also for the field of the `reshaped photon' and 
gives the real charge distribution. Notice that in this case, 
there is no way to distinguish whether or not the `reshaped photon' 
is massive, for the $V(r)$ becomes the constant $M^2c^4-E^2$ 
asymptotically, and a massless photon with a positive energy $E$ 
is equivalent to an energy-less photon with the mass $Mc^2:=iE$. 
Thus any photon to feel the same asymptotic nonzero $V(\infty )$ 
can create the same Yukawa-type electric field and therefore can be 
taken to be massless. Indeed, the Klein-Gordon equation 
for a massless photon with the 
$\frac{1}{r}$-potential $\phi_G$ (of gravity) can be solved 
from (\ref{ko31 }) to give 
\beqn 
\frac{\Delta R(r) }{R(r)} & = & -\frac{(E_\gamma -\phi_{G\gamma})^2}{{\hbar}^2c^2} 
\nonumber \\ 
& \sim & -\frac{(E_W-\phi_{GW}-e\phi_{EM})^2-{M_W}^2c^4}{{\hbar}^2c^2} 
\nonumber \\ & \sim & m^2-2mn\frac{1}{r}+n(n-1)\frac{1}{r^2,} 
\nonumber \\
R(r) & \sim & e^{-mr}r^n. \label{C7} \eeqn 
This is in fact the definition of the mass, i.e., 
for a massless and chargeless photon to feel gravity, 
some universal unit for the mass is needed. Here we define 
the unit by the fine structure constant $\alpha :=\frac{e^2}{\hbar c}$. 
Then, we can write $\phi_G:=\frac{e^2 {\cal M}}{r}$, where $\cal M$ 
is some constant proportional to the mass the photon feels. 
In the same way, the electric potential that the $W^+$ feels 
can be written as $\phi_{EM}:=\frac{e{\cal Q}}{r}$, where $\cal Q$ 
is some constant proportional to the electric charge the $W^+$ feels. 
Thus, the initial energy $E_\gamma$ of the massless photon feeling the
`universal potential of gravity' $\phi_G$ 
is equal to the potential that the identified $W^+$ boson 
feels, i.e., asymptotically ${E_\gamma}^2={E_W}^2-{M}_W^2c^4$ 
with $E_W$ and $M_W$ the energy and mass of the $W^+$ boson respectively. 
From the {\bf line 9} of (\ref{ko31 }) with $i=-1$, 
$E_W$ is equal to ${b_{-1}}^2=:-m^2$, which 
in turn creates `reshaped photon' potential $\phi_W \sim\frac{e^{-mr}}{r}$.
In the low energy limit of the photon $E_\gamma =0$, this means that a 
`stopped photon' is just a static electric field. 

{\bf This is the origin of the Higgs mechanism.} 
Therefore, from (\ref{C7}) 
\beqn 
\begin{array}{lll}
{\left \{ \begin{array}{l}
\alpha ({\cal M+Q}) E_W =-mn\hbar c\hspace{14mm}\\
(m\hbar c)^2=-{E_\gamma}^2={M}_W^2c^4-{E_W}^2\end{array}\right.} & 
\Rightarrow & 
{\left \{\begin{array}{l}
-{E_\gamma}^2=(m\hbar c)^2\\
=\frac{{M}_W^2c^4}{(1+\frac{n^2}{{e'}^2})},
\end{array}\right.}

\end{array} \nonumber \\
\mbox{with } {e'}^2:=\{\alpha ({\cal M+Q})\}^2 
\label{G7} 
\eeqn
\\
for $W^\pm$. In the same way, 
\beqn 
\begin{array}{lll}
{\left \{ \begin{array}{l}
\alpha ({\cal M}) E_Z =-mn\hbar c\hspace{14mm}\\
(m\hbar c)^2=-{E_\gamma}^2={M}_Z^2c^4-{E_Z}^2\end{array}\right.} & 
\Rightarrow & 
{\left \{\begin{array}{l}
-{E_\gamma}^2=(m\hbar c)^2\\
=\frac{{M}_Z^2c^4}{(1+\frac{n^2}{{e''}^2})},
\end{array}\right.}
\end{array} \nonumber \\
\mbox{with } {e''}^2:=\{\alpha ({\cal M})\}^2 
\label{G8} 
\eeqn
\\
for $Z^0$. But, wait, the mass of $W^\pm$ must always be pure imaginary! 
Something is wrong$\cdots$
\footnote{In fact, the difficulty is not so serious but rather technical. 
We shall make use of constant ambiguity of a potential to solve the paradox. 
Our construction of a generalized Taylor expansion of the type (g) in previous 
section \ref{Sing} allows no imaginary coefficients of an exponent. However in some special cases, the expansion is also valid (Appendix \ref{L2}).
}.
\subsection{Unification}
Let us keep away the problem of imaginary mass, and consider the original 
eigen function of the photon. The $\frac{1}{r^2}$-rule of gravity may 
not be altered much, for it is always attractive. Then, from 
{\bf Theorem 3} the photon can not create an exactly $\frac{1}{r^2}$-like 
long range force. But it can create {\bf almost $\frac{1}{r^2}$-like} 
medium range force, by taking $n=1+\epsilon$ in the line $8$ 
of (\ref{ko31 }), without violating {\bf (NC-0BL)}. Maybe a radical, but 
not so contradictory solution is just to expand a most general eigen 
function $R(z)$ in a shape like (\ref{ko27' })
\beqn R(z)=a+bz  & + & \sum_{n=1 }^\infty 
a_nz^n\sim\cdots +\sum_{i<0 }(\pm )e^{-b_iz^i\sim\cdots } 
\cdots  \nonumber \\ 
 & + & \sum_{j<0 }(\pm )e^{-e^{c_jz^j\sim\cdots }. 
.. }\cdots +\sum_{k<0 }(\pm )e^{-e^{e^{d_kz^k\sim\cdots }\cdots } 
\cdots }\cdots , \label{U1} \eeqn 
{\footnote{The only deference is that the first infinite sum can start 
from $z^{1+\epsilon}$, which means to assume $C^1$ class $\phi (z)$ 
instead of $C^2$, but causes no problem for a moment.}}
and assume the shape of a graviton $G(z)$ as the first infinite sum 
part of (\ref{U1}) with $a=0\neq b$ and no $i, j, k,\cdots$. 
Then, we can always choose either of the signature of $b$ for a 
graviton without losing generality, and redefine the gravity potential by $\phi_G:=g_{EM}G(z)$, where $g_G$ is the positive coupling constant of gravity. 
Then, the gravity becomes {\it by definition} attractive. 
Let us consider a virtual world in which only gravitons exist. 
Because $\phi_G$ is universal to any particles, it must also be 
satisfied for the graviton. Therefore, 
\beqn\frac{\Delta G(z) }{G(z)} & = & -\frac{(E_G -\phi_G )^2}
{{\hbar}^2c^2}.  \label{U2} \eeqn 
This is a self-consistent condition that resembles the Einstein 
equation in a sense, but from {\bf Theorem 3} there is no solution 
for (\ref{U2}) to create a $\frac{1}{r^2}$-like long range force, 
and (\ref{U2}) leads to $b=0$. This indicates that a graviton must be 
accompanied with another field, say photon. From above discussion, 
we can naturally assume the shape of a photon $A(z)$ as the first 
infinite sum part of (\ref{U1}) with $a=b=0$ (may start from 
$z^{1+\epsilon}$) and no $i, j, k,\cdots$. 
Then, we can redefine the electric potential and gravity potential 
by $\phi_{EM}:=g_{EM}A(z)$ and $\phi_{G}:=g_{EM}bz$, 
where $g_{EM}$ is the positive coupling constant of electricity. 
But in this case, there is no way to 
make $\phi_{EM}$ always positive, as it depends on the relative 
sign of $a_n$ to $b$ in $G(z)$. With these redefinition, 
a graviton and a photon suit well for (\ref{ko31 }) and (\ref{C7}).
Let us abandon (\ref{U2}) and interpret $G(z)$ as a {\it virtual} 
field of the first term $bz$. Then, the only equation for $A(z)$ to satisfy is 
\beqn\frac{\Delta A(z) }{A(z)} & = & -\frac{(E_\gamma -\phi_G )^2}
{{\hbar}^2c^2}, \nonumber \\
\mbox{where } \phi_G & = & g_{EM}bz. \label{U3} \eeqn 
This has a consistent solution for $E_\gamma =0$ as in the 
{\bf (line 8)} of (\ref{ko31 }). 
{\footnote{(\ref{U1}) is closed in this shape of iterative 
expansion and no $i, j, k, \cdots$ appears.}}
\\
Let us come back again to ({\ref{C7}}). Exactly $\frac{1}{r}$-like 
potential comes only from the first $g_{EM}\frac{b}{r}$ term in this equation. 
For $W^\pm$ to make a short 
range force, the solution $W^\pm (z)$ must be the {\bf (line 9)} case of 
(\ref{ko31 }). For almost all the unbounded states, $Mc^2<E_W$. Then, from 
the previous discussion we must take 
$i=-1, |m|=b_i=E_\gamma=\sqrt{{E_W}^2-M^2c^4}>0$ in 
the {\bf (line 9)}. From footnote in subsection \ref{3.2} this `$W^\pm$ boson with an imaginary mass' is just a free photon. Notice that $E_\gamma$ can always be taken 
positive {\it by definition} and negative energy means just a complex 
conjugate. Let us denote these eigen functions of unbounded photons 
(those were not included in the previous expansion because of imaginary 
coefficients) as $A^\pm(z)$, and previous one (with $0$ energy) $A^0(z)$. 
Contrastingly, for a bounded state $E_W<Mc^2$, 
$i=-1, m=|b_i|=|E_\gamma |=\sqrt{{M^2c^4}-{E_W}^2}>0$ and this $W^\pm$ boson 
is really massive. But for $M$ to be real, only pure imaginary $E_\gamma$ 
is allowed, and $m$ must be positive to satisfy normalization conditions. 
With this $i$, $W^\pm (z)$ is identified to be the second infinite 
sum appears in the general expansion (\ref{U1})
\setcounter{footnote}{1}{\footnote{This expansion is also closed in itself if we take $i=-1$, 
in such a sense that no other $i$ appears in the iterative expansion 
$\displaystyle W^\pm(z)=\sum_{k=1,2,3\cdots }^\infty\sum_{l_k} 
e^{-k\frac{m}{z}}z^{l_k}$, except if we consider the special case 
$E_W=Mc^2$ when a half integer $i=-\frac12$ must also be included.
If we assume (\ref{U1}) to be Taylor expanded in $r$, we can avoid 
other eccentric exponential decays with a non-integer $i\neq -1$ 
or appearance of $j,k,\cdots$. }}
. Experimentally, $M_W<M_Z$. Then from (\ref{G7}) and (\ref{G8}) 
${\cal Q}<-2{\cal M}$ or $0<{\cal Q}$, the latter of which 
is in good accordance with the interpretation that gravity is a 
virtual force induced by electric polarization. Suppose that 

there are two metal balls, one is neutral and the other with positive 
or negative electric charge. Then, in both case the two balls will 
attract each another by the induced surface charge. Then, what is 
the meaning of ${\cal M}$ and ${\cal Q}$ ? $\cdots$ well, say, 
the mass and the charge induced on the surface of the universe, 
for they are universal constants and every matter is bounded by gravity 
in the universe. Isn't it an interesting idea? 

 Then, let us consider the eigenfunctions $W^\pm (z)$ and $Z^0(z)$. The only 
difference of them are that $W^\pm$ can feel electricity but $Z^0$ 
can not. Therefore in a theory that includes no graviton like the standard 
model, from {\bf Theorem 4} a massive $Z^0$ boson must be 
written as $Z^0(z)=b_Zz+d_Zz^{1-n''}e^{-\frac{m}{z}}+\cdots$, where 
$b_Z, d_Z\neq 0$, while $W^\pm (z)=\sum_{n=1}^\infty {a_W}_nz^n\sim +\cdots 
+d_W\sum_{k=1,2,3\cdots}^\infty z^{1-n'}e^{-k\frac{m}{z}}+\cdots$, where 
${a_W}_n, d_W\neq 0$ and $n', n''$ are the $n$ s which satisfy (\ref{G7}) and 
(\ref{G8}) respectively. But from (\ref{C7}) this is an altered shape of 
a massless photon only feeling gravity and with energy 
$E_\gamma :=\pm im\hbar c$. 
Above discussion is valid only if $E_\gamma$ can take a specific 
pure imaginary value. This is artificially accomplished by the redefinition of 
$\phi_G$, i.e., taking $G(z)\to G(z)\pm i\frac{m\hbar c}{g_E}$ in
(\ref{U3}). 
In the standard model, neutrinos are exactly massless because their eigen 
functions must be 
\eq \nu (z)=a_\nu +d_\nu z^{-n''}e^{-\frac{m}{z}}+\cdots\label{nu}
\en 
to avoid divergence. 
Strictly speaking, $Z^0(z)$ must also have this shape, i.e., 
\beqn Z^0(z) & = & a_Z +d_Z z^{-n}e^{-\frac{m}{z}}+\cdots \nonumber \\ 
(b=^\forall a_n & = & 0 \;\;\mbox{because it can not feel an electric field) 
and then }
\label{Z} \nonumber \\
W^\pm(z) & = & a_W +\sum_{j=1}^\infty {a_W}_jz^j\sim +\cdots 
+d_W z^{-n}e^{-\frac{m}{z}}+\cdots\nonumber \\
(^\exists{a_W}_j & \neq 0 & \;\;\mbox{because it can feel an electric field), } 
\label{W} \nonumber \\
\mbox{where} & n & \mbox{is the solution of (\ref{G7}) with }
\phi_G=g_{EM}bz
\nonumber \eeqn 
to avoid logarithmic divergence and thus must be massless. 
The only way to allow massive $Z^0$ is then to subtract $a_Z$ from every particles that feel weak force by using the constant ambiguity of a potential. 
This requires the Higgs vacuum energy $<v>\neq 0$. Then, for $Z^0$ 
and $W^\pm$ to feel weak force of the same strength, 
$d_W=d_Z$ and we must replace $-n\to 1-n$ in the term 
$z^{-n}e^{-\frac{m}{z}}+\cdots$. Then, at last gauge bosons [and neutrinos 
and charged leptons (and quarks)] in the 
standard model can be unified (or decomposed) in the following form:
{\footnote{
Usage of these equations: If you want to calculate the two-body 
problem of $X$ and $Y$, use the reduced mass $\frac{M_XM_Y}{M_X+M_Y}$
 for the last term $M_Z,\;\;M_\nu\cdots$ of the numerators. 
On the other hand, for the gravitational potential, we should 
use the total mass ${\cal M}\propto M_X+M_Y$. This is because even 
massless particles can feel gravity and from the symmetry requirement.
Then, the particles obey the Klein-Gordon equation such that both 
particles feel the same potential. Therefore, the eigen function for 
this system can be obtained from $X(z)$, $Y(z)$ such that 
 both expansions contain the same order terms.
$A^0(z)$ must start from just $n=2$ for a massive $W^\pm$ to stop.
${a_W}_n$ and ${a_L}_n$ and ${a_Q}_n$ must start from at least 
$2+\epsilon$ to avoid logarithmic divergence.}} 
\beqn 
R(z) & = & A^0(z)+A^\pm (z)+Z^0(z)+W^\pm (z)+\nu (z)+L^\pm (z)(+G(z)+Q^\pm (z)),\;\;
\mbox{where} \nonumber \\
A^0(z) & := & \sum_{n=2}^\infty {a_A}_nz^n\sim\cdots\;\;\mbox{and}
 \nonumber \\
A^\pm(z) & := & \sum_{k=1,2,3\cdots }^\infty\sum_{l_k=1}^\infty
{d_A}_{kl_k}^\pm e^{\pm ik\frac{\omega}{z}}z^{l_k}\;\;\mbox{satisfy} 
\nonumber \\
\frac{\Delta A(z) }{A(z)} & = & -\frac{(\pm E_\gamma -\phi_{G})^2}
{{\hbar}^2c^2} \;\;\mbox{with} \nonumber \\
E_\gamma & = & 
{\left \{ \begin{array}{l}
im\hbar c\;\; (\mbox{for}A^0(z))\\
im\hbar c+\omega\;\;\mbox{, where }\omega\mbox{ is any positive number}\;\; 
(\mbox{for}A^\pm(z))\\
\end{array}\right.}, \nonumber \\
Z^0(z) & := & (a_Z-bz)\left (\delta_{E_ZM_Z}\pm\sum_{k=1,2,3\cdots }^\infty
\sum_{l_k=1}^\infty 
{d_Z}_{kl_k} e^{-k\frac{m}{z}}z^{l_k}+\cdots\right )\;\;\mbox{and}\nonumber \\
\nu^0(z) & :\approx & a_\nu\left (\delta_{E_\nu M_\nu}\pm\sum_{k=1,2,3\cdots }^\infty
\sum_{l_k=1}^\infty {d_\nu}_{kl_k}e^{-k\frac{m}{z}}z^{l_k}+\cdots\right )
\;\;\mbox{and}\nonumber \\
W^\pm (z) & := & a_W\left (\left (\pm 1\;\;\mbox{or}\;\;\sum_{n=2}^\infty 
{a_W}_nz^n\sim\right )\delta_{E_WM_W}+\sum_{k=1,2,3\cdots }^\infty
\sum_{l_k=1}^\infty {d_W}_{kl_k} e^{-k\frac{m}{z}}z^{l_k}+\cdots\right )
\;\;\mbox{and}\nonumber \\
L^\pm (z) & :\approx & a_L\left (\left (\pm 1\;\;\mbox{or}
\;\;\sum_{n=2}^\infty 
{a_L}_nz^n\sim\right )\delta_{E_LM_L}+\sum_{k=1,2,3\cdots }^\infty
\sum_{l_k=1}^\infty {d_L}_{kl_k} e^{-k\frac{m}{z}}z^{l_k}
+\cdots\right )\;\;\mbox{and}\nonumber \\
Q^\pm (z) & :\approx & a_Q\left (\left (\pm 1\;\;\mbox{or}\;\;
\sum_{n=2}^\infty {a_Q}_nz^n\sim\right )\delta_{E_QM_Q}+
\sum_{k=1,2,3\cdots }^\infty\sum_{l_k=1}^\infty 
{d_Q}_{kl_k} e^{-k\frac{m}{z}}z^{l_k}+\cdots\right )\nonumber \\
 &   & \mbox{or}\;\;
\sum_{k'=1,2,3\cdots }^\infty \sum_{l'_{k'}=1}^\infty {f_Q}_{k'l'_{k'}} 
e^{-k'\frac{M}{z^2}}z^{l'_{k'}}+\cdots\;\;\mbox{satisfy} \nonumber \\
\frac{\Delta Z^0(z) }{Z^0(z)} & = & -\frac{(\pm E_Z-\phi_{GZ}
-\vec{I_Z}\cdot\vec{\phi_W})^2-{M_Z}^2c^4}
{{\hbar}^2c^2}\qquad (\phi_{GZ}=\alpha ({\cal M})bz)\;\;\mbox{and}\nonumber \\
\frac{\Delta \nu(z) }{\nu(z)} & \approx & -\frac{(\pm E_\nu -\phi_{G\nu}
-\vec{I_\nu}\cdot\vec{\phi_W})^2-{M_\nu}^2c^4}
{{\hbar}^2c^2}\;\;\mbox{and}\nonumber \\
\frac{\Delta W^\pm (z) }{W^\pm (z)} & = & 
-\frac{(\pm E_W-\phi_{GW}-e\phi_{EM}-\vec{I_W}\cdot\vec{\phi_W})^2-{M_W}^2c^4}{{\hbar}^2c^2},\qquad ((\phi_{GW}+e\phi_{EM}):=\alpha ({\cal M+Q})bz)
\;\;\mbox{and}\nonumber \\
\frac{\Delta L^\pm (z) }{L^\pm (z)} & \approx & 
-\frac{(\pm E_L-\phi_{GL}-e\phi_{EM}-\vec{I_L}\cdot\vec{\phi_W})^2-{M_L}^2c^4}{{\hbar}^2c^2}
\;\;\mbox{and}\nonumber \\
\frac{\Delta Q^\pm (z) }{Q^\pm (z)} & \approx & 
-\frac{(\pm E_Q-\phi_{GQ}-q\phi_{EM}-\vec{I_Q}\cdot\vec{\phi_W}\pm C_G\phi_S)^2-{M_Q}^2c^4}{{\hbar}^2c^2}, 
\mbox{where}\nonumber \\
\phi_{G} & := & g_{EM}bz+im\hbar c,\qquad
\phi_{G\nu},\;\;
\phi_{GL},\;\;
\phi_{GQ} \propto g_{EM}bz\;\;\mbox{and}\nonumber \\
\phi_{EM} & := & g_{EM}A^0(z)\;\;\mbox{and}\nonumber \\
\vec{\phi_W} & := & g_{EM}
(W^+(z)\delta_{E_WM_W}-a_W, W^-(z)\delta_{E_WM_W}-a_W, 
Z^0(z)\delta_{E_ZM_Z}-a_Z)\;\;\mbox{and}\nonumber \\
\vec{d} & := & ({d_{W^+}}_{11}, {d_{W^-}}_{11}, {d_Z}_{11}),\;\;
{I^3}_Xa_Zg_{EM}:=im\hbar c, \nonumber \\
\frac{{M_Z}^2c^4}{1+(n^2/e''^2)}& = & (m\hbar c)^2 = \frac{{M_W}^2c^4}{1+(n^2/e'^2)}\;\;\mbox{and}
\nonumber \\
M_Z\frac{\vec I_Z\cdot\vec d}{d_Z}_{11} & = & \frac{(m\hbar
c)^2}{2g_{EM}} 
= M_W\frac{\vec I_W\cdot\vec d}{d_W}_{11}
\approx M_\nu\frac{\vec I_\nu\cdot\vec d}{d_{\nu}}_{11}
\approx M_L\frac{\vec I_L\cdot\vec d}{d_L}_{11}
\approx M_Q\frac{\vec I_Q\cdot\vec d}{d_Q}_{11}
\;\;\mbox{and}\nonumber \\ 
\frac{n(n-1)}{eM_W} & = & 2\frac{g_{EM}}{\hbar^2 c^2}{a_A}_2\approx 
\frac{n(n-1)}{eM_L}\approx\frac{n(n-1)}{qM_Q},\;\;\mbox{where}\nonumber \\
{a_W}_nz^n & \sim & \;\;\mbox{is the first term for the stopped} 
\;\;W^\pm ,\;\;\mbox{etc., the gluon field}\nonumber \\ 
G^\pm (z) & := &  \sum_{n=-1}^\infty {a_G}_nz^n\sim +\cdots 
+\sum_{k'=1,2,3\cdots }^\infty \sum_{l'_{k'}=1}^\infty 
{f_G}_{k'l'_{k'}}e^{-k'\frac{M}{z^2}}z^{l'_{k'}}\sim +\cdots
\;\;\mbox{satisfies} \nonumber \\
\frac{\Delta G^\pm (z)}{G^\pm (z)} & = & -\frac
{(\pm E_G-\phi_G-{\vec{C}_G}\cdot\vec{\phi_S})^2}{\hbar^2c^2}, 
\;\;\mbox{where}\nonumber \\
\vec\phi_{S} & := & g_{S}\vec G(z), \;\; 
-\left (\frac{g_s\vec C_G\cdot\vec{a_G}_{-1}}{\hbar c}
\right )^2=(2M)^2, \;\; (\frac{g_{EM}b}{\hbar c})^2=-2.
\label{SSM}\eeqn
Thus, the God made the light at first, or a man can, by defining all the 
coupling constants after $g_{EM}\cdots$.
\subsection{Comparison with the usual standard model}
Let us compare (\ref{SSM}) with the usual standard model Lagrangian 
Appendix \ref{SM}. We can find from (\ref{ZW}) that the Weinberg 
angle $\theta_W$ satisfies 
\beqn \cos^2\theta_W&=&\frac{1+(n^2/e'^2)}{1+(n^2/e''^2)} \qquad
\mbox{and from (\ref{G7}) and (\ref{G8})} \\ 
&=& \frac{({\cal M})^2}{({\cal M+Q})^2}
\frac{\{\alpha ({\cal M+Q})\}^2+n^2}{\{\alpha ({\cal M})\}^2+n^2}, 
\qquad \alpha :=\frac{g'^2}{\hbar c}\eeqn
if our derivation is valid. Here $n$ is the common power of 
the original eigen function (\ref{C7}) of a photon 
\beqn 
R(r) & \sim & e^{-mr}r^n\\ 
\mbox{such that}\qquad\frac{\Delta R(r) }{R(r)} & = & -\frac{(E_\gamma -\phi_{G
\gamma})^2}{{\hbar}^2c^2} \qquad (\phi_{G\gamma}=g'bz)
\nonumber \\ 
& \sim & -\frac{(E_W-\phi_{GW}-e\phi_{EM})^2-{M_W}^2c^4}{{\hbar}^2c^2} 
\qquad ((\phi_{GW}+e\phi_{EM}):=\alpha ({\cal M+Q})bz)\nonumber \\ 
& \sim & -\frac{(E_Z-\phi_{GZ})^2-{M_Z}^2c^4}{{\hbar}^2c^2} 
\qquad (\phi_{GZ}=\alpha ({\cal M})bz)
\nonumber \\ & \sim & m^2-2mn\frac{1}{r}+n(n-1)\frac{1}{r^2.} 
 \label{C7'} \\ 
\mbox{and further from (\ref{defW})}\qquad
\tan\theta_W&=&\frac{g'}{g}.\eeqn 

 As for $W^\pm$, the charge was included from the start by assuming a 
complex boson field. Contrastingly, we must take special linear 
combinations $A^+(z)+A^-(z)$ and $(A^+(z)-A^-(z))/(2i)$ for an eigen 
energy $E_\gamma =\hbar\omega$ to make photons real as observed. 

 It is known that a parity transformation $P$ does not have a 
definite eigen value $\pm 1$ for the weak interaction. However, 
this effect is not observed for bosons and therefore does not 
lead to a contradiction in the present approximation of 
neglecting spins. 

 We introduced imaginary constant potential $(im\hbar c,\; 0,\; 0,\; 0)$
 to explain the Higgs mechanism. This violates the Lorenz covariance. 
However, There are many values not Lorenz covariant in the real world 
(for example, the angle between two coordinate axes).  Indeed, it is natural that inclusion of the mass term into the Yang-Mills 
field violates it. Such a mass term can be regarded as a gauge fixing term 
$-\frac{1}{2\xi}(\part^\mu A_\mu )^2$\cite{TH, FJ} and causes no problem. 
Notice that our main contribution is to derive $SU(2)_L\times U(1)$ unification without assuming a phase transition nor a Higgs particle. In this view, the 
`vacuum energy' $im\hbar c$ plays the role of that of a Higgs field $v$ in 
(\ref{VEV}), and can be treated merely as a c-number, without mentioning 
the Lagrangian of $\Phi$ in (\ref{LH}). Rather, (\ref{LH}) is required to 
provide terms necessary for renormalization of other terms. $v$ is defined as the absolute value of the broken direction, any linear combination of the 4 real components in $\Phi$ coordinates, and thus always positive. 

 As for the vacuum energy and gravity, there is the famous cosmological constant problem. The experimental observations show that $\Lambda$ is less than or on the order of $10^{-120}E_{\mbox{Planck}}^4$. Therefore, the vacuum energy as a quantum effect does not contribute to gravity. This contradiction with the general principle of equivalence\cite{BH, WeinC, Brout, Milt} can be solved in our framework to treat it as the origin of gauge boson masses. See also Appendix \ref{Spe} concerning our interpretation of the Higgs mechanism without assuming a phase transition. 

 As for the validity of {\bf Theorem 1}, we can apply it to a photon in 
the gravitational field, in special case of assuming time independent 
and spherical symmetric potentials $(\phi (r),\; 0,\; 0,\; 0)$, $(\phi '(r),\; 0,\; 0,\; 0)$ respectively for gravity and electromagnetics. We shall review in section \ref{Mom} the fact that even a massless particle can feel gravity. 
In an usual quantum field theory, gauge fields are considered to be created by 
source terms of the corresponding charge, and in particular such effects of 
charged fermions can not be neglected in the real world. However, in our view 
such $\delta (x)$-function like sources do not affect the gauge field in the 
long distance limit. Rather, a charge distribution is subsidiary and derived 
as the singular points of $F_{\mu\nu}$ from the original gauge configuration 
$F_{\mu\nu}$ by solving the Klein-Gordon equation (\ref{KG}) (in general time dependent and not spherical symmetric). Then, $\delta (x)$-function like singularities of a potential $\phi (r)$ and their derivatives do not have a physical meaning, for only distant behavior of $F_{\mu\nu}$ is important. 
Experiments can `observe' from only a finite nonzero distance from a 
charge, and we can not distinguish a quadrapole from a 4 charged particles. Indeed in renormalization theory we neglect constant or higher order terms in momentums of a potential\cite{Wein}, which is equivalent to neglecting $\delta (x)$-function like singularities in the configuration space. 

 An angular momentum $l$ is neglected for above discussion. However, 
it is natural that any field has a $l=0$ solution and the result 
has physical meaning. If we take $l\neq 0$, it has the same effect for 
a photon as decreasing the attractive force of gravity by the repulsive 
centrifugal force. In special case that both the two forces balance and 
$E_\gamma =0$, a photon can not feel any $1/r^2$-like long range 
force and thus behaves exactly like $\sim 1/r+Const.$. 

 In addition, even a chargeless particle never feels `a dominant Yukawa-type force' in this case, for the centrifugal force dominates. 
\subsection{Conclusion for this section}
In this section we classified possible singularities of a potential for 
the spherical symmetric Klein-Gordon equation in the long distance limit, 
assuming that $\phi (r)$ of a time independent spherical symmetric $U(1)$ potential $A^\mu :=(\phi (r), 0, 0, 0)$ has at least one normalizable eigen 
function $R(r)$. In this approximation to neglect angular momentums, $R(r)$s are considered as the eigen function of 2 body problems for every pairs of the elementary particles. Previous analysis in section \ref{Sing} indicates that possible shapes of the potential and the eigen functions of particles are restricted by the consistency condition of this simple model. In particular, a photon that feels a dominant $r^{-2}$ like long range force can not create a dominant long range force. Then we discussed a natural possibility that gravity and weak coupling constants $g_G$ and $g_W$ are defined after $g_{EM}$. 
In this point of view, gravity and the weak force are subsidiary and derived 
from electricity. 
 The fact that the iterative solution inevitably includes several  
infinite series of different order in one expansion may be 
the origin of the non-commutative gauge invariance. 
\sect{The origin of a $U(1)$ gauge phase $\theta (x^\mu )$} \label{Sup}
\subsection{Introduction to this section}
 In this section we shall deal with 2 body problems. In particular, it is important to clarify what kind of degrees of freedom are proper to two point particles. 
 For example, in classical mechanics, a force proportional to the mass of a particle, say gravity, can be divided into two forces: The former is internal and proportional to the reduced mass, and the latter external and proportional to the sum of the two masses. Is this valid for quantum gravity? 
 Another example is spins, where Dirac succeeded in natural description of the interactions between spins and orbital angular momentums in his equation. Spinor representations of the Lorenz group $SO(1,3)$ are double covered, because of its not simply connected spatial rotation subgroup $SO(3)$\cite{Onukispin}, and at least 4 linearly independent and mutually anti-commuting matrices are necessary to represent $SO(1,3)$\cite{Ngspin}. 

 Since we concentrate on pure 2 body problems in an isotropic and uniform spacetime, spatial dependence of all interactions must reduce only to relative positions of the two particles. In this sense, an eigen function of the Klein-Gordon equation is a function of their relative coordinates, and the mass of the same equation is a reduced mass
. 
 So what is the meaning of `relative spins' ?
An accurate physical interpretation of spins and other variables in the Poincar\'{e} group is important, and worth studying.
\subsection{Superposing coordinates}
\begin{figure}
\epsfxsize=450pt \epsfbox{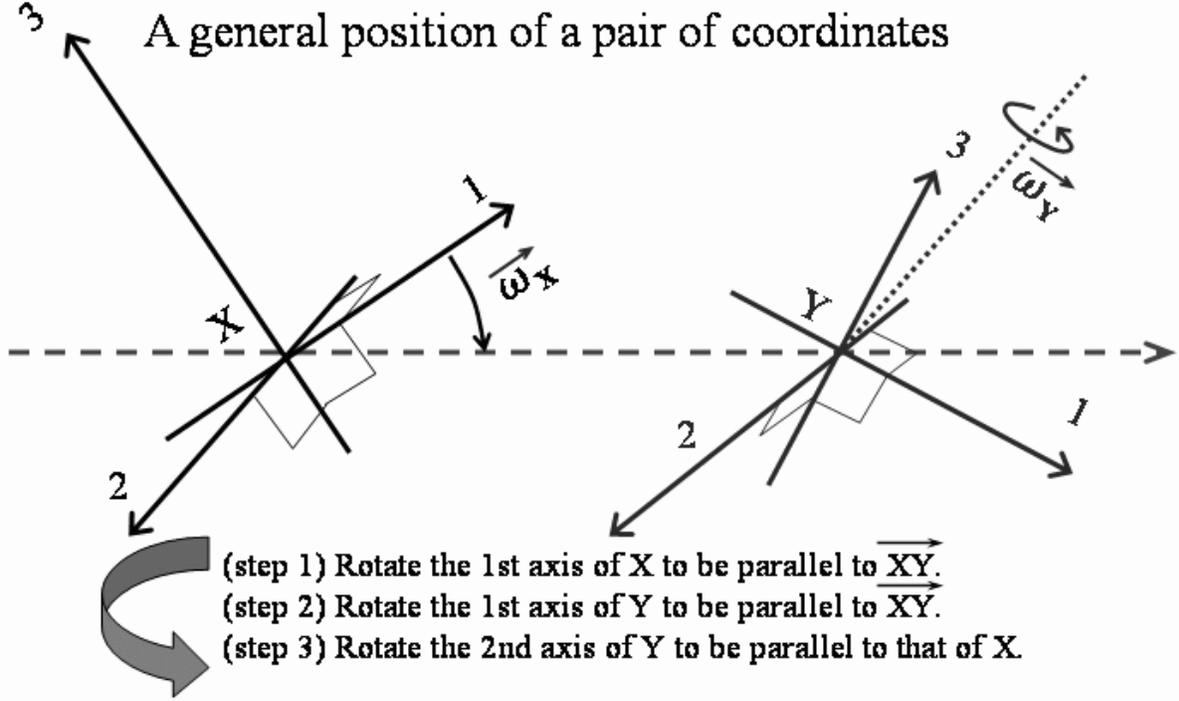}\\
\caption{A general position of two frames.}
\end{figure} 
 Let us begin with two left handed rectilinear frames $X$ and $Y$ in a general but timelike position, $X$ in the past of $Y$. Each of them has an arbitrary direction, velocity and spin (FIG.1). We are observers on a inertial frame. As is well known\cite{Onukispin}, a continuous Lorenz transformation matrix $\Lambda_{\mu}^{\;\;\nu}$ defined by 
\beqn \Lambda_{\rho}^{\;\;\mu}\eta_{\mu\nu}\Lambda^{\nu}_{\;\;\sigma} & = & 
\eta_{\rho\sigma}=\left (\begin{array}{cccc}
1 & 0 & 0 & 0 \\ 0 & -1 & 0 & 0 \\0 & 0 & -1 & 0 \\ 0 & 0 & 0 & -1 \end{array}
\right )\qquad \mbox{(and so, }\Lambda_{\rho\mu}\Lambda^{\mu\sigma}
=\eta_{\rho}^{\;\;\sigma}=\delta_{\rho}^{\;\;\sigma}\mbox{),} \\
\det (\Lambda) & = & 1, \\
\Lambda_{00} & \geq & 1 \label{Lo}\eeqn 
always reduces to the standard form 
\eq \Lambda_{\mu}^{\;\;\nu}=R_{\mu}^{'\;\;\rho}B_{\rho}^{\;\;\sigma}
R_{\sigma}^{\;\;\nu}, \label{R'BR}\en
\begin{figure}
\epsfxsize=450pt \epsfbox{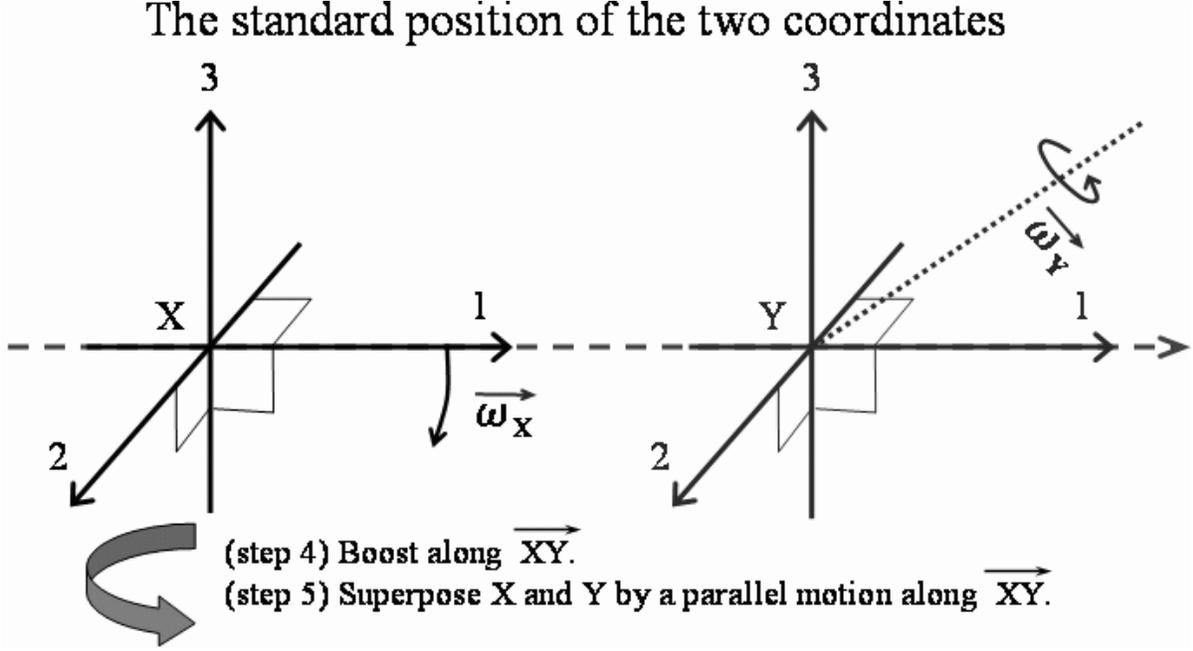}\\
\caption{A transformation procedure to the standard position.}
\end{figure} 
where $R, R'$ are spatial rotation matrices and $B$ is a boost 
along the first axis. Notice that the order of 3 matrices allows us to move $X, Y$ dynamically to the `standard position' illustrated in FIG.2\cite{Uchuyama}. In particular, each of the two rotation matrices acts on $X$ and $Y$ respectively. Not the usual meaning for inertial frames
that `rotate $X$ until its first axis becomes parallel to $\overrightarrow{V_Y}-\overrightarrow{V_X}$ (the relative velocity of $Y$ toward $X$), then boost $X$ to cancel $\overrightarrow{V_Y}-\overrightarrow{V_X}$, and again rotate $X$, then it becomes parallel and static to $Y$' but that `rotate $X$ and $Y$ until both of their first axes become parallel to $\overrightarrow{XY}$ (the vector combining the origins of $X$ and $Y$), then boost $X$ or $Y$ until their relative velocity $\overrightarrow{V_Y}-\overrightarrow{V_X}$ projected onto their first axes becomes zero, and they become parallel and mutually going round'
\footnote{A rotation needs an axis and it is natural to take the axis to be the origin of the rotating frame, in absence of a special reference frame.}. 

 This reminds us of the moon always facing us, in contrast to the sun spinning rapidly to us. If we can not turn it around beforehand nor run faster than light, there exists a threshold time limit to see the profile of a moon keeping the same distance. In the special case that $X, Y$ are `mutually static', the last step for superposing $X$ on $Y$ is a parallel motion to dock them. Notice that this parallel motion has a relation to the boost $B$ of velocity $v(t)$
\footnote{For convenience, we assume $0<v(t)$ during this motion $t(X)<t<t(Y)$.} along the first axis through
\eq \overline{XY}=\int_{X(t)}^{Y(t)}dx=\int_{t(X)}^{t(Y)}v
(t)dt. \label{parallel}\en
 Thus the time interval of $X$ and $Y$ is definite, even if only boundary conditions 
\eq B(t(X))=B(t(Y))=1 \en 
are necessary for $X$ and $Y$ to be mutually static at both ends of the parallel motion. We can perform other rotational transformations $R, R'$ independently in advance, for there is no speed of light restriction for the angular velocity of a spinning point particle. Thus by restricting the order of transformations, we can avoid the problem of noncommutativity.

 In a more general `mutually going round' case, we can always divide the two velocities $\overrightarrow{V_X}, \overrightarrow{V_Y}$ into two components, the center of mass velocity $\overrightarrow{V_G}:=(m_X\overrightarrow{V_X}+m_Y\overrightarrow{V_Y})/(m_X+m_Y)$ and the relative velocity $\overrightarrow{V_r}:=\overrightarrow{V_Y}-\overrightarrow{V_X}$. 
Then, we can express the total angular momentum $\overrightarrow{L}$ 
in a symmetric form as 
\beqn 
\overrightarrow{L}
&:=&\Frac12\;\overrightarrow{XY}\times\{ m_Y
\overrightarrow{V_Y}-m_X\overrightarrow{V_X}\}
\to \frac{\; m_Xm_Y}{m_X+m_Y}\overrightarrow{XY}
\times\overrightarrow{V_r},\nonumber \\
\mbox{where }\quad\overrightarrow{V_X}
&=& \overrightarrow{V_G}-\frac{m_Y}{m_X+m_Y}\overrightarrow{V_r} 
\to -\frac{m_Y}{m_X+m_Y}\overrightarrow{V_r}, \nonumber \\
\overrightarrow{V_Y} &=& 
\overrightarrow{V_G}+\frac{m_X}{m_X+m_Y}\overrightarrow{V_r}
\to \frac{m_X}{m_X+m_Y}\overrightarrow{V_r}
\quad (\overrightarrow{V_r}\cdot\overrightarrow{XY}
=0), 
\eeqn
and thus neglect the not internal, center of mass velocity $\overrightarrow{V_G}$. 
\subsection{Supersymmetry as the Mach's principle to mix spin and orbital angular momentums}
\begin{figure}
\epsfxsize=450pt \epsfbox{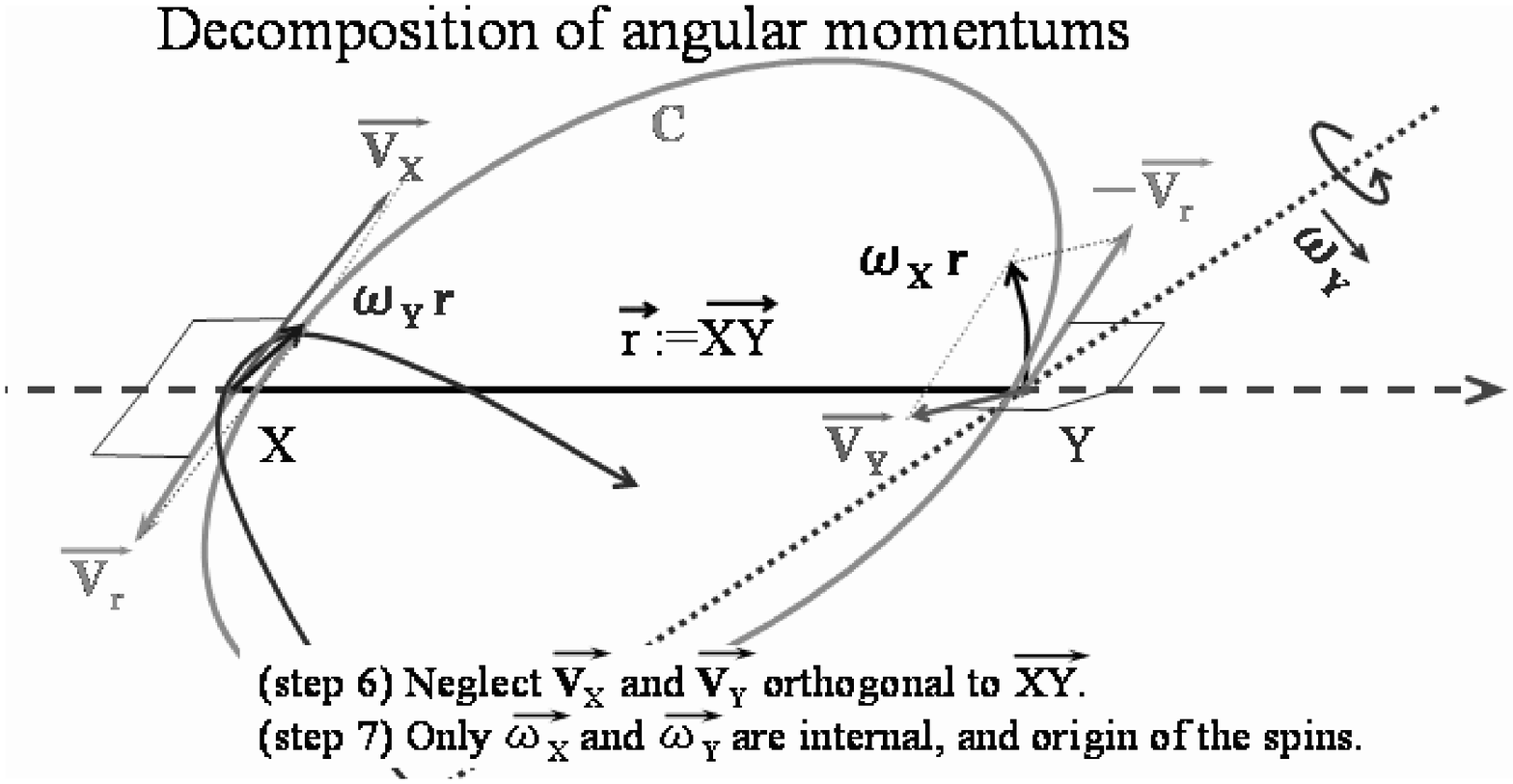}\\
\caption{Decomposition of rotational motions to an orbital angular momentum
 and two spins.}
\end{figure}
We can always divide the two velocities on the 2 dimensional parallel planes orthogonal to $\overrightarrow{XY}$ into three components, 
one around a tangent circle $C$ with the diameter $XY$, and the rest two. To do this, we have only to project out mutually parallel but opposite components of $\overrightarrow{V_X}, \overrightarrow{V_Y}$ orthogonal to $\overrightarrow{XY}$ (FIG.3). The former velocity around $C$, i.e., the pair components of $\overrightarrow{V_X}, \overrightarrow{V_Y}$ rounding along $C$ respectively at the same speed, is just the origin of an orbital angular momentum. The rest two components of $\overrightarrow{V_X}$ and $\overrightarrow{V_Y}$ orthogonal to $\overrightarrow{XY}$ are respectively the origin of the spins of $Y$ and $X$. This can answer the mystery why the quantum number of an orbital angular momentum is integer, while for a spin is half integer. The former comes from a pair motion of $X$ and $Y$, while the latter, the spin of $X$ is equivalent to a motion of $Y$ around the static $X$ and vice versa. We can take any angle $\theta$ between  $C$ and $\overrightarrow{V_Y}-\overrightarrow{V_X}$
\footnote{Here all vectors are orthogonal to $\overrightarrow{XY}$, 
for having been Lorenz boosted along $\overrightarrow{XY}$ in the previous step.}
, for general relativity admits no difference between the rotating frames and the static frames
\footnote{We assumed an isotropic and uniform spacetime, where no external electromagnetic field can exist.}. However, the angular velocity $\omega_X$ of the rotating $\overrightarrow{XY}$ observed from $X$, i.e. time derivative of the angle between $\overrightarrow{XY}$ and the first axis of $X$, is internal and equivalent to the spin $S_X$. 
\begin{figure}
\epsfxsize=450pt \epsfbox{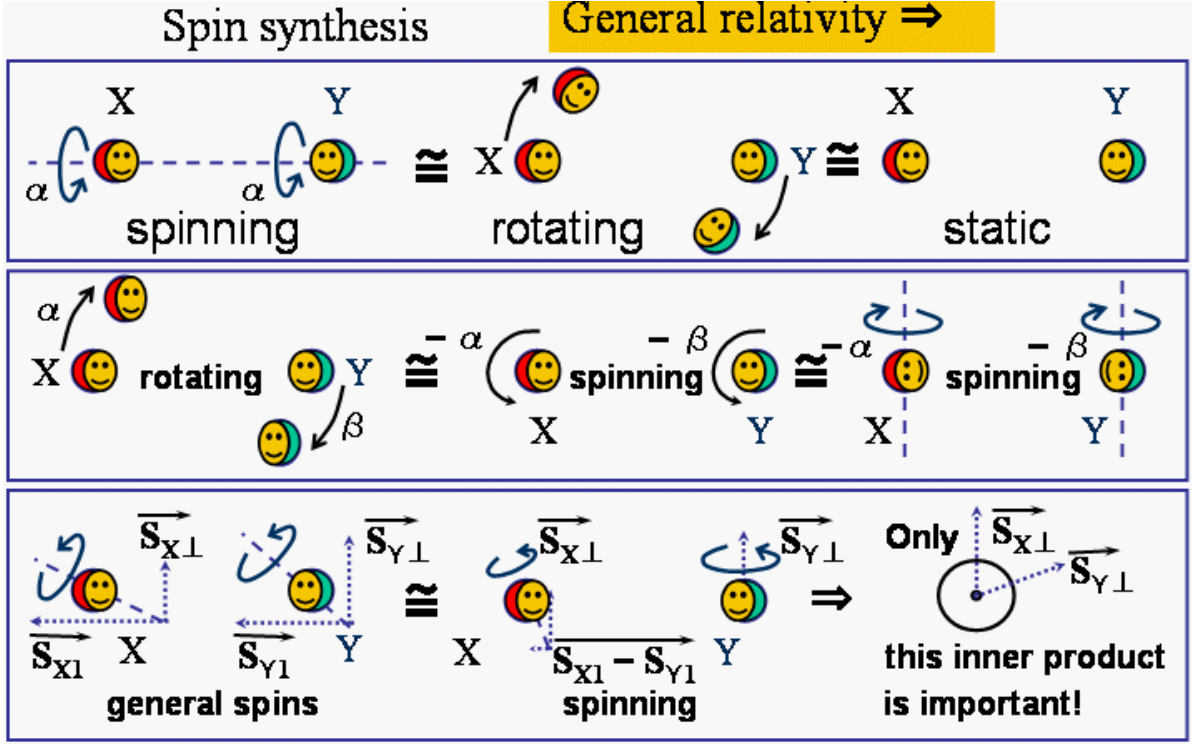}\\
\caption{Equivalent rotational motions of the two frames.}
\end{figure} 
 Then, the (3 dimensional) angular velocity $\overrightarrow{\omega}$ of the rotating $\overrightarrow{XY}$ observed from $Y$, is internal and equivalent to the spin $S_Y$. Thus the only internal scalar product of angular momentums for a 2 body system is $\overrightarrow{S_X}\cdot\overrightarrow{S_Y}$, which observed from $X$ looks like $\overrightarrow{L}\cdot \overrightarrow{S_Y}$ and vice versa. As illustrated in FIG.4, this product is reduced to $S_{X2}S_{Y2}+S_{X3}S_{Y3}$. Furthermore, if both $X$ and $Y$ are spin eigen states along the first axis, the external eigen values $S_{X1}, S_{Y1}$ are reduced to the internal ones $S_{XY1}\in{|S_{X1}-S_{Y1}|, |S_{X1}-S_{Y1}|+1, |S_{X1}-S_{Y1}|+2, \cdots|S_{X1}+S_{Y1}|}$ by acting the ladder operators $S_{\pm}:=S_1\pm iS_2$. The height of a ladder step is $1$, for $S_{\pm}$ acts on $X, Y$ at a time. Such particles, for example two spin 1 particles running along the first axis have a conserved total angular momentum for an external (not spinning) observer, but each of them behaves like a spin 2 or a spin 1 or a spin 0 particle when viewed from the other. $S_{X1}, S_{Y1}\in\{\pm 1, 0\}$ (Exception: Massless particles such as photons have only 2 possible spin eigen values along their momentum direction, namely their absolute spin $S$ or $-S$. 
In this case, total spin degrees of freedom are only $4$, those of a graviton and two massless scalars.)\\

 In summary, the internal (plus external) spatial degrees of freedom for a 2 body problem, both rested on inertial [or possibly spinning] frames, are total 19[+5](+11[+1]):\\
(1) $5 (+1)$ for rotation angles of $R, R'$
\footnote{$2+2$ for making the first axes of $X$ and $Y$ parallel to 
$\overrightarrow{XY}$, $1 (+1)$ for making the second axis of $X$ parallel to 
that of $Y$.}, \\
(2) $1$ for a Lorenz boost along the first axis, \\
(3) $1 (+3+4)$ for the parallel motion of the Poincar\'{e} group, \\
(4) $2$ for two masses, \\
(5) $2[+5] (+[1]+3)$ for two spins [possibly not inertial] or an angular momentum ([except for the component around the axis $\overrightarrow{XY}$] and the center of mass velocity), 
if not quantized
\footnote{Quantization follows from the usual commutation relation $[S_i , S_j]=i\sum_{k=1}^3\epsilon_{ijk}S_k$. A point particle has no moment around itself and so, may have a mass equivallent to its angular velocity: $2mc^2=\hbar\omega$. Then, the corresponding angular momentum around a photon is $mV_r^2/\omega =\hbar /2$. A natural geometric interpretation of the quark flavor mixing angle like this is added in the Conclusion of this thesis.}, \\
(6) $4+4$ for optional rescaling of each of the $4$ axes. \\
 Notice that for a spacetime neither isotropic nor uniform, the total degrees of freedom are $2\times (4+4+3+2+1)+2=30$ for posing each of the origins and 4 axes of the 2 frames and 2 masses, if $6$ of not inertial spins of both frames are neglected. 
If rotational reductions in FIG.4 (Mach's principle) are physically wrong, $2$ for a relative angular momentum are not external but internal, and total internal and inertial degrees of freedom become $19$. This is the case, for in special relativity both transverse and parallel momentums affect the Lorenz contraction. Contrastingly, spinning frames are not inertial frames and do not cause Lorenz contractions. We can test if we are spinning or not by observing the direction and Doppler shifts of sunlight. Thus the spin of the earth can not reduce to the rotation of the universe. In quantum mechanics, these $2$ are internal and appear as a centrifugal force. 

 Let two persons be spinning and rotating to us. Then, identification of the 
nearly equal states in FIG.4 changes spins of each of 
the two person by the Planck constant $\hbar$, times an half integer, 
without changing the relative spin of one person to the other. 
The total angular momentum to us changes by the Planck constant $\hbar$, 
times an integer. This transformation is indeed supersymmetric. 
However, a rotating frame is not an inertial frame. As Galileo 
Galilei verified that the Copernican (or heliocentric) theory 
is correct, the earth is spinning. The Ptolemaic (or geocentric) 
theory is incorrect. Now everyone knows that the earth is spinning, 
not that the universe is rotating around the earth. 
Thus, supersymmetry seems to be a wrong theory to confuse these two 
physically different pictures. 

However, a spin around the axis $\overrightarrow{XY}$ in FIG.4 may not be distinguished in an isotoropic spacetime for point particles. It is interesting to 
interpret this $1$ as the longitudinal component of a photon. In the same way, 
other external 1 of (1) and 7 of (3) as 8 gluons, 3 center of mass velocity as $W^\pm$ and $Z^0$ bosons. Then, each of the 12 external degrees of freedom has the corresponding gauge boson of the standard model! Particulary, 3 center of mass velocity are special in a sense that they influence the apparent angle of spins, i.e., $\bf L\cdot S$ interactions. This can be origin of the broken symmetry and nonzero masses of these 3 bosons, $Z^0$ along $\overrightarrow{XY}$. 7 of (3) are simply parallel motions and thus, commutative in themselves.

In the same way, 19[+5] internal degrees of freedom may have a relation to the 18 parameters of the model. Of which $4+4$ of (6) are probably not physical, for an observer can not accomplish the scale transformation of (6) in a usual motion, unless changing a measure by, say, heating it to expand (which is not realistic for frames generated by the light)! 
\subsection{Spinor representations} 
 Let us consider the infinitesimal transformations of the Poincar\'{e} group. 
We can formally expand an arbitrary function $F(x^\rho )$ of the $4$ dimensional coordinate variables $x^\rho$:
\beqn F(x^\rho +\delta x^\rho )=F(x^\rho )+\delta x^\mu \part_\mu F(x^\rho )+\frac{\delta x^\mu\delta x^\nu}{2}\part_\mu\part_\nu F(x^\rho )+\frac{\delta x^\mu\delta x^\nu\delta x^\sigma}{6}\part_\mu\part_\nu\part_\sigma F(x^\rho )+\cdots
 =e^{\delta x^\mu\part_\mu}F(x^\rho ).\nonumber\eeqn
Then, discussion in the previous section allows us to write 
\beqn B_1(v, t)R_2(\theta_2)R_3(\theta_3)\Sigma_X(x^\mu )&=&R_1(\theta '_1)R_2(\theta '_2)R_3(\theta '_3)\Sigma_Y(x^\mu ), \\
B_1(v, \delta t):=e^{v\delta t\part_1+c\delta t\part_0}\quad (v\leq c),
& &R_i(\delta\theta_d):=e^{\delta\theta_d\epsilon_{ijk}x_j\part_k},\quad (\delta\theta_2, \delta\theta_3, \delta\theta '_1, \delta\theta '_2, \delta\theta '_3)
=(\overrightarrow{\omega_X}, \overrightarrow{\omega})\delta t,
\nonumber \label{sup}\eeqn
where $\Sigma_X, \Sigma_Y$ are respectively the bases of coordinates $X,Y$, $B_1(v, t)$ is the uniform parallel motion along the first axis with a velocity $v$ and a distance $r:=\int_{t(X)}^{t(Y)}v(t)\delta t$, and $R_i(\theta_d)$ is the rotation around the axis $i$ of an angle $\theta_d$. Notice that $B_1(v, t)$ is rather the definition of a time dilation and evolving a spatial symmetric system to Lorenz symmetric one. (\ref{sup}) is valid to order $\delta t$. 

 As usual, we can represent any Lorenz transformation matrix $\Lambda_{\mu}^{\;\;\nu}$ in (\ref{Lo}) by the Pauli matrices\cite{Onukispin}:
\beqn & & 
\sigma_0:=\left (\begin{array}{cc} 1& 0 \\ 0 & 1 \\ \end{array}\right ),\quad
\sigma_1 :=\left (\begin{array}{cc}0 & 1 \\ 1 & 0 \\ \end{array}\right ),\quad 
\sigma_2:=\left (\begin{array}{cc}0 & -i \\ i & 0 \\ \end{array}\right ),\quad 
\sigma_3:=\left (\begin{array}{cc}1 & 0 \\  0 & -1 \\ \end{array}\right ),\\
& & g\sigma_\mu g^\dagger =\Lambda_{\mu}^{\;\;\nu}\sigma_\nu ,\quad\det (g)=1,
\quad g\mbox{ is continuously connected to the unit element }\sigma_0,
\label{sym}\\
& & \mbox{where with the identity } 
M=\frac12 Tr (\overline{\sigma^\mu}M)\sigma_\mu ,\quad 
\overline{\sigma_\mu}:=(\sigma_0, -\sigma_1, -\sigma_2, -\sigma_3),\quad
\sigma^\mu :=\sigma_\nu\eta^{\nu\mu}\\
& & \mbox{ we can expand any }2\times 2 \mbox{ matrix } M \mbox{ and thus}\quad
\Lambda_{\mu}^{\;\;\nu}=\frac12Tr (\overline{\sigma^\nu} g\sigma_\mu g^\dagger ).\eeqn 
The spinor $\sigma_\mu$ itself is a $SO(1, 3)$ covariant vector. Then the infinitesimal boost $\delta v^i$ and rotation $\delta\theta^i$ along or around the axis $i$ turn into a concise form 
\beqn g(\delta v^i, \delta\theta^i)
=\sigma_0+\frac{\pm\delta v^i/c\pm i\delta\theta^i}{2}\sigma_i 
\qquad\mbox{(and so, } g^\dagger\rightleftharpoons\Lambda^\dagger\mbox{)},
\label{inf}\eeqn 
where two $\pm$s correspond respectively to T/P transformations
\footnote{From now on, we take the vacuum velocity of light as $c=1$.}.
This enables us to change the $x^\mu$ dependence of a function into that of the real coordinates. We enjoyed the identities $\{\sigma_i, \sigma_j\} =2\delta_{ij}$ and $[\sigma_i, \sigma_j]=2i\epsilon_{ijk}\sigma_k$ for deriving the boost and rotation in (\ref{inf}) respectively. 
However with $\sigma_i\sigma_j=\delta_{ij}+i\epsilon_{ijk}\sigma_k$ we can represent (\ref{sym}) in an asymmetric form 
(Appendix \ref{sigma}):
\beqn \mbox{\bf Re } g(\delta v^i+\delta u^i, \delta\theta^i+\delta\alpha^i )
\sigma_\mu g^\dagger (\delta v^i-\delta u^i, \delta\theta^i-\delta\alpha )
&=&\nonumber \\
\mbox{\bf Re } g(\delta v^i\pm i\delta\alpha^i, \delta\theta^i\mp i\delta u^i)
\sigma_\mu g^\dagger (\delta v^i\mp i\delta\alpha^i, 
\delta\theta^i\pm i\delta u^i)
&\simeq& g(\delta v^i, \delta\theta^i)\sigma_\mu g^\dagger (\delta v^i, \delta\theta^i) \eeqn 
by neglecting `an imaginary boost or rotation coefficient of $\sigma_\mu$'. 
(\ref{R'BR}) allows us to write 
\beqn \delta\Lambda_{\mu}^{\;\;\nu}\sigma_\nu &=& 
r'(-\theta '^i)b(v^i)r(\theta^i)\sigma_\mu r^\dagger (\theta^i)
b^\dagger (v^i)r'^\dagger (-\theta '^i)\simeq
g\left (\delta v^i, \delta\theta^i-\delta\theta '^i\right )
\sigma_\mu g^\dagger\left (\delta v^i, \delta\theta^i
-\delta\theta '^i\right ) \quad \mbox{and so, }
\nonumber \\ 
&\simeq&\mbox{\bf Re }[\mbox{\bf  Re}\{ b\;\;\mbox{\bf Re }(r\sigma_\mu )\} 
r'^\dagger ]=\mbox{\bf Re  }g\left (2\delta v^i, 2\delta\theta^i\right )
\sigma_\mu g^\dagger\left (0, -2\delta\theta '^i\right )=:\sigma_\mu +\delta\sigma_\mu \label{dR'BR}\eeqn 
for the infinitesimal form of (\ref{R'BR}) and the spin interpretation via coordinates rotations. Here $b, r, r'$ are infinitesimal Lorenz transformation 
generators of a boost $B(v^i)$ and rotations $R(\theta ^i), R'(\theta '^i)$ respectively, and in particular with (\ref{sup}), we can write $\delta v^i=(\delta v, 0, 0), \delta\theta^i=(0, \delta\theta^2, \delta\theta^3)$. This approximation is valid up to the first order of infinitesimal parameters. Notice that because 
\beqn \delta\sigma_\mu &:=&
\left (\sigma_i\delta v^i, \sigma_0\delta v^i+\epsilon_{ijk}\sigma_j(\delta\theta +\delta\theta ')^k\right )+i\left (\sigma_i(\delta\theta +\delta\theta ')^i, \sigma_0(\delta\theta +\delta\theta ')^i+\epsilon_{ijk}\sigma_j\delta v^k\right )\label{imsigma}\\
&=&\left (\begin{array}{cccc}
0 & \delta v^1+i(\delta\theta^1+\delta\theta '^1) & \delta v^2+i(\delta\theta^2+\delta\theta '^2) & \delta v^3+i(\delta\theta^3+\delta\theta '^3)\\
\delta v^1+i(\delta\theta^1+\delta\theta '^1) & 0 & \delta\theta^3-\delta\theta '^3+i\delta v^3 & -\delta\theta^2+\delta\theta '^2-i\delta v^2\\
\delta v^2+i(\delta\theta^2+\delta\theta '^2) & -\delta\theta^3+\delta\theta '^3-i\delta v^3 & 0 & \delta\theta^1-\delta\theta '^1+i\delta v^1\\
\delta v^3+i(\delta\theta^3+\delta\theta '^3) & \delta\theta^2-\delta\theta '^2+i\delta v^2 & -\delta\theta^1+\delta\theta '^1-i\delta v^1 & 0
\end{array}\right )
\;_\nu\;^\mu 
\sigma_\nu ,\nonumber \eeqn 
the neglected imaginary part {\bf Im }$\delta_\mu$ involves a term proportional to $\delta\theta^i+\delta\theta '^i$. Thus, not only a real boost $\delta v^1$ and a rotation $\delta\theta^i-\delta\theta '^i$ but also the second component of two imaginary boosts $\delta\theta^2+\delta\theta '^2$, $\delta\theta^3+\delta\theta '^3$ are necessary to decide all internal degrees of freedom. We can propose this to be the origin of a $U(1)$ gauge phase.

 We can naturally regard $Y$ as the coordinates to represent time development of $X$, or in other words, a final state of the initial state $X$. Then, a symmetric spinor transformation (\ref{sym}) stands for a common motion of the whole system $X, Y$ toward the observer, which can equivalently be reduced to the motion of the observer. While an asymmetric transformation (\ref{asym}) stands for the motion of each of $X, Y$ or the relative motion of $X$ toward $Y$
\footnote{The situation is like a domino game arranged on a circle. If a perfect set of states $|n><m|=\delta_{mn}$ are given, the action of a unitary group element $g=(g^{-1})^\dagger$ on an operator matrix element $<m|AB|n>$ is $<m|g^\dagger Agg^\dagger Bg|n>=<m|g^\dagger ABg|n>=<gm|AB|gn>$. On laying down the first token of dominos by hand, all other tokens turn around at the same angle. This process does not necessarily be restricted by the vacuum velocity of the light, for the motion can be absorbed to that of an external observer $g|n><m|g^\dagger =|n><m|$. And if $g_A, g_B$ are different group elements respectively acting on $A, B$, the result is $<m|g_A^\dagger Ag_Ag_B^\dagger Bg_B|n>=<g_Bm|g_Bg_A^\dagger Ag_Ag_B^\dagger B|g_Bn>$. Thus $g_A, g_B$ are reduced to act only on $A$ after the transformation of an external observer. This corresponds to a case the tokens turn around at respectively different angles. It may spend no time in case of a nonphysical gauge group, while for the Poincar\'{e} group with nonzero parallel motion must spend. As Lorenz boosts are not unitary, do not satisfy $g^\dagger =g^{-1}$ and we must decompose a scalar operator $AB$ into a form like $A'^\mu g_{\mu\nu}B'^\nu$ and rewrite $|n><m|=\eta_{mn}$, when any element $g$ of the Poincar\'{e} group acts like $<m|g^\dagger A^\mu gg_{\mu\nu}g^\dagger B^\nu g|n>=<m|g^\dagger ABg|n>=<gm|AB|gn>$. In general, the result depends on what types of a Lorenz tensor $AB$ is. 

Notice that (\ref{imsigma}) involved an imaginary term proportional to $\delta v_X^i-\delta v_Y^i$ if we decomposed $2\delta v^i=:\delta v_X+\delta v_Y$ and calculated $g\left (\delta v_X^i, 2\delta\theta^i\right )\sigma_\mu g^\dagger\left (\delta v_Y^i, -2\delta\theta '^i\right )$ instead. The difference between the superfluous term $\delta v_X^i-\delta v_Y^i$ and $U(1)$ gauge generator $\delta\theta^i+\delta\theta '^i$ is that the former can be absorbed to the motion of an external observer, while the latter can not (cf.FIG.1), as long as for a nonzero distance $\overline{XY}\neq 0$.}
. An observer $Y$ can locally Lorenz transform to become parallel and static to $X$, and next reduce the transverse or parallel motion of $X$ toward $Y$ to a spin or boost of $Y$, and finally, the spin of $X$ to a transverse motion and a spin around the axis $XY$ of $Y$. Then, the only possible accelerated motions of $X$ are $\frac{d}{dt}(\omega_2, \omega_3)=\frac{d^2}{dt^2}(\theta_2, \theta _3)$
\footnote{Strictly speaking, $\omega_2, \omega_3$ are the components of $\omega_{\bot}$ in FIG.4 measured by the $2, 3$ axes of the rest frame of $X$. $\theta_2$ is not a scalar, but we can correct the value $\theta_{Y2}$ of $\theta_2$ observed by $Y$ to that by $X$ as follows and thus calculated $\theta '_2$ is a scalar. Let a scientist $Y$ find the angle between the first axis of the coordinates $X$ and $\overrightarrow{XY}$, both on the same plane $p$, to be $\theta_{Y2}$. Then, $\sin\theta_{Y2}$ is the ratio of the lengths of light rays from $X$ emitted in the following 2 ways: The former along the first axis of $X$, and the latter along the vector on $p$ orthogonal to $\overrightarrow{XY}$ (so that we will call the direction `height'). Both lights are assumed to turn back at the same height ($X$ probably does not `know' the direction of future $\overrightarrow{XY}$, so $X$ should be enclosed in a half mirror sphere with the light source!). This ratio is Lorenz invariant under the local motion of $Y$.}, which from the idea of interactions through medium, must be due to local fields at $X$. All other motions of $X$ must be subsidiary and derived from them by a local Lorenz transformation of $Y$ and due to local fields at $Y$. Furthermore, we can naturally assume that the interactions for $X$ are described by the second order linear differential equations with local internal positional variables of $X$ and the external fields initially generated at future of $X$ and propagated toward $X$. The former possible variables are only $(\omega_2, \omega_3), (\theta_2, \theta_3)$. The latter parameters for the fields are $t$. 

 Then, possible shapes of the classical equations of motion for $Y$ are
\beqn \frac{d^2}{dt^2}\left (\begin{array}{c}
\theta_2 \\ \theta_3\end{array}\right )=\left (\begin{array}{cc}
a_{11} & a_{12} \\ a_{21} & a_{22}\end{array}\right )
\frac{d}{dt}\left (\begin{array}{c}
\theta_2 \\ \theta_3\end{array}\right )&+&
\left (\begin{array}{cc}
b_{11} & b_{12} \\ b_{21} & b_{22}\end{array}\right )
\left (\begin{array}{c}
\theta_2 \\ \theta_3\end{array}\right )=:A(t)\dot{\vec{\theta}}+B(t){\vec{\theta}}\label{omegaeq}\\
\hspace{-20mm}\mbox{with the initial conditions}\quad\dot{\overrightarrow{\theta_0}}:=\left (\begin{array}{c}\frac{d}{dt}\theta_2(0) \\ \frac{d}{dt}\theta_3(0)\end{array}\right )&,&
\overrightarrow{\theta_0}:=\left (\begin{array}{c}\theta_2(0) \\ \theta_3(0)\end{array}\right ),\label{init}\eeqn 
which are just equations of simple oscillations. Taking $\overrightarrow{\theta_0}=\overrightarrow{0}$, and $\alpha , \beta$ to 
be the eigenvalues of $A(t)$, an eigen vector ${\overrightarrow{\omega_1}}$ satisfies 
$\frac{d\overrightarrow{\omega_1}}{dt}=\alpha{\overrightarrow{\omega_1}}
={\overrightarrow{C_1}\alpha}e^{\alpha t}$ and the other 
$\frac{d\overrightarrow{\omega_2}}{dt}=\beta{\overrightarrow{\omega_2}}(+{\overrightarrow{\omega_1}})
=(\overrightarrow{C_2}(+\overrightarrow{C_1}))e^{\beta t}$.
For $\theta_2, \theta_3$ to be always real, the $2\times 2$ complex matrix $A$ is reduced to 
\begin{description}
\item{(1): if not always $\overrightarrow{\omega_1(t)}\propto\overrightarrow{\omega_2(t)}$,} real i.e., $a_{ij}\in$ {\bf R},\quad $i,j\in\{ 1, 2\}$, 
\item{(2): if always $\overrightarrow{\omega_1(t)}\propto\overrightarrow{\omega_2(t)}$,} taking $\overrightarrow{\omega}=(1, C)^Te^{\alpha t}$, \quad $C, \alpha $ must be real and \\
$A=\left (\begin{array}{cc}a +ixC & b-ix \\ aC+ixC^2 & bC-iCx\end{array}\right ), \quad a, b, x\in {\bf R}, \qquad a+bC=\alpha$ .
\end{description}
 Both (1) and (2) involves 4 real degrees of freedom and satisfies tr $A, \;\;\det A\in$ {\bf R}. Therefore, (i) $\alpha =\beta^*\not\in$ {\bf R} or (ii) $\alpha , \beta\in$ {\bf R}, and solutions are exponentially growing or decaying, superpositions of two modes conversely rotating around the first axis. Let us examine more precise geometrical properties of the solutions for (\ref{omegaeq}). First assume $\det A=\pm 1$ for (i). Then $A$ can be diagonalized and the rest 3 degrees of freedom correspond respectively to the rotation angle, the deformation angle and the side ratio from a square lattice to a parallelogram. Therefore, in this case general trajectories of solutions are ellipses. If $\det A\neq\pm 1$, they are logarithmic spirals. Second, for case (ii) with $\alpha \neq\beta$, the solutions are simple superpositions of the expansion or the shrink along the two eigen directions. If $\det A=1$, the trajectories are hyperbolics. 
The physical meaning of the special mode like $\propto te^{\alpha t}$ possible in degenerate cases of (ii) is not clear, but the above results for (i) remind us of a spin rotating around a magnetic field $\vec{B}$. 
If $\det A=1$, the spin vector $\vec{S_X}$ is constantly rotating around $\vec{B}$, keeping the length of itself and the angle to $\vec{B}$. This same system, if viewed from the rest frame of $X$, is equivalent to $Y$ rotating around the spin axis of $X$ which is also constantly rotating. If $\det A\neq 0$, the decay rate $\alpha$ corresponds to a time dependence of $\vec{B}$ or equivalently, to $Y$ accelerated along the direction orthogonal to $\overrightarrow{XY}$, and the accelerating field itself is rotating around $\overrightarrow{XY}$. 
\subsection{Poincar\'{e} invariant equations} 
The only possible second order linear differential equation, Lorenz invariant for a scalar field $\phi$ is the Klein-Gordon equation ($\hbar =c=1$)
\eq (\part_\mu\part^\mu +m^2)\phi =0, \label{KG!}\en 
where each solution $\phi$ has a phase ambiguity, comes from the fact only the absolute value $\int |\phi (x^\mu )|^2d^4x^\mu$ has a physical meaning\cite{Dirac}, if the phase $\theta (x^\mu )$ is a real function. As is well known, this $U(1)$ gauge transformation $\phi\to\phi ':=\phi e^{i\theta},\;\; \theta\in{\bf R}$ gives another equation
\beqn (\part_\mu\part^\mu +m^2)\phi ' & = & e^{i\theta}\{ (\part_\mu +i\part_\mu\theta )^2+m^2\}\phi \nonumber \\ 
 & = & e^{i\theta}\{\part_\mu\part^\mu +i(\part_\mu\part^\mu\theta )
+2i\part_\mu\theta\part^\mu-\part_\mu\theta\part^\mu\theta +m^2\}\phi =0. \label{KG!'}\eeqn 
Such an unitary transformation does not affect the distribution of a solution $\phi$ of (\ref{KG!'}), but arises the concept of a gauge field $A_\mu :=\part_\mu\theta$ acting as a vector potential on $\phi$ and only after differentiated induces real observable fields.

 If we use the $\gamma$ matrices 
\beqn \gamma_
\mu:=\left (\begin{array}{cc}
0 & \sigma_\mu\\ \overline{\sigma_\mu} & 0\end{array}\right )\mbox{ and then, }
\quad\{\gamma_\mu , \gamma_\nu\}:=\gamma_\mu \gamma_\nu +\gamma_\nu \gamma_\mu =2\eta_{\mu\nu}, \label{anticom} \eeqn
cross terms of $(\part_\mu\gamma^\mu )^2$ vanish to give a d'Alambertian $\part_\mu\part^\mu$ of (\ref{KG!}) and 
\beqn \{(\part_\mu\gamma^\mu )^2+m^2\}\phi &=&(i\part_\mu\gamma^\mu +m)(-i\part_\nu\gamma^\nu +m)\phi =(-p_\mu\gamma^\mu +m)(p_\nu\gamma^\nu +m)\phi =0, \nonumber \\
&\Leftarrow & (p_\mu\gamma^\mu -m)\phi =0\mbox{ or }(p_\mu\gamma^\mu +m)\phi =0.\label{KG!''}\eeqn 
where the anticommutation rule of (\ref{anticom}) is essential and $\gamma_\mu\;\; (\mu\in\{0, 1, 2, 3\})$ do not necessarily be $4\times 4$ matrices representations. However the $\phi$ in (\ref{KG!'}) is no longer a scalar but divided into a pair classes of $n$ component spinors, each corresponds to the negative or positive energy solution of (\ref{KG!'}), where $n$ is the dimension of the matrix representation $\gamma_\mu$. The number of linearly independent solutions in each class is $2$ for $SO(1, 3)$, which just says that the spin $\frac12$ particle or antiparticle can have any momentum $p_\mu$ and up or down spin eigen states along it (i.e. be left or right handed), and any solution of (\ref{KG!'}) with $p_\mu$ is a superposition of the 2 ($\times$ 2 for $\pm E$) eigen states. In particular if $m=0$, a particle and its antiparticle satisfy the same equation, but (anti) Hermicity of $\gamma$ matrices guarantees the existence of $4$ independent solutions. 
For the standard model, a photon and a $Z^0$ boson 
does not have any charge, which are necessary conditions for the equivalence of the particle and its antiparticle. 

 An explicit solution of (\ref{KG!'}) is then
\setcounter{footnote}{1}
\footnote{The sum on spins $\pm$ allows for any linear combination of $\pm$ spins.}
\beqn \phi (x^\mu )&=&\sum_\pm\int_{-\infty}^\infty \left (u(p^\mu ,\pm )e^{p_0x^0+ip_j x^j} +v(p^\mu ,\pm )e^{p_0x^0-ip_j x^j}\right )\delta (p^\mu p_\mu )d^4p^\mu ,\nonumber \\
\mbox{where } (u(p_\mu ,+), u(p_\mu ,-))&:=&\left (\begin{array}{l}\sqrt{p_\mu\overline{\sigma}^\mu}\\ \sqrt{p_\mu\sigma^\mu}\end{array}\right ), \quad (v(p_\mu ,+), v(p_\mu ,-)):=\left (\begin{array}{l}\sqrt{p_\mu\overline{\sigma}^\mu}\\ -\sqrt{p_\mu\sigma^\mu}\end{array}\right ) \mbox{ and thus satisfy }\nonumber \\
\mbox{ the orthonormal conditions }&&\overline{u(p_\mu ,s)}u(p_\mu ,s')=2m\delta_{ss'}, \quad\overline{v(p_\mu ,s)}v(p_\mu ,s')=-2m\delta_{ss'}, \nonumber \\
&&\overline{u(p_\mu ,s)}\gamma^\mu u(p_\mu ,s')=\overline{v(p_\mu ,s)}\gamma^\mu v(p_\mu ,s')=2p^\mu\delta_{ss'}, \nonumber \\
&&u^\dagger (p_\mu ,s)v(-p_\mu ,s')=v^\dagger (p_\mu ,s)u(-p_\mu ,s')=0,\quad\mbox{ and }\nonumber \\
\mbox{the perfectness conditions } &&\sum_\pm u(p_\mu ,s)\overline{u(p_\mu ,s)}=\not p+m,\quad\sum_\pm v(p_\mu ,s)\overline{v(p_\mu ,s)}=\not p-m, \nonumber \\
&&\hspace{-56mm}\mbox{where }\left (\begin{array}{l}\overline{u(p_\mu ,+)}\\
\overline{u(p_\mu ,-)}\end{array}\right )
:=\left (\begin{array}{l}u^\dagger (p_\mu ,+)\\
u^\dagger (p_\mu ,-)\end{array}\right )
\gamma_0=\left (\begin{array}{l}\sqrt{p_\mu\sigma^\mu}\\ \sqrt{p_\mu\overline{\sigma}^\mu}\end{array}\right ), \\
\left (\begin{array}{l}\overline{v(p^\mu ,+)}\\ \overline{v(p^\mu ,-)}\end{array}\right )&:=&\left (\begin{array}{l}v^\dagger (p_\mu ,+)\\ v^\dagger (p_\mu ,-)\end{array}\right )\gamma_0=\left (\begin{array}{l}\sqrt{p_\mu\sigma^\mu}\\ -\sqrt{p_\mu\overline{\sigma}^\mu}\end{array}\right ).
\eeqn
A general integer spin $n$ particle $\phi_{\mu_1\mu_2\cdots\mu_n}$ of mass $m$ satisfies the following  covariant
\beqn\mbox{\bf Fierz-Pauli equations: }\qquad\left (\part_\mu\part^\mu +m^2\right )\phi_{\mu_1\mu_2\cdots\mu_n}&=&0,\label{F-P}\\ 
\part_\mu \phi_{\mu\mu_2\cdots\mu_n}&=&0,\\ 
\phi_{\mu\mu\mu_3\cdots\mu_n}&=&0.\eeqn
In the same way, a general half-integer spin $n+\frac12$ particle $\phi_{\mu_1\mu_2\cdots\mu_n}$ of mass $m$ satisfies the following  covariant \vspace{-6mm}
\beqn\mbox{\bf Rarita-Schwinger equations: }\qquad\left (i\gamma^\mu\part_\mu -m\right )\phi_{\mu_1\mu_2\cdots\mu_n}&=&0,\label{R-S}\\ 
\part_\mu\phi_{\mu\mu_2\cdots\mu_n}&=&0,\\ 
\phi_{\mu\mu\mu_3\cdots\mu_n}&=&0.\eeqn
\subsection{Conclusion for this section}
 We focused on a 2 body problem and examined the meaning of internal and external degrees of freedom in view of the position of two frames. Then we found two special degrees of freedom that can not reduce to the local motion of the other frame. These phases $\theta_2, \theta_3$ naturally arises from a Poincar\'{e} transformation. We can formally express $\theta_2, \theta_3$ in terms of an asymmetric spinor representation, and thus interpret them as imaginary rotations. This interpretation seems more explicit than the traditional explanation for the origin of a gauge phase\cite{Berry}.
\sect{A derivation of fundamental fields via the U(1) gauge symmetry}
\subsection{Introduction to this section} 
In this section, we try to derive all types of interactions between the 2 body system in a uniform and isotropic and flat spacetime, assuming they are determined by the relative position of coordinates and the fields there, created somewhere of the system and propagated through medium. The unique phase $\theta$ in previous section \ref{Sup} plays a crucial role.
\subsection{The expansion of a complex scalar angle $\theta (x^\mu )$ by vectors and tensors}
We can formally expand a complex scalar angle $\theta (x^\mu )$ around $x^\mu =0$ as 
\beqn \theta (x^\mu )=\theta (0)+\frac{\part\theta (0)}{\part x^\mu}x^\mu 
+\frac12\frac{\part^2\theta(0)}{\part x^{\mu}\part x^{\nu}}x^\mu x^\nu +\cdots , \eeqn
where the expansion can have an essential singularity as in the sense of 
section 2. 

For the first derivative $\frac{\part\theta (0)}{\part x^\mu}$, we could neglect it by choosing the center of mass system, if $\theta$ was strictly Lorentz invariant. Then, a general covariant tensor of rank 2 
\beqn \frac{\part^2\theta}{\part x^{\mu}\part x^{\nu}} & := & 
\left (\begin{array}{cccc}
\frac{\part^2}{\part t^2} & \frac{\part^2}{\part t\part x} & \frac{\part^2}{\part t\part y} & \frac{\part^2}{\part t\part z} \\ 
\frac{\part^2}{\part x\part t} & \frac{\part^2}{\part x^2} & \frac{\part^2}{\part x\part y} & \frac{\part^2}{\part x\part z} \\ 
\frac{\part^2}{\part y\part t} & \frac{\part^2}{\part y\part x}  & \frac{\part^2}{\part y^2} & \frac{\part^2}{\part y\part z} \\ 
\frac{\part^2}{\part z\part t} & \frac{\part^2}{\part z\part x}  & \frac{\part^2}{\part z\part y} & \frac{\part^2}{\part z^2} \end{array}\right )\theta \qquad \mbox{can be divided into }\nonumber \\ 
& = & \frac12\left (\begin{array}{cccc}
2\frac{\part^2}{\part t^2} & \frac{\part^2}{\part t\part x}+\frac{\part^2}{\part x\part t} & \frac{\part^2}{\part t\part y}+\frac{\part^2}{\part y\part t} & \frac{\part^2}{\part t\part z}+\frac{\part^2}{\part z\part t} \\ 
\frac{\part^2}{\part t\part x}+\frac{\part^2}{\part x\part t} & 2\frac{\part^2}{\part x^2} & \frac{\part^2}{\part y\part x}+\frac{\part^2}{\part x\part y} & \frac{\part^2}{\part z\part x}+\frac{\part^2}{\part x\part z} \\ 
\frac{\part^2}{\part t\part y}+\frac{\part^2}{\part y\part t} & \frac{\part^2}{\part x\part y}+\frac{\part^2}{\part y\part x}  & 2\frac{\part^2}{\part y^2} & \frac{\part^2}{\part z\part y}+\frac{\part^2}{\part y\part z} \\ 
\frac{\part^2}{\part t\part z}+\frac{\part^2}{\part z\part t} & \frac{\part^2}{\part x\part z}+\frac{\part^2}{\part z\part x}  & \frac{\part^2}{\part y\part z}+\frac{\part^2}{\part z\part y} & 2\frac{\part^2}{\part z^2} \end{array}\right )\theta\nonumber \\ 
&   & \hspace{14mm} 
+\frac12\left (\begin{array}{cccc}
0 & \frac{\part^2}{\part t\part x}-\frac{\part^2}{\part x\part t} & \frac{\part^2}{\part t\part y}-\frac{\part^2}{\part y\part t} & \frac{\part^2}{\part t\part z}-\frac{\part^2}{\part z\part t} \\ 
\frac{\part^2}{\part x\part t}-\frac{\part^2}{\part t\part x} & 0 & \frac{\part^2}{\part x\part y}-\frac{\part^2}{\part y\part x} & \frac{\part^2}{\part x\part z}-\frac{\part^2}{\part z\part x} \\ 
\frac{\part^2}{\part y\part t}-\frac{\part^2}{\part t\part y} & \frac{\part^2}{\part y\part x}-\frac{\part^2}{\part x\part y} & 0 & \frac{\part^2}{\part y\part z}-\frac{\part^2}{\part z\part y} \\ 
\frac{\part^2}{\part z\part t}-\frac{\part^2}{\part t\part z} & \frac{\part^2}{\part z\part x}-\frac{\part^2}{\part x\part z} & \frac{\part^2}{\part z\part y}-\frac{\part^2}{\part y\part z} & 0 \end{array}\right )\theta\nonumber \\ 
& =: & G_{\mu\nu}+F_{\mu\nu}, \qquad E^{EMWg}_i:=F_{0i}, \quad B^{EMWg}_i:=\epsilon_{ijk}F_{jk},  \eeqn 
each of them proportional to a gravitational and an ($SU(3)\times SU(2)_L\times U(1)$ unified) electromagnetic field, respectively. The noncommutative gauge groups result from the original $U(1)$ symmetry by regarding each of different order expansions as a basis, i.e., via $U(1)\times U(1)\times U(1)$. Notice that $F_{\mu\nu}\neq 0$ only if the second partial derivative of the operand has a singularity at the origin. Our aim is to unify gravity into $U(1)$ symmetry as well as the electroweak force, and we contract Lorenz indices only with the flat spacetime metric $\eta_{\mu\nu}$ throughout this section. $G_{\mu\nu}$ is identified with the gravitational field, but it has nothing to do with the metric. 
\begin{description}
	\item[Example 1. ] $\left (\frac{\part^2}{\part y\part x}-\frac{\part^2}{\part x\part y}\right )\frac{1}{r}|_{x=y=z=0}=\left (\frac{\part}{\part y}\left (-\frac{x}{r}\frac{1}{r^2}\right )_{x=0}\right )_{y=z=0}-\left (\frac{\part}{\part x}\left (-\frac{y}{r}\frac{1}{r^2}\right )_{y=0}\right )_{x=z=0}=0$.\\
 From this we can learn that functions of the only variable $r:=\sqrt{x^2+y^2+z^2} $ or $ \tilde{r}:=\sqrt{r^2-t^2} $ do not contribute to $F_{\mu\nu}$. 
	\item[Example 2. ] $\left (\frac{\part^2}{\part y\part x}-\frac{\part^2}{\part x\part y}\right )\frac{x\pm y}{r}|_{x=y=z=0}$\\ 
$=\left (\frac{\part}{\part y}\left (-\frac{x}{r}\frac{x\pm y}{r^2}+\frac{1}{r}\right )_{x=0}\right )_{y=z=0}-\left (\frac{\part}{\part x}\left (-\frac{y}{r}\frac{x\pm y}{r^2}\pm\frac{1}{r}\right )_{y=0}\right )_{x=z=0}$\\
$=\lim_{y, z\to 0}\left (-\frac{y}{r}\frac{1}{r^2}\right )_{x=0}\mp\lim_{x, z\to 0}\left (-\frac{x}{r}\frac{1}{r^2}\right )_{y=0}$ is singular. From this we can learn that singular terms, even if symmetric with respect to $x, y$, can contribute. 
	\item[Example 3. ] $\left (\frac{\part^2}{\part y\part x}-\frac{\part^2}{\part x\part y}\right )(x\pm y)r|_{x=y=z=0}$\\ 
$=\left (\frac{\part}{\part y}\left (-\frac{x}{r}(x\pm y)+r\right )_{x=0}\right )_{y=z=0}-\left (\frac{\part}{\part x}\left (-\frac{y}{r}(x\pm y)\pm r\right )_{y=0}\right )_{x=z=0}$\\
$=\lim_{y, z\to 0}\left (\frac{y}{r}\right )_{x=0}\mp\lim_{x, z\to 0}\left (\frac{x}{r}\right )_{y=0}$ is finite but singular. From this we can learn that $F_{\mu\nu}$ can be finite at the origin but the value depends on the approach to the point. 
	\item[Example 4. ] $\left (\frac{\part^2}{\part y\part x}-\frac{\part^2}{\part x\part y}\right )(x^2\pm y^2)\ln r|_{x=y=z=0}=\left (\frac{\part}{\part y}\left (\frac{x}{r}\frac{x^2\pm y^2}{r}+2x\ln r\right )_{x=0}\right )_{y=z=0}$\\ 
$-\left (\frac{\part}{\part x}\left (\frac{y}{r}\frac{x^2\pm y^2}{r}\pm 2y\ln r\right )_{y=0}\right )_{x=z=0}=0$. From this we can learn that the terms higher than either of the differential variables do not contribute. 
	\item[Example 5. ] $\left (\frac{\part^2}{\part y\part x}-\frac{\part^2}{\part x\part y}\right )xyf(r)|_{x=y=z=0}=\left (\frac{\part}{\part y}\left (\frac{x}{r}xyf'(r)+yf(r)\right )_{x=0}\right )_{y=z=0}$\\ 
$-\left (\frac{\part}{\part x}\left (\frac{y}{r}xyf'(r)+xf(r)\right )_{y=0}\right )_{x=z=0}=0$ if $f(r), f'(r)$ is not singular at $r=0$.
	\item[Example 6. ] $\frac{\part}{\part w}:=\frac12\left (\frac{\part}{\part x}-i\frac{\part}{\part y}\right ),\quad\frac{\part}{\part\overline{w}}:=\frac12\left (\frac{\part}{\part x}+i\frac{\part}{\part y}\right )$ allows us to write\\ 
$\frac{\part^2}{\part y\part x}-\frac{\part^2}{\part x\part y}=2i\left (\frac{\part^2}{\part w\overline{w}}-\frac{\part^2}{\part\overline{w}\part w}\right )$ and so, analytic functions of $w$ or $\overline{w}$ and their linear combinations (harmonic functions) do not contribute. 
	\item[Example 7. ] $\left (\frac{\part^2}{\part y\part x}-\frac{\part^2}{\part x\part y}\right )f(x^2, y)|_{x=y=z=0}=\left (\frac{\part}{\part y}\left (2x\frac{\part f(x^2, y)}{\part x}\right )_{x=0}\right )_{y=z=0}$\\ 
$-\left (\frac{\part}{\part x}\left (\frac{\part f(x^2, y)}{\part y}\right )_{y=0}\right )_{x=z=0}=0$ if $\lim_{x\to 0}x\frac{\part f(x^2, y)}{\part x}=\lim_{x\to 0}x\frac{\part^2f(x^2, y)}{\part x\part y}=0$ near $(x, y)=(0, 0)$, where $f(x^2, y)$ can depend on other variables. 
\end{description}
 Thus, the terms like $x/r, xy/r$ can but not $z^2/r$ can contribute to $F_{\mu\nu}$. 
From previous section \ref{AHU} we can conclude that only gravitational part of the phase $\theta (x, y, z, t)$ is not polarized. Notice that in our view of unifying all interactions into one phase $\theta$, $U(1)$ gauge invariance is valid only if $\theta$ is not singular. Indeed, the singularity of $\theta$ is the origin of every field. Then, from the above example 4 we can learn that a gluon potential $V_G$ is at most linear order in coordinate variables. 

 Let us identify the complex angle $\theta$ as $\theta_2+i\theta_3\in{\bf C}$ with the notation of previous section. Then $\theta$ can be treated as the function of only $t$, the proper time difference between $X, Y$ observed from $X$. $Y$ would observe the same $\theta$ as the function of $x'^\mu$, the time and spatial difference between $X, Y$ observed from $Y$. It can also be written as $\int\omega_2+i\omega_3 dt$, in terms of the 
angular velocity of $X$ observed from itself, integrated with respect to the time variable of $X$. Then, 
\beqn \frac{d\theta}{dt}&=&\frac{dx^\mu}{dt}\frac{\part\theta}{\part x^\mu}\label{ge1}\\
\frac{\part}{\part x^i}\frac{d\theta}{dt}&=&\frac{dx^\mu}{dt}\frac{\part^2\theta}{\part x^i\part x^\mu}=v^\mu G_{i\mu}+\overrightarrow{E^{EMWg}}+\overrightarrow{v}\times\overrightarrow{B^{EMWg}}\label{ge2}\\
\frac{d^2\theta}{dt^2}&=&\frac{dx^\mu}{dt}\frac{dx^\nu}{dt}\frac{\part^2\theta}{\part x^\mu\part x^\nu}=v^\mu G_{\mu\nu}v^\nu ,\quad \mbox{where }v^\mu :=\frac{dx^\mu}{dt}.\label{ge3}\eeqn 
 $G_{\mu\nu}$ is symmetric and $F_{\mu\nu}^{EMWg}$ is skew-symmetric. For $e^{i\theta}$ to be unitary, $\theta$ must be real and both are real. 
Indeed, $G_{\mu\nu}, E_i^{EMWg}$, and $B_i^{EMWg}$ are real fields. In particular, the symmetry of boosts and rotations $L^T=L$ and $R^T=-R$ allows us to simplify $G_{\mu\nu}, F_{\mu\nu}^{EMWg}$ via Lorenz transformations: 
\beqn (R'^TL^TR^TGRLR')^T=R'LRGRLR'
, \quad (R'^TL^TR^TFRLR')^T=-R'LRFRLR'.&&\\
\mbox{Furthermore,}\quad G_{\mu\nu}G^{\mu\nu}=G_{ij}^2-2G_{0i}^2+G_{00}^2,
\quad F_{\mu\nu}^{EMWg}F^{{\mu\nu}\; EMWg}=2({\bf B}^{EMWg\; 2}-{\bf E}^{EMWg\; 2}),&&\\
G_{\mu\nu}G_{\mu\nu}^*=4(G_{01}G_{23}+G_{02}G_{31}+G_{03}G_{12}),
\mbox{ and } F_{\mu\nu}^{EMWg}F_{\mu\nu}^{EMWg\; *}
=4{\bf B}^{EMWg}{\bf\;\cdot\; E}^{EMWg}&&\eeqn
 are invariant under such Lorenz transformations.
Then,  (\ref{ge3}) 
reads
\beqn \mbox{for } \;\; G_{\mu\nu}=-\eta_{\mu\nu }\theta, \quad \frac{d^2\theta}{dt^2}&=&v^\mu G_{\mu\nu}v^\nu =
-\theta , \label{ge4}\eeqn 
or equivalently, 
\beqn A(t)=\left (\begin{array}{cc}0&0\\ 0&0\end{array}\right )
\mbox{ and }B(t)=\left (\begin{array}{cc}-1&0\\ 0&-1\end{array}\right )\eeqn 
in previous notations. 
We would like to interpret this situation as follows: If gravity is absent, 
$\theta$, defined as the spin of $X$, is rotating constantly, regardless of the velocity of $X$ toward $Y$.
Furthermore, a massless particle feels only time independent potential force grad $(d\theta / dt)$. This follows from the vanishing R.H.S. of (\ref{ge4}), substituted into the equation of motion (\ref{ge2}). Equivalently, {\it a time dependent force for a massless particle comes only from gravity}. This result is interesting, for usually general relativity has the effect of losing time development.  
\subsection{Gauge invariance induced by Lorenz transformations}
\setcounter{footnote}{1}
It is known in electrodynamics\cite{kyoritu} that we can Lorenz transform any of electromagnetic fields $\bf B$ and $\bf E$ 
\begin{enumerate}
\item if ${\bf E\cdot\bf B}={\bf B}^2-{\bf E}^2=0$, into an inertial frame where $\bf E$ and $\bf B$ are perpendicular and of the same length. For massless particles like photons and gravitons, the boost along their running direction does not flip their polarization and thus not change their fields.
\item if ${\bf E\cdot\bf B}=0$ and ${\bf B}^2-{\bf E}^2\neq 0$, into an inertial frame where only either of $\bf E$ or $\bf B$ exists. This $\bf E$ or $\bf B$ is invariant under a boost along and a rotation around itself. 
\item if ${\bf E\cdot\bf B}\neq 0$, into an inertial frame where $\bf E\propto\bf B$. These $\bf E$ and $\bf B$ are invariant under a boost along and a rotation around themselves. 
\end{enumerate}
 Above invariant transformations are the origin of gauge invariance\cite{WeinG}. These gauge transformations for electromagnetic fields are in general not the same as those for gravity. Rather, this boost can naturally change the kinematical energy of a charged particle in the electromagnetic field and thus influence its gravitational potential energy. In the same way, gauge transformations for gravity can influence the electromagnetic fields. In other words, a graviton and a photon can run not parallel to each other. 

In general we can not diagonalize $G_{\mu\nu}$ and $F_{\mu\nu}$ by the same Lorenz transformation. However, we can always diagonalize the spatial part of $G_{\mu\nu}$ by rotations, without concern about $F_{\mu\nu}$. 
If we are lucky enough to diagonalize $G_{\mu\nu}$ including time components by Lorenz transformations, we can define the inertial frame of $X$ as follows: If we require initial conditions (\ref{init}) for $d\overrightarrow{\theta_0}/dt$ to hold for any $v^\mu$, it corresponds to vanishing (\ref{ge1}), i.e. the initial momentum becomes zero. The initial phase $\overrightarrow{\theta_0}$ is still arbitrary. From ({\ref{ge2}), the gravitational field is parallel to $v^i$ in this frame. 
Further if we require $X$ to keep 
stable rotations (i.e. spin), $G_{\mu\nu}$ must have the same signature as of $-\eta_{\mu\nu}\theta$. 
Then we can accelerate the frame to have $G_{\mu\nu}\propto\eta_{\mu\nu}$. 
\subsection{Conclusion for this section}
We tried to interpret every field via the singularity of an original (complex) $U(1)$ phase $\theta$. Noting that gravity has spin 2, while 3 other fields have spin 1, it is natural to take them respectively as symmetric and asymmetric part of the second derivative of $\theta$. For spin 1 fields not to vanish, $\theta (t,\; x,\; y,\; z)$ must be at most linear in $t,\; x,\; y,\; z$, which may explain why the QCD gluon potential behaves linearly at long distance. $U(1)$ gauge transformations for photons may influence gravity, which may explain why we can make use of the constant term of gravity potential (literal vacuum energy) to the Higgs mechanism in section \ref{AHU}. This term is interpreted as the constant part of $\part\theta (0)/\part x^0$ i.e., energy.
\sect{Quantum gravity in the Minkowski spacetime}\label{Grav}
\subsection{Introduction to this section}
 In this section, we discuss the treatment of quantum gravity in the flat spacetime (see Appendix \ref{QGMA} for the historical review of approaches to treat quantum gravity with minimal assumptions)
\footnote{Busy readers can skip this rather supplementary section.}. This idea dates back to S. Weinberg\cite{WeinG}, where he derived the Einstein equation only by assuming the Lorenz invariance of the S matrix. His results are summarized as follows: 
\begin{enumerate}
	\item Invariance under general coordinate transformations and the geometric interpretation of gravity are subsidiary properties derived by a classical approximation.
	\item It is not necessary to assume a curved spacetime.
	\item He does not give a special treatment to a graviton. A graviton and a photon are both massless particles and behave the like. The only difference lies in that the former is spin 2 and the latter is spin 1.
	\item Potentials necessarily exist for both gauge particles to create long range forces and to satisfy, respectively the Einstein and Newton equations.
	\item He classified all particle fields by their spins. Therefore Lorenz invariance and causality are evident. Particularly, higher spin $s=3, 4, 5\cdots$ particles can not create a $\frac{1}{r^2}$-like long range force.
\end{enumerate}
 Good agreement between theory and experiment for the Lamb shift\cite{Wein} shows that special relativity is precisely valid in the quantum world. However, for general relativity, some precise experiments are difficult to perform. For example, gravitational waves are theoretically predicted. Many people have tried to observe them for almost 60 years\cite{Wheeler}, yet none has succeeded in direct observation. To examine the necessity of general relativity, we should try to explain experiments within the range of special relativity and compare the results with those of general relativity.
\subsection{Scalar-vector potentials of gravity}
 Let us review the electromagnetic like expression of general relativity\cite{kyoritu}. 
\beqn\mbox{The action }S:=\int_Q^Pds=\int_Q^P\sqrt{g_{\mu\nu}\frac{dx^\mu}{ds}\frac{dx^\nu}{ds}}ds&,&\mbox{where}s\mbox{ is the length of the curve }ds^2:=g_{\mu\nu}dx^\mu dx^\nu=:-c^2d\tau^2\nonumber\\
\mbox{taken infinitesimal deviation with respect to }x^\rho&,&\mbox{ gives us the Euler-Lagrange equation }\nonumber \\
\mbox{ (the equation of geodesics)}\quad
\frac12\frac{\part g_{\mu\nu}}{\part x^\rho}\frac{dx^\mu}{ds}\frac{dx^\nu}{ds}&=&\frac{d}{ds}\left (g_{\mu\rho}\frac{dx^\mu}{ds}\right )\label{geo}.\\
(\ref{geo}) \mbox{ with } \rho =k \mbox{ gives }\qquad g_{kl}\frac{d^2x^l}{d\tau^2}&+&\frac{d}{d\tau}\left (g_{k0}\frac{dx^0}{d\tau}\right )\simeq\frac12\frac{\part g_{00}}{\part x^k}\left (\frac{dx^0}{d\tau}\right )^2\quad (dx^k\ll cd\tau\simeq dx^0).\label{geod}\\
\mbox{Then, the identity }ds^2=g_{kl}dx^kdx^l&-&g_{k0}g_{l0}dx^kdx^l/g_{00}+(g_{k0}dx^k+g_{00}dx^0)^2/g_{00}\nonumber \\
=(g_{kl}+\gamma_k\gamma_l)dx^kdx^l&+&g_{00}(dx^0)^2\left \{1-\frac{\gamma_k}{\sqrt{-g_{00}}}\cdot\frac{dx^k}{dx^0}\right \}^2\qquad \left (\gamma_k:=\frac{g_{k0}}{\sqrt{-g_{00}}}\right )\\ 
&&\hspace{-56mm}\mbox{reads }\qquad c=\frac{dx^0}{d\tau}\;\cdot\;\sqrt{(-g_{00})\left (1-\frac{\gamma_k}{\sqrt{-g_{00}}}\frac{dx^k}{dx^0}\right )^2-(g_{kl}+\gamma_k\gamma_l)\frac{dx^kdx^l}{dx^0dx^0}}.\label{identity}\\
\mbox{ We can rewrite }(\ref{geod})\mbox{ with }(\ref{identity})\mbox{ to give } &&
(g_{kl}+\gamma_k\gamma_l)\frac{d^2x^l}{dt^2}\simeq -\frac{\part\phi_G}{\part x^k}-c\sqrt{1+\frac{2\phi_G}{c^2}}\frac{\part\gamma_k}{\part t}, \label{accel}\\
\mbox{where }\quad \phi_G :=-(1+g_{00})c^2/2&&\mbox{and}\;\;\gamma_k:=\frac{g_{k0}}{\sqrt{-g_{00}}}\label{phiG} \eeqn 
are respectively called scalar and vector potentials of gravity. L.H.S. of (\ref{accel}) is the spatial component of a covariant acceleration induced by $\phi_G$, and $\gamma_k$. In particular, $\gamma_k$ vanishes when spatial axes are orthogonal to the time axis:\vspace{-5mm}
\eq \gamma_k=0.\en
 It is valuable to compare famous general relativistic effects with potential induced effects in the Minkowski spacetime. The following subsections are devoted to such careful examinations close to experiments. 
\subsection{The Schwarzschild's exterior solution in curved spacetime}
 The spherical symmetric, static solution of the Einstein equation 
\beqn R_{\mu\nu}-\frac12Rg_{\mu\nu}&=&\frac{8\pi G}{c^4}T_{\mu\nu}\\ 
\mbox{in the vacuum }\quad T_{\mu\nu}&=&0\quad\mbox{ around the source center 
mass }M,\;\mbox{ written in polar coordinates }\nonumber \\
(x^0, x^1, x^2, x^3)&=&(ct, r\sin\theta\cos\varphi , r\sin\theta\sin\varphi , r\cos\theta )\quad\mbox{ is }\nonumber \\
ds^2=\frac{1}{1-2a/r}(dr)^2&+&r^2\{(d\theta )^2+\sin^2\theta (d\varphi )^2\} -(1-2a/r)(dx^0)^2,\quad \mbox{ where }\quad  a:=\frac{GM}{c^2}\label{Sch}\eeqn 
is consistent with the Newtonian potential in the weak limit of the gravitational field. 

 Let us study the trajectories in the Schwarzschild spacetime\cite{kyoritu}. (\ref{geo}) with (\ref{Sch}) reads 
\beqn \frac{d^2x^0}{ds^2}&+&\frac{d}{ds}\log\left (1-2\frac{a}{r}\right )\frac{dx^0}{ds}=\left (1-2\frac{a}{r}\right )^{-1}\frac{d}{ds}\left \{\left (1-2\frac{a}{r}\right )\frac{dx^0}{ds}\right \}=0,\label{Sorbit0}\\ 
\frac{d^2r}{ds^2}&+&\frac14\frac{d}{dr}\left (1-2\frac{a}{r}\right )^2\left (\frac{dx^0}{ds}\right )^2-\frac12\frac{d}{ds}\log\left (1-2\frac{a}{r}\right )\frac{dr}{ds}=(r-2a)\left [\left (\frac{d\theta}{ds}\right )^2+\sin^2\theta\left (\frac{d\varphi}{ds}\right )^2\right ],\label{Sorbit1}\\ 
\frac{d^2\theta}{ds^2}&+&\frac{2}{r}\frac{dr}{ds}\frac{d\theta}{ds}-\sin\theta\cos\theta\left (\frac{d\varphi}{ds}\right )^2=0,\label{Sorbit2}\\ 
\frac{d^2\varphi}{ds^2}&+&\frac{2}{r}\frac{dr}{ds}\frac{d\varphi}{ds}+2\cot\theta\frac{d\theta}{ds}\frac{d\varphi}{ds}=\frac{1}{r^2}\frac{d}{ds}\left (r^2\frac{d\varphi}{ds}\right )+2\cot\theta\frac{d\theta}{ds}\frac{d\varphi}{ds}=0,\label{Sorbit3}
\eeqn
(\ref{Sorbit2}) ensures initial conditions $\quad\theta=\frac{\pi}{2}, \quad\frac{d\theta}{ds}=0 \quad$ to hold whenever after, so without loss of generality, we can assume that the motion is in the  $\quad\theta =\frac{\pi}{2}\quad$ plane. Then, we can integrate (\ref{Sorbit0}) and (\ref{Sorbit3}) to give 
\beqn \mbox{Energy conservation law:}\qquad\left (1-2\frac{a}{r}\right )\frac{dx^0}{d\tau}&=:&E=const. \qquad\mbox{and}\label{ene}\\ 
\mbox{Angular momentum conservation law:}\qquad r^2\frac{d\varphi}{d\tau}&=:&L=const.. \label{ang}\eeqn
\beqn \mbox{(\ref{Sch}) with (\ref{ene}) and (\ref{ang}) reads}\quad
-c^2&=&\left (1-2\frac{a}{r}\right )^{-1}\left (\frac{dr}{d\tau}\right )^2+\frac{L^2}{r^2}-\left (1-2\frac{a}{r}\right )^{-1}E^2\label{MGR}\\ 
\mbox{or with the change of variable}\qquad \frac{dr}{d\tau}&=&
\frac{dr}{d\varphi}\frac{d\varphi}{d\tau}=\frac{dr}{d\varphi}\frac{L}{r^2}
=-L\frac{dz}{d\varphi}\qquad (z:=\frac{1}{r})\\
-(1-2az)c^2&=&L^2\left \{\left (\frac{dz}{d\varphi}\right )^2+(1-2az)z^2\right \}-E^2\label{dzdv}\\
\mbox{and then, differentiating (\ref{dzdv}) with }\varphi, &&
\frac{d^2z}{d\varphi^2}-\frac{ac^2}{L^2}+z-3az^2=0.\label{d2zdv2}\eeqn
(\ref{d2zdv2}) together with (\ref{ang}) gives the equation of trajectories
\setcounter{footnote}{1}\footnote{(\ref{Sorbit1}) with (\ref{Sch}) and (\ref{ang}) gives (\ref{d2zdv2}) more directly:
\beqn\frac{d^2r}{ds^2}&+&\frac{a}{r^2}\left (1-2\frac{a}{r}\right )\left (\frac{dx^0}{ds}\right )^2-\frac{a}{r^2}\left (1-2\frac{a}{r}\right )^{-1}\left (\frac{dr}{ds}\right )^{2}-(r-2a)\left (\frac{d\varphi}{ds}\right )^2=0\nonumber \\
\iff\quad \frac{d^2r}{ds^2}&-&\frac{a}{r^2}-(r-3a)\left (\frac{d\varphi}{ds}\right )^2=0\quad\iff\quad\frac{d^2r}{d\tau^2}=-ac^2z^2+(z^3-3az^4)L^2.\label{geocal2}\eeqn}.
\subsection{Quantum gravity in the Schwarzschild spacetime: Two more paradoxes}
Apart from an ambiguity or information loss problem related to integral constants mentioned in section \ref{Sing}, there are two more paradoxes in general relativity: 
\begin{itemize}
	\item{\bf Paradox 1 } 
Notice that (\ref{MGR}) implies to replace the usual Einstein relation 
\beqn E^2=p^2c^2+m^2c^4&&\mbox{or the Klein-Gordon equation}\label{Emc}\\
\left [-\frac{1}{r^2}\frac{d}{dr}\left (r^2\frac{d}{dr}\right )
+\frac{l(l+1)}{r^2}\right ]\Phi (r)
&=& \frac{(E-e\phi )^2-m^2c^4}{{\hbar}^2c^2}\Phi (r)\qquad\mbox{with}\\
\left [-\left (1-2\frac{a}{r}\right )^{-1}\frac{1}{r^2}\frac{d}{dr}\left (r^2\frac{d}{dr}\right )+\frac{l(l+1)}{r^2}\right ]\Phi (r)
&=& \frac{\left (1-2\frac{a}{r}\right )^{-1}(E-e\phi )^2-m^2c^4}{{\hbar}^2c^2}\Phi (r),\label{KGGR}\eeqn
while natural application of the metric (\ref{Sch}) to the 4 momentum operator 
$p_\mu =-i\hbar\frac{\part}{\part x^\mu}$ of quantum mechanics implies 
\vspace{-2mm}\eq \left [-\sqrt{1-2\frac{a}{r}}\frac{1}{r^2}\frac{d}{dr}\left (r^2\sqrt{1-2\frac{a}{r}}\frac{d}{dr}\right )+\frac{l(l+1)}{r^2}\right ]\Phi (r)
= \frac{\left (1-2\frac{a}{r}\right )^{-1}(E-e\phi )^2-m^2c^4}{{\hbar}^2c^2}\Phi (r),\label{GRKG}\en
instead of (\ref{KGGR}). This is because in quantum mechanics, contrary to classical mechanics, a momentum operator involves inverse of the distance. 

\indent Which equation is correct? Of cause only experiments can judge. Both (\ref{KGGR}) and (\ref{GRKG}) involve further difficulty in unclear behavior of the $U(1)$ potential $\phi$ under the gravitational field
\footnote{A potential is not even Lorenz invariant\cite{WeinG}\cite{Mu}.}. 
 This paradox encourages us to carefully examine the above derivation of general relativity. 
  (\ref{geocal2}) reads the equation of motion
\beqn -\frac{ac^2}{r^2}&=&\frac{d^2r}{d\tau^2}-\frac{L^2}{r^3}
+3\frac{aL^2}{r^4}
\iff\frac{d^2r}{d\tau^2}=-\frac{dV(r)}{dr},\label{GRM}\\ V(r)&:=&const.-\frac{ac^2}{r}+\frac{L^2}{2r^2}-\frac{aL^2}{r^3}=\frac12\left (\frac{ac^2}{L}-\frac{L}{r}\right )^2-\frac{aL^2}{r^3}+const.\label{GRV}\eeqn 
The first term in (\ref{GRV}) is just an integral constant, with (\ref{Emc}) or (\ref{MGR}) related to the definition of a rest mass. The second term is simply the Newtonian gravity, always attractive but with a few room of ambiguity in the definition of mass appearing in $a:=\frac{MG_0}{c^2}$. Discussion of the section \ref{Sup} implies to replace it with a reduced mass: $M\to \frac{Mm}{M+m}$. The third term is the repulsive centrifugal force, automatically derived by a coordinate transformation to a rotating system. It is reduced to the motion of an external world toward the 2 body system. Finally, the fourth term is the multiplication of the second and the third terms. Though coefficients are confusing and somehow suppressed by $2/c^2$, this attractive force is unique to general relativity and probably means that a graviton couples to the centrifugal force potential as well as the rest mass in the rest frame of the rotating celestial body. 
A physical interpretation of (\ref{MGR}) is as follows:
\begin{enumerate}
	\item Let us begin with an observer $X$ on a celestial body of mass $m$ in a gravity-free space. The rest mass energy is $mc^2$.
	\item Consider $X$ noticed that the system is rotating around a fixed point (the center of mass of $X$ and another celestial body $Y$ of mass $M$) with the angular velocity $\omega$. Then, the rotation component of the momentum of $X$ is $m\overline{XY}\omega =:p_{XL}$. The corresponding kinematical energy is $\frac{p_{XL}^2}{2m}$.
	\item Let us take account of gravity. The above two kinds of energy are both influenced by gravity and suppressed by a factor $\sqrt{1-2\frac{a}{r}}$.
	\item The radial component of motion for $X$ is solved by above [(rest mass $+$ centrifugal force)$\times$gravity] potential $V$, while the angular momentum $L$ is invariant under such a central force ($V=V(r)$). Then we can obtain another constant of motion $E$, interpreted as the potential $V$ plus the radial component of kinetic energy. 
	\item To make $E$ covariant under Lorenz transformations and rotations with constant angular momentums, we must adopt the metric (\ref{Sch}) instead of the definition in special relativity $d\tau :=\sqrt{1-v^2/c^2}dt$. 
\end{enumerate}
 Above interpretation implies that (\ref{KGGR}) is more natural than (\ref{GRKG}). However, if we redefine the radial component of momentum operator as $\displaystyle p_r:=-i\hbar\sqrt{1-2\frac{a}{r}}\frac{\part}{\part r}$, (\ref{GRKG}) is more natural. This paradox is related to the problem of quantization that in quantum mechanics, the usual law of function preservation no longer holds. For example\cite{Isham}, if we try to define the symmetric product operator as 
\beqn fg&\mapsto&\frac12(\widehat{f}\widehat{g}+\widehat{g}\widehat{f})\qquad 
\mbox{for any functions}\quad f,\; g\in L^2(\mbox{\bf R}),\label{fp}\\
x(xp)&\mapsto& \frac12(\widehat{x}\widehat{xp}+\widehat{xp}\widehat{x})
=\frac14(\widehat{x}(\widehat{xp}+\widehat{px})+(\widehat{xp}+\widehat{px})\widehat{x})=\frac14(\widehat{x}^2\widehat{p}+\widehat{p}\widehat{x}^2+2\widehat{x}\widehat{p}\widehat{x}),\\
\mbox{while}\quad&\not\hspace{-4mm}{\iff}&(x^2p)\mapsto\frac12(\widehat{x^2}\widehat{p}+\widehat{p}\widehat{x^2})=\frac12(\widehat{x}^2\widehat{p}+\widehat{p}\widehat{x}^2).\eeqn 
Thus the quantization formula (\ref{fp}) is no longer valid as an operator identity. 
	\item{\bf Paradox 2 } 
This paradox is related to the mass-free definition of an orbital angular momentum $L$ in a 2 body problem. In relativistic quantum mechanics, the total angular momentum is conserved, as the result of usual momentum conservation law by the microscopic view of interactions via exchanges of gauge particles. In particular, photons can have finite Lorenz covariant momentums and angular momentums. However, the general or special relativistic definition of an orbital angular momentum (\ref{ang}) or (\ref{angl}) contradicts to the fact. That is, the Newton's third law of motion requires both the 2 bodies to feel equal strength but opposite forces. This is valid in quantum physics, but not in the above general relativistic definition of an angular momentum.
\end{itemize}

\indent The following subsections are devoted to review briefly the traditional derivation of general relativity effects\cite{Kyoritu}. 
\subsection{Evidence 1: Perihelion precession}
 Let us solve (\ref{d2zdv2}) substituted $a=\frac{GM_0}{c^2}$ ($M_0$ is the mass of the sun) 
\beqn\frac{d^2z}{d\varphi^2}+z&=&\frac{GM_0}{L^2}+3\frac{GM_0}{c^2}z^2
\label{prehel}\\
\mbox{with the ansatz }
\quad z&=&(1+\epsilon\cos\gamma\varphi )/l.\label{ansz}\eeqn 
 Notice that $L$, defined in (\ref{ang}), is real for a timelike orbit $(ds)^2<0$ and the second term in the R.H.S. of (\ref{prehel}) is suppressed by an order $c^2$. Then, expanding 
\beqn
\epsilon &=&\epsilon_0+\epsilon_1/c^2+\epsilon_2/c^4+\cdots ,\\
\gamma &=&\gamma_0+\gamma_1/c^2+\gamma_2/c^4+\cdots ,\\
l &=&l_0+l_1/c^2+l_2/c^4+\cdots ,\\
\mbox{we obtain }\quad l_0&=&\frac{L^2}{GM_0},\quad\gamma_0=1,\\
-\frac{l_1}{l_0^2}+\frac{\epsilon_0}{l_0}(-2\gamma_1\gamma_0)\cos\gamma_0\varphi &=&\frac{3GM_0}{l_0^2}(1+\epsilon_0\cos\gamma_0\varphi )^2 
\iff l_1=-3GM_0,\quad\gamma_1=-3\frac{GM_0}{l_0}=-3\left (\frac{GM_0}{L}\right )^2\nonumber \eeqn
i.e., the angle of perihelion precession during a period is $-2\pi\gamma_1/c^2= 6\pi\left (\frac{GM_0}{Lc}\right )^2$, where $L$ is the total angular momentum, divided by the total mass $m+M_0$, of a celestial body (say, Mercury) rotating around the sun plus the sun. This quantity is known to be $6$ times as large as that of special relativity. Careful readers know that this is just the multiplication of the two correction coefficients appearing in the R.H.S of (\ref{prehel})$\times c^2$. However, the term $3\frac{GM_0}{c^2}z^2$ replaced with $3\frac{GM_0z}{c^2l_0}$ would give the same result, for $L^2\ll l_0$. (\ref{GRV}) with (\ref{ang}) is enough for the correct derivation of a perihelion precession. However, it is not necessary. 
 We can neglect the $O(z^3)$ term in (\ref{GRV}), and also describe the equivalent potential to give the same $\gamma_1$: 
\beqn V(r)&:=&const.-\frac{ac^2}{r}+\frac{L^2}{2r^2}\left (1-3\frac{a^2c^2}{L^2}\right )^2\simeq \frac12\left \{\frac{ac^2}{L}\left (1+3\frac{a^2c^2}{L^2}\right )-\frac{L}{r}\left (1-3\frac{a^2c^2}{L^2}\right )\right \}^2+const.\nonumber\eeqn 
Notice that in such an alternative potential second order in $z$, all precession effects reduce to the coefficient of the $z^2$ term, for the ansatz (\ref{ansz}) is strictly valid then and a pure deviation in the $z$ term coefficient only affects $l_0$. 

 Let us briefly review the prediction of special relativity for the perihelion precession\cite{kyoritu}. The fundamental equations of motion in polar coordinates on a $2$ dimensional flat plane are 
\beqn\frac{d}{dt}\left (\frac{m}{\sqrt{1-(v/c)^2}}\frac{dr}{dt}\right )&=&\frac{mr}{\sqrt{1-(v/c)^2}}\left (\frac{d\theta}{dt}\right )^2+F_r,\label{eqm1}\\ 
\frac{d}{dt}\left (\frac{mr^2}{\sqrt{1-(v/c)^2}}\frac{d\theta}{dt}\right )
&=&rF_\theta\label{eqm2}, \\
\mbox{with the flat metric }\quad (ds)^2&=&(dr)^2+(rd\theta )^2-(cdt)^2=-(cd\tau )^2.\label{flat}\\
\mbox{For Newtonian gravity }\quad F_r=-\frac{mac^2}{r^2}\quad \mbox{ and }&&F_\theta =0,\quad\mbox{(\ref{eqm1}) and (\ref{eqm2}) are integrated to give }\nonumber\\
\mbox{Energy conservation law:}\quad\frac{1}{\sqrt{1-(v/c)^2}}&-&\frac{a}{r}
=const.=:E' \quad\mbox{ and}\label{energ}\\
\mbox{Angular momentum conservation law:}\quad&&\frac{r^2}{\sqrt{1-(v/c)^2}}\frac{d\theta}{dt}=r^2\frac{d\theta}{d\tau}=const.=:L'.\label{angl}\eeqn
Notice that we omitted constant coefficients $m,\; c$ in above definitions of $E,\; L$. Notations might be a little confusing, but definitions of $a,\; s,\; E',\; L'$ are the same or parallel as in previous sections. (\ref{angl}) is the same as (\ref{ang}), while (\ref{energ}) is not the same as (\ref{ene}).
(\ref{eqm1}) with (\ref{angl}) gives
\beqn \frac{d}{dt}\left (-\frac{L'\sqrt{1-(v/c)^2}}{\sqrt{1-(v/c)^2}}\frac{dz}{d\theta}\right )=\frac{L'^2}{r^3}\sqrt{1-(v/c)^2}-\frac{ac^2}{r^2}
&\iff&-L'^2\sqrt{1-(v/c)^2}\; z^2\frac{d^2z}{d\theta^2}=L'^2z^3\sqrt{1-(v/c)^2}-ac^2z^2\nonumber\\
\iff\frac{d^2z}{d\theta^2}+z=\frac{ac^2}{L'^2\sqrt{1-(v/c)^2}}
=\mbox{ (with (\ref{energ})) }&&\hspace{-3mm}\frac{ac^2}{L'^2}(az+E')
=\frac{a^2c^2}{L'^2}(E'/a+z).\hspace{-11mm}\label{spepr}\eeqn
 Comparing (\ref{spepr}) with (\ref{ansz}) results in $\displaystyle\quad\gamma_1/c^2\simeq\frac{(GM_0)^2}{2(L'c)^2},\quad\frac16$ times as large as that of general relativity. 
\subsection{Evidence 2: Gravitational lens}
 Let us consider the orbit of the light passing near the sun. As a photon is massless, we must adopt the vanishing line element: \vspace{-5mm}
\beqn ds&=&0\qquad
\mbox{and }\qquad\frac{d^2z}{d\varphi^2}+z=3\frac{GM_0}{c^2}z^2\qquad
\label{lens}\mbox{instead of (\ref{prehel}).}\\
\mbox{Neglecting the }O(1/c^2)\mbox{ term in }&&\hspace{-5mm}\mbox{(\ref{lens}) reads the approximate solution }\qquad z\simeq z_0\cos\varphi .\label{soll}\\
\mbox{Substituting this to (\ref{lens}) reads}&\mbox{at}&\mbox{order }O(1/c^2)\qquad\frac{d^2z}{d\varphi^2}+z\simeq 3\frac{GM_0}{c^2}z_0^2\cos^2\varphi =\frac{3GM_0}{2c^2}z_0^2(1+\cos 2\varphi ),\nonumber\\
\mbox{ and thus }\qquad z&\simeq& z_0\cos\varphi +\frac{3GM_0}{2c^2}z_0^2\left (1-\frac13\cos 2\varphi\right )\nonumber \\ 
&=&z_0\cos\varphi +\frac{GM_0}{c^2}z_0^2(2-\cos^2\varphi )\to 0\qquad \left (z:=\frac{1}{r}\to 0 \right ).\label{solens}\\
\mbox{The asymptotic equation (\ref{solens})}&&\cos^2\varphi -\frac{c^2}{GM_0z_0}\cos\varphi 
\simeq 2\qquad\mbox{ in the long distance limit gives }\nonumber \\
\cos\varphi&\simeq&-\frac{2GM_0z_0}{c^2},\quad \frac{c^2}{GM_0z_0}\qquad (2\ll\frac{c^2}{GM_0z_0}).\label{cosol} \\ 
\mbox{ The second solution of (\ref{cosol}) gives}&&\hspace{-3mm}\mbox{imaginary  }\varphi\mbox{ and the first solution corresponds to the two asymptotic lines } 
\nonumber \\
x=\frac{1}{z_0}-\frac{GM_0z_0}{c^2}\frac{x^2+2y^2}{\sqrt{x^2+y^2}}&\to& 
\frac{1}{z_0}-\frac{2GM_0z_0}{c^2}|y|\qquad (|y|\to\infty )\quad\mbox{ i.e., incoming and outgoing orbits.}\nonumber\\
\mbox{Then, the deviation angle of the}&&\hspace{-6mm}\mbox{ two lines  (in general relativity) is }\qquad\delta_{GR} :\simeq\frac{4GM_0z_0}{c^2}.\label{GR}
\eeqn
\indent Contrastingly in the Newtonian gravity dealt with special relativity, there is no gravitational lens, for in the velocity of light limit $(v\to c)$, the two terms of R.H.S. appearing in (\ref{spepr}) vanish (cf. (\ref{energ}), (\ref{angl})) and the solution reduces to (Appendix \ref{Lens})
\beqn \quad z=z_0\cos\varphi ,&\quad&
\mbox{which vanishes at }\quad\cos\varphi =0.\\
\mbox{Therefore, the orbit does not curve:}&&\quad\varphi =\pm\frac{\pi}{2},
\qquad \delta_{NG}:=0.\label{NG}\eeqn 
\indent However, as is well known, we can derive gravitational lens effects only from Newtonian gravity plus the equivalence principle as follows. Consider a particle $P$ with a momentum $p$ passing near along the straight line a distance $r$ away from the mass $M_0$. $P$ earns the transverse impulse $\displaystyle F_\bot (t)dt:=\frac{GM_0p_{//}dt}{vr^2}$ during an infinitesimal time $dt$, while the parallel momentum $p_{//}$ is invariant
\footnote{$_\bot =\sin\theta$ is the projection to the radial direction, and $_{//}=\cos\theta$ is the projection to the argument direction.}.
Therefore, total transverse momentum deviation from the nearest distance $r=r_0$ to $r=\infty$ is 
\beqn \Delta p_\bot&:=&\int_{t_0}^\infty\frac{GM_0p_{//}}{vr^2}dt= \int_{r_0}^\infty\frac{GM_0p_{//}}{vv_\bot r^2}dr\simeq\int_{r_0}^\infty\frac{GM_0pr_0}{v^2\sqrt{r^2-r_0^2}r^2}dr\nonumber \\
&\simeq&\frac{GM_0p}{v^2}\int_0^{z_0}\frac{z}{\sqrt{z_0^2-z^2}}dz\qquad (z_0:=\frac1{r_0})
\quad =\frac{GM_0p}{v^2}\left [(z_0^2-z^2)^\frac12\right ]_{z_0}^0
=\frac{GM_0pz_0}{v^2}.\qquad\\
\mbox{Therefore, the deviation}&&\hspace{-6mm}\mbox{ angle of the two asymptotic lines  (in Newtonian gravity + the equivalence principle) is }\nonumber \\
\delta_{Neq}:&\simeq&2\frac{\Delta p_\bot}{p}=\frac{2GM_0z_0}{v^2}\to \frac{2GM_0z_0}{c^2}\quad (v\to c),\label{Neq}\eeqn
which is one half as large as $\delta_{GR}$. 
\subsection{Extension of special relativistic equations of motion to massless particles}
 What is the difference of the two derivations for Newtonian gravity?
Careful examination shows that the limit of an angular momentum $L'\to\infty$, again, plays a trick. This is not a surprise, for (\ref{eqm1}) and (\ref{eqm2}) are no longer valid for a massless particle. Instead, the correct equations in the same notation are 
\beqn\frac{dp_{//}r}{dt}&=&rF_\theta\label{eqm2'}, \quad\mbox{which reads}\quad \quad p_{//}r =const.=:L''\;\; (<\infty)\label{Angla}\quad\mbox{ for a central force }\overrightarrow{F}:=-\mbox{grad }\phi (r), \mbox{ and}\nonumber\\ 
\frac{dp_\bot}{dt}&=&p_{//}\left (\frac{d\theta}{dt}\right )+F_r,\label{eqm1'}
\quad\mbox{which reads}\quad (p_\bot )^2+\frac{d\phi (r)}{dp}\frac{p^2}{v}=0
\quad\mbox{via }\\\sin\theta\frac{dp}{dt}&=&F_r\iff 
p^2\sin^2\theta\frac{dp}{dt}=F_rp^2\sin\theta =F_r\frac{p^2}{v}\frac{dr}{dt}\iff \int p^2\sin^2\theta dp=\int F_r\frac{p^2}{v}dr\nonumber \\ 
&\iff& \int\left \{p^2-\left (\frac{L''}{r}\right )^2\right \}dp=\int F_r\frac{p^2}{v}dr=\int F_r\frac{p^2dr}{vdp}dp
=\int F_rp^2\sin\theta dt. 
\eeqn 
Notice that here $p_\bot ,p_{//}, p, F_\theta , F_r$ are spatial components of Lorenz covariant 4 vectors, but $v, r, \theta$ are not. In particular, vectors transverse to the momentum $\overrightarrow{p}$ along the orbit  never Lorenz contract. Here integrals are taken always counterclockwise direction. Then, $dr, dp$ change signs while the integrands do not to give the opposite value of the incoming and outgoing integrals $\int_{-\infty}^0\;\; =\;\; -\int_0^{\infty}$. In the last integral $dt$ is always positive, while the integrand change signs to give the opposite value. 

Thus, we have successfully derived the Lorenz invariant law of energy conservation:
\beqn p_\bot^2=\left \{p^2-\left (\frac{L''}{r}\right )^2\right \}
&=& -\frac{d\phi}{dp}\frac{p^2}{v}
=-\frac{d}{dp}\left (\phi\frac{p^2}{v}\right ) 
+\phi\frac{d}{dp}\left (\frac{p^2}{v}\right ), \label{Pote}\eeqn
where the first term in the R.H.S. is a total derivative. 
Therefore, the transverse momentum depends only on $p$ and the potential $\phi(r)$ in the neighborhood of  the point, and the mass $m$ of the object. However, this derivation is equivalent to (\ref{spepr}) and if we naively substitute the Newtonian potential $\phi (r)=\phi_{NW}:=-\frac{GmM}{r}$ in (\ref{Pote}), the $z$ order term of the corresponding potential $V(z)$ in (\ref{spepr}) still 
 vanishes, which immediately loses the gravitational lens effect. Our results are rather summarized as:
\begin{enumerate}
	\item Possible alternative potentials second order in $z:=\frac{1}{r}$ to give the correct value of perihelion precession are restricted by $\gamma$, the angle factor, to have $\displaystyle\left \{1-3\left (\frac{GM}{Lc}\right )^2\right \}$ term. Special relativity can explain only $\frac16$ of the correction. 
	\item In general relativity, this effect comes from the $z^3$ order terms of the potential (\ref{GRV}) . 
	\item Naive treatment of special or general relativity gives the wrong result of a vanishing gravitational lens. However, physically natural classical treatment of momentum with the equivalence principle, under the Newtonian potential gives one half of the correct value.
	\item The wrong result is due to the naive treatment of $v\to c$ limit, which results in the divergent, mass-free orbital angular momentum $L\to\infty$ in (\ref{angl}). Instead, we should naturally extend the equations of motion (\ref{eqm1}), (\ref{eqm2}) in special relativity to be valid both for massive and massless particles. 
	\item The equations of geodesics (\ref{Sorbit0})-(\ref{Sorbit3}) in general relativity are already in this mass-independent form. However, a natural requirement to symmetrize the equations to both two objects leads to the wrong result. 
	\item Existence of the gravitational lens requires the $z$ order term not to vanish in possible alternative potentials, second order in $z$. This contradicts to the ordinary shape of the Newtonian potential $\phi_G=-\frac{GMm}{r}\to 0$ as $m\to 0$. 
	\item (\ref{eqm1'}), (\ref{eqm2'}) are a natural extension of (\ref{eqm1}), (\ref{eqm2}) to a photon. Then, the energy conservation law (\ref{Pote}) of a transverse momentum $p\bot$ for a central force is automatically derived by effort of obtaining $p_\bot$ in the L.H.S.. 
	\item The differentiation $d\phi /dp$ in the energy conservation law (\ref{Pote}) of the original potential $\phi (r)$ requires that $\phi (r)$ depend on $p$, the spatial radius of the total $4$ momentum of a particle. This opens the possibility for a massless particle to feel gravity. 
	\item (\ref{Pote}) must be a scalar equation. Then, the extra factor $p/v$ requires $\phi$ to behave under Lorenz transformations like the time component of a contravariant $4$ vector, and the force $F_r$ to be like the time-spatial components of a $4$ tensor of contravariant rank $1$ and covariant rank $1$.
	\item Here possible different properties of the force under parity or time conjugation are ignored. However, above discussion is valid to both electric and gravitational fields, thus proving the conception of a force mediated by a vector field. In particular, a central force can not Lorenz transform as a whole tensor of a higher rank. It must be decomposed in a (contracted tensor product of) (pseudo) spatial vector(s).
	\item In this sense, a potential $\phi$ of a central force contributes to the energy only through $p$, but not through the total energy $E=:p^0$, $(E-$potentials$)^2=p^2c^2+m^2c^4$. In other words, it couples neither to itself nor other potentials. \label{Rem}
	\item This picture is definitely different from general relativity in which the metric is generated by the energy momentum tensor including the gravitational field itself. However in the Schwarzschild solution, this self coupling effect is neglected by assuming the vanishing energy-momentum tensor: $T^{\mu\nu}=0$.
	\item For a massless photon to feel gravity, $\phi_G(r)\neq -G\frac{Mm}{r}$ but $\phi_G (r)=-G\frac{p^\mu p'_\nu}{c^2r}+Constant.$, where $p_\mu$, $p'_\nu$ are respectively Lorenz covariant $4$ momentums of the 2 bodies and the metric is flat $g_{\mu\nu}=\eta_{\mu\nu}$.
	\item In the same way, the electromagnetic potential is $\phi_{EM}(r)=g_{EM}\frac{qQ p^\mu p'_\mu }{Mmc^2r}+Constant.$. This expression is valid only for massive particles and if either particle is massless, it reduces to $\phi_{EM}(r)=g_{EM}\frac{qQ p^\mu p'_\mu}{EE'c^2r}+Constant.$
\footnote{However, from the classical theory of electromagnetism, a charged particle naturally obtains an electromass\cite{EM}.}. 
\end{enumerate}
\subsection{Can a gravitational lens make a rainbow?}
The modified equation, substituted $\phi_G(r)=-G\frac{M_0p}{rc}$, corresponding to (\ref{spepr}) is
\beqn rp\sin\theta &=&L'',\qquad \sin^2\theta =\frac{GM_0}{crv}
\label{spepr'}\\
\Rightarrow\frac{dr}{dt}&=&v\sin\theta =\frac{GM_0\sin\theta}{cr\sin^2\theta}\qquad\Rightarrow \qquad 2\cos (2\theta )=\frac{GM_0d^2(z/v)}{cd\theta^2}=2-4\frac{GM_0z}{c^2}\qquad (v\to c).\nonumber \eeqn
(\ref{spepr'}), obtained from the energy conservation law (\ref{Pote}) is already describing a trajectory $z (\theta )$ and shows that in our theory the incoming and outgoing light rays approach the same straight line $r\sin\theta =0$, only evading $M_0$ near $r\to r_0$. This result is unexpected for an attractive force of gravity and seems strange, but an observer at $r=r_1,\;\; \theta =\theta_1$ would consider that the ray vent angle $\delta$ is 
\beqn\tan\delta =\frac{d(r\sin\theta )}{d(r\cos\theta )}&=&\frac{dr\sin\theta +r\cos\theta d\theta}{dr\cos\theta -r\sin\theta d\theta}=\frac{r^2cv\sin (2\theta )\sin\theta -GM_0r\cos\theta}{r^2cv\sin (2\theta )\cos\theta +GM_0r\sin\theta}\nonumber \\
&=&\frac{2r_1c^2\sin^2\theta_1 -GM_0}{2r_1c^2\cos^2\theta_1  +GM_0}\to 
-\frac{GM_0}{2r_1c^2}\quad (\theta_1\to 0, \pi ),\eeqn 
where $M_0$ is not negligible if only $\theta_1\simeq\pi$. 
The trajectories do not depend on the asymptotic momentum $p(\sin\theta =0)$, but the distance to the asymptotic straight line is $\frac{L''}{p}=\frac{cp_0}{p}$, i.e., the red shift ratio, where $p_0$ is the momentum at the nearest point of the ray $r=r_0$. In other words, the gravitational lens can make a rainbow, in a sense that the light from a distant star is observed red shifted by $M_0$, but if emitted from another star that is not on the same straight line combining $M_0$ and the former star (nor on its axial symmetric cone to the observer), the light would be red shifted differently, even if the distances between the stars and $M_0$ are the same. This results only from the finite angular momentum conservation law (\ref{Angla}). Our theory discussed in this subsection is probably wrong, but a natural extension of special relativity, and makes an experimentally probable prophet: In general relativity the red shift comes only from acceleration, without mentioning whether it is caused by gravity or a rocket engine, so not depends on the angle between the light ray and the acceleration. Contrastingly, in our theory it depends on the angle of, still without mentioning the cause of, the acceleration. The difference comes from the different definitions of an orbital angular momentum (cf.(\ref{ang}), (\ref{Angla})).

 Above discussion shows that $z^3$ order terms of a potential are indispensable to explain famous general relativity effects.
\subsection{A possible approach to the solution: The retarded Newtonian potential}
This subsection is only a review \cite{graphoto2} of the classical argument that general relativity effects are also derived by the special relativity. 
 The derivation is parallel to that of Lienard-Wiechelt potentials in electromagnetic theory. The corresponding potentials are 
\beqn \phi_G&:=&-\frac{GM_0}{r\sqrt{1-\frac{v^2\sin^2\theta}{c^2}}},\label{G1}\\
\gamma_k&:=&-\frac{GM_0v_k}{c^2r\sqrt{1-\frac{v^2\sin^2\theta}{c^2}}}.\label{G2}\eeqn
Then, with the classical definition of a (mass dependent) orbital angular momentum
\beqn L_{Cl}&:=&mrv\sin\theta\\
\mbox{(\ref{G1}) can be written as }\quad
V_{GLW}&=&-\frac{GM_0}{r}-\frac{GM_0L_{Cl}^2}{2m^2c^2r^3},\label{GLW}\eeqn
which involves one half of the required $z^3$ term. As explained in \cite{graphoto2}, taking account of the gravitational Thomas precession gives the more correct form of the $z^3$ term: 
\beqn \Delta V&=& -\frac{h}{r^3},\qquad h:=\frac{GM_0L_{Cl}^2}{m^2c^2}
\left [1+\frac{\vec{L_{Cl}}\cdot\vec{S}}{2L_{Cl}^2}\right ].\eeqn
Here no symmetry between $M_0$ and $m$ is required, for 
the system is not written in the center of mass frame. Symmetrization gives instead of (\ref{GLW})
\beqn \frac{Mm}{M+m}V_{GLW}&=&-\frac{GMm}{r}-\frac{G(M^2+2Mm+m^2)L_{Cl}^2}{2mMc^2r^3}.\eeqn
Experimentally, the term $\displaystyle\frac{\vec{L_{Cl}}\cdot\vec{S}}{2L_{Cl}^2}$ is the order of $10^{-10}$ for Mercury and quite negligible. 
\footnote{By the way, the relation of a spin and revolution frequencies as in the moon is fulfilled classically from the tidal force after many years. If this is the case, quantization of angular momentums in quantum mechanics also is due to some kinematical balance as the final result of many times of scattering process. Then, the quantization effect can naturally become stronger as the particle state becomes lower. This is another interpretation of $\Delta E\Delta t\sim\hbar$. In this sense, quantum gravity is the theory of the final shape of the universe, rather than initial.}

\subsection{Conclusion for this section}
We discussed some paradoxes in general relativity and its possible solution. The crucial point is how to derive the $O(1/r^3)$ term of a potential. This can be derived in flat spacetime \cite{graphoto2} from the scalar retarded Newtonian potential by taking account of the Thomas precession. 

 An angular momentum is in fact not a Lorenz covariant 4 `momentum'. If we adopted a covariant definition $L$ of (\ref{ang}), equivalently $L'$ of (\ref{angl}), the result would lead to a self contradiction, for the terms in the $\sqrt{\qquad}$ of (\ref{GLW}) would be negative for relativistic $L,\; L'\to \infty$. Indeed, in relativistic quantum mechanics orbital angular momentums and spins are intrinsic properties of the 2 body system, and the change of their lengths (Lorentz contractions) by the motion of an observer does not affect physics. For instance, the $\bf L\cdot S$ interaction is a scalar product of $3$ dimentional vectors.

 If otherwise we extend $L$ to photons, $L''$ of (\ref{Angla}) is natural,
which implies the connection of an orbital angular momentum with a central force. However, naive treatment of the conservation law of angular momentum for $L''$ would contradict the experimental lens effect, for $r\to\infty\iff\sin\theta =0$.
\sect{Unification revisited: Including angular momentums}\label{Mom}
\subsection{Introduction to this section}
In this section, we shall apply the results of previous sections \ref{Sup} and \ref{Grav} to make more strict the previous discussion in the section $\ref{AHU}$. Spins and orbital angular momentum effects come into consideration. 
In particular, we assume that the equation has a solution also for $E=mc^2$ i.e., a rest particle. 
\subsection{Asymptotic behavior of the Dirac equation for central forces}
Let us solve the Dirac equation 
\eq\left [i\gamma^\mu\left (\frac{\part}{\part x^\mu}-i\frac{q}{\hbar}A_\mu\right )-\frac{mc}{\hbar}\right ]\psi =0 \label{Dir} \en 
for the type of potentials $A^\mu :=(\phi(r), 0, 0, 0)$ discussed in section \ref{Sing}, \ref{AHU}\cite{Kyoritu}. 
\beqn\mbox{If we take }\qquad \psi &=&\chi e^{-iEt/\hbar}\quad\left (cp^0:=i\hbar \frac{\part}{\part t}-q\phi (r)\right )\\
\mbox{(\ref{Dir}) reads }\qquad
\left (\begin{array}{cc}
p^0+mc & -i\hbar\part_i\sigma^i \\ 
-i\hbar\part_i\sigma^i & p^0-mc\end{array}\right )
\psi &=& 0\label{Dir2},\qquad\mbox{which commutes with}\\
\mbox{the total angular momentum}\qquad J^i&:=&\left (\begin{array}{cc}-i\hbar \epsilon^{ijk}r_j\part_k & 0\\ 
0 & -i\hbar \epsilon^{ijk}r_j\part_k \end{array}\right )
+\frac{\hbar}{2}\left (\begin{array}{cc}
\sigma^i & 0\\ 0 & \sigma^i \end{array}\right ).\\
\mbox{We take (the eigen value of }J^2/\hbar^2)&=:&j(j+1)\quad\mbox{ and}\label{J1}\quad\mbox{(the eigen value of }J_z/\hbar )=:m.\label{J2}\\
\mbox{There are two eigen states }\qquad 
\Phi_{j, m}^{(+)}&:=&\frac{1}{\sqrt{2(j+1)}}\left (\begin{array}{c}
\sqrt{j-m+1}\; Y_{j+1/2}^{m-1/2}\\ -\sqrt{j+m+1}\; Y_{j+1/2}^{m+1/2}
\end{array}\right ) \quad\mbox{ and}\\
\Phi_{j, m}^{(-)}&:=&\frac{1}{\sqrt{2j}}\left (\begin{array}{c}
\sqrt{j+m}\; Y_{j-1/2}^{m-1/2}\\ \sqrt{j-m}\; Y_{j-1/2}^{m+1/2}
\end{array}\right ) 
\mbox{ for (\ref{J1}), where}\\ 
Y_l^m(\theta , \phi ):=\epsilon\sqrt{\frac{(2l+1)(l-|m|)!}{4\pi (l+|m|)!}}\; &e^{im\phi}&\;\sin^{|m|}\theta\frac{d^{|m|}}{d(\cos\theta )^{|m|}}\left \{\frac{1}{2^ll!}\frac{d^l}{d(\cos\theta )^l}(\cos^2\theta -1)^l\right \} ,\\
\epsilon&:=&\left \{\begin{array}{cc}(-1)^m & (0<m)\\
1 & (m\leq 0)\end{array}\right.\quad\mbox{ are spherical harmonic functions.}\\
\mbox{As }\qquad\frac{\sigma^ir_i}{r}
\left (\begin{array}{cc}\Phi_{j, m}^{(+)} & \Phi_{j, m}^{(-)}\end{array}\right )&=&\left (\begin{array}{cc}
\cos\theta & e^{-i\phi}\sin\theta \\ e^{i\phi}\sin\theta & -\cos\theta \end{array}\right )\left (\begin{array}{cc} \Phi_{j, m}^{(+)}&\Phi_{j, m}^{(-)}\end{array}\right )=\left (\begin{array}{cc} \Phi_{j, m}^{(-)}&\Phi_{j, m}^{(+)}\end{array}\right ),\nonumber\\
\mbox{we can take }\qquad\psi (x^i)&=&\left (\begin{array}{c} f(r)\Phi_{j, m}^{(+)}\\ g(r)\Phi_{j, m}^{(-)}\end{array}\right )e^{-iEt/\hbar}\qquad\mbox{ as a solution of (\ref{Dir2}). Then, }\nonumber \\
(p^0+mc)f(r)\Phi_{j, m}^{(+)}-i\hbar\part_\mu\sigma^\mu 
g(r)\Phi_{j, m}^{(-)}\quad =&0&=\quad (p^0-mc)g(r)\Phi_{j, m}^{(-)}-i\hbar\part_\mu\sigma^\mu f(r)\Phi_{j, m}^{(+)}.\label{Dir3}\\ 
\mbox{With the identity }\quad \sigma^i\part_i=\frac{\sigma^ir_i\sigma^jr_j}{r^2}\sigma^k\part_k&=&\frac{\sigma^ir_i}{r^2}(r^j\part_j-i\epsilon^{jkl}\sigma_jr_k\part_l)=\frac{\sigma^ir_i}{r^2}\left (r\frac{\part}{\part r}-\frac{J^jJ_j-L^jL_j-S^jS_j}{\hbar^2}\right )\nonumber \\
&&\hspace{-77mm}\mbox{(for }\Phi_{j, m}^{(+/-)},\;\; l=j\pm\frac12\mbox{)}\; =\frac{\sigma^ir_i}{r^2}\left \{r\frac{\part}{\part r}-j(j+1)+(j\pm\frac12)(j+1\pm\frac12)+\frac34\right \}=\frac{\sigma^ir_i}{r^2}\left \{r\frac{\part}{\part r}+1\pm (j+\frac12)\right \},\label{sigr}\\
\mbox{(\ref{Dir3}) reads }\qquad 0=(p^0+mc)f(r)\Phi_{j, m}^{(+)}&-&i\hbar\frac{\sigma^ir_i}{r^2}\left \{r\frac{\part}{\part r}+1-(j+\frac12)\right \} g(r)\Phi_{j, m}^{(-)}\label{neu} \\ 
&&\hspace{-39mm}=(p^0-mc)g(r)\Phi_{j, m}^{(-)}-i\hbar\frac{\sigma^ir_i}{r^2}\left \{r\frac{\part}{\part r}+1+(j+\frac12)\right \} f(r)\Phi_{j, m}^{(+)}\label{neu2}\\ 
&&\hspace{-63mm}\mbox{ and so, }\qquad 0=\left [(p^0+mc)+\hbar^2\left (\frac{d}{dr}+\frac{1-2j}{2r}\right )\frac{1}{p^0-mc}\left (\frac{d}{dr}+\frac{2j+3}{2r}\right )\right ]f(r)\label{DKG1}\\ &&\hspace{-39mm}=\left [(p^0-mc)+\hbar^2\left (\frac{d}{dr}+\frac{2j+3}{2r}\right )\frac{1}{p^0+mc}\left (\frac{d}{dr}+\frac{1-2j}{2r}\right )\right ]g(r).\label{DKG2}\\
&&\hspace{-63mm}(\ref{DKG1})\iff \left [m^2c^4-(E-q\phi (r))^2-\hbar^2c^2\left (\frac{d^2}{dr^2}+\frac{2}{r}\frac{d}{dr}-\frac{j(j+1)-3/4}{r^2}\right )\right ]f(r)\nonumber \\ 
&&\hspace{-28mm}=-\hbar^2c^2\left \{\frac{-q}{E-q\phi (r)-mc^2}\frac{d\phi (r)}{dr}\left (\frac{2j+3}{2r}+\frac{d}{dr}\right )+\frac{2j+3}{2r^2}\right \} f(r),\label{DKG1'}\\
&&\hspace{-63mm}(\ref{DKG2})\iff \left [m^2c^4-(E-q\phi (r))^2-\hbar^2c^2\left (\frac{d^2}{dr^2}+\frac{2}{r}\frac{d}{dr}-\frac{j(j+1)-3/4}{r^2}\right )\right ]g(r)\nonumber \\ 
&&\hspace{-28mm}=-\hbar^2c^2\left \{\frac{-q}{E-q\phi (r)+mc^2}\frac{d\phi (r)}{dr}\left (\frac{1-2j}{2r}+\frac{d}{dr}\right )+\frac{1-2j}{2r^2}\right \} g(r).\label{DKG2'} 
\eeqn
If we take $j=\frac12$, (\ref{DKG2'}) reduces to the usual Klein-Gordon equation (\ref{KG}) with $l=0$. 

The first term of R.H.S. of (\ref{DKG1'}) ($\bf L\cdot S$ term) for a rest particle $E=mc^2$ 
 is unique in a sense that it always acts as an attractive force, whether or not the potential $\phi (r)$ is attractive or repulsive, as long as $\phi (r)$ is a negative power of $r$ or higher order (including the Yukawa type) interaction $\phi (r)=O(\frac{1}{r^\epsilon})$. 
In particular, if $\phi (r)\to r^{-n}$ $(0<n)$ as $r\to\infty$, the centrifugal force in (\ref{DKG1'}) is deformed by the R.H.S. to give 
\beqn \frac{j(j+1)-3/4}{r^2}\begin{array}{c}\mbox{the third term}\\ 
\overrightarrow{\qquad\qquad\qquad\qquad\qquad}\end{array}\frac{(j+1)^2-1/4}{r^2} \begin{array}{c}\mbox{the first term}\\ 
\overrightarrow{\qquad\qquad\qquad\qquad\qquad}\end{array}\frac{(j+1)^2-1/4-(2j+3)n/2}{r^2}.\nonumber
\eeqn
\begin{enumerate}
	\item If $n=1$, this effect vanishes and the centrifugal force does not change. The solution is well known for hydrogen atoms\cite{Schiff,Wein}.
	\item If $n< 2$, the third term $\sim 2mc^2q\phi f(r)$ of L.H.S dominates R.H.S.. 
	\item If $n=2$, these terms balance to give the dominant term $\displaystyle (2mc^2q\phi +\hbar^2c^2\frac{j^2-9/4}{r^2})f(r)$. This case is physically interesting for positroniums and photons (if they had spin $3/2$). 
	\item If $2<n=j+\frac12$, this effect is strong enough to cancel the dominant centrifugal force and $O(1/r^3)$ term is dominant. 
	\item If $2,\;\; j+\frac12<n$, the attractive $\hbar^2c^2\frac{(j+1)^2-1/4-(2j+3)n/2}{r^2}f(r)$ term is dominant. 
\end{enumerate}

The Yukawa type potential $\phi (r)\sim e^{-Mr}/r^{n'}$ is particularly interesting, when the first and second terms of R.H.S. can be dominant to give $\displaystyle M\hbar^2c^2\left (\frac{2j+3}{2r}+\frac{d}{dr}\right )f(r)$. This means that for a stopped spin $\frac12$ particle that feels a dominant Yukawa type force, $\bf L\cdot S$ term acts like a Newtonian potential. 
The possible cases are as follows.
\begin{itemize}
	\item The second term dominates the first. This can occur when $f(r)$ vanishes more rapidly than a polynomial of $1/r$ and 
\beqn\left \{\frac{d^2}{dr^2}+\left (\frac{2}{r}+M\right )\frac{d}{dr}\right \}f(r)&\sim& 0
\iff f(r)\sim A+Be^{-Mr},\eeqn 
and then, for $\propto\frac{1}{r}$ term to vanish, $A=0$. In addition, from (\ref{neu}) $g(r)$ begins with $r^{j-1/2}$ or exponential decay, but with (\ref{neu2}) $g(r)\sim \frac{iM\hbar}{q}r^{n'-2}$. Thus $n'=j+\frac32$ for consistency. 
	\item The second term does not dominate the first. This can occur only when $f(r)\sim r^{-j-3/2}$ and further 
\beqn\left (\frac{d^2}{dr^2}+\frac{2}{r}\frac{d}{dr}-\frac{(j+1)^2-1/4}{r^2}\right )f(r)&\sim& 0\\ 
\iff\qquad (j+3/2)(j+5/2)-(2j+3)&\sim& (j+1)^2-1/4.
\eeqn 
This consistency condition is an identity. Thus, the radial component eigen function $f(r)$ of a spin $\frac12$ particle can feel a dominant Yukawa type force of L.H.S. and still behave as a polynomial of $1/r$. 
%
\end{itemize}
 \indent Contrastingly. for $g(r)$ in (\ref{DKG2'}), the third term $2mc^2q\phi g(r)$ of L.H.S always dominates the two terms of R.H.S., except for the only possiblity the second term balance it when $\phi (r)\sim e^{-Mr}/r^{n'}$ and $g(r)\sim e^{-M'r}/r^{n''}$ give the effect of $\displaystyle 2mc^2q\phi g(r)\to (2mc^2 -\frac{\hbar^2MM'}{2m})q\phi g(r)$. However, in this case clearly the kinetic term is dominant. \\
Thus, $g(r)$ can feel a dominant Yukawa type potential iff $j=\frac12$, in the sense of section \ref{AHU} i.e., 
\beqn g(r)&=&a+\frac{b}{r}+\sum_{k=1, 2, 3\cdots}^\infty\sum_{n=l_k}^\infty 
d_{kn}e^{-kMr}r^{-n}+\cdots\qquad (a\mbox{ or }b\neq 0), \\
\mbox{ where }\qquad&&\frac{2qmc^2}{\hbar^2c^2}=\left \{\begin{array}{lll}
d_{1l_1}M^2/a,\quad & l_1=n'\quad& (a\neq 0)\\ d_{1l_1}M^2/b,\quad&l_1=n'+1\quad& (a=0\neq b)\end{array}\right. .\\
\mbox{Then, from (\ref{neu}) }&& \nonumber \\
f(r)&\propto& \frac{i\hbar r^{n'}}{2mc^2r^{n'}-qe^{-Mr}}\left \{-\frac{b}{r^2}+\sum_{k=1, 2, 3\cdots}^\infty\sum_{n=l_k}^\infty d_{kn}e^{-kMr}r^{-n}\left (-kM-\frac{n}{r}\right )+\cdots \right \}\nonumber.
\eeqn
\beqn\mbox{The inverse parity spinor }\quad\psi (x^i)&=&\left (\begin{array}{c} f(r)\Phi_{j, m}^{(-)}\\ -g(r)\Phi_{j, m}^{(+)}\end{array}\right )e^{-iEt/\hbar}\quad\mbox{ is also the (right handed) solution of (\ref{Dir2})}.\nonumber \\ 
&&\hspace{-56mm}\mbox{We can also take }\quad\psi (x^i)=\left (\begin{array}{c} g'(r)\Phi_{j, m}^{(-)}\\ f'(r)\Phi_{j, m}^{(+)}\end{array}\right )e^{-iEt/\hbar},\quad\left (\begin{array}{c} -g'(r)\Phi_{j, m}^{(+)}\\ f'(r)\Phi_{j, m}^{(-)}\end{array}\right )e^{-iEt/\hbar}\quad\mbox{ as other solutions},
\nonumber \eeqn 
where $f'(r), g'(r)$ satisfy the same equation as (\ref{neu}) and (\ref{neu2}), respectively with $m\to -m$.

Therefore, the mass term (Appendix \ref{SM}) for a $E=mc^2$ spin $\frac12$ particle with $j=\frac12$ that feels a dominant Yukawa type potential is 
\beqn -m\int_0^\infty \left (|f(r)|^2+|g(r)|^2\right )r^2dr\to -\infty ,\eeqn
which violates the normalization condition and leads to a contradiction(Appendix \ref{L2}).
 For a general spin $s=n/2$, 
\beqn &&\sigma^i\part_i=\frac{\sigma^ir_i}{r^2}\left \{r\frac{\part}{\part r}+s^2+s+(u\pm 1/2)(u\pm 1/2-2j-1)\right \}\qquad\\ 
\mbox{for }&&\Phi_{j, m},\;\; j=l+u\pm 1/2\in\{ |l-s|,\; |l-s|+1, \cdots |l+s|\},\qquad \mbox{ instead of (\ref{sigr}).}\nonumber\eeqn
\subsection{The relativistic electric and Newtonian potentials}
 According to S. Weinberg{\cite{WeinG}, the charge and gravitational mass in a two body problem behave as follows. 

 Let us consider a massless particle with the 3 momentum ${\bf q}$ and helicity $\pm j$, exchanged by the two particles $a,\; b$ with a 4 momentum $p_a^\mu :=(E_a,\; {\bf p_a}),\; p_b^\mu :=(E_b,\; {\bf p_b})$, a mass $m_a,\; m_b$, a spin $J_a,\; J_b$, a coupling constant to the photon $e_a,\; e_b$, and a coupling constant to the graviton $f_a,\; f_b$, respectively. For $j=1,\; 2$, the vertex amplitude of these interactions can be written as 
\beqn \frac{2ie(2\pi )^4\delta_{\sigma\sigma '}p_\mu e_\pm^{\mu *}(\hat 
q)}{(2\pi )^{9/2}[2E({\bf p})](2|{\bf q}|)^{1/2}}\qquad&& (j=1), \label{defe}\\
\frac{2if(8\pi G)^{1/2}(2\pi )^4\delta_{\sigma\sigma '}(p_\mu e_\pm^{\mu *}(\hat q))^2}{(2\pi )^{9/2}[2E({\bf p})](2|{\bf q}|)^{1/2}} &&\;\; (j=2)\; ,\label{deff}  \eeqn
where $G$ is the Newton constant and $e_\pm^{\mu}(\hat q)$ ($e_\pm^{\mu *}(\hat q)$) are (conjugate) polarization vectors of the two massless particles, orthogonal to $q^\mu$ and respectively satisfy 
\beqn \sum_\pm e_\pm^{\;\;\mu}(\hat q)e_\pm^{\;\;\nu *}(\hat q) & = & 
\Pi^{\mu\nu}(\hat q):=g^{\mu\nu} +(\bar q^\mu q^\nu +\bar q^\nu q^\mu)/(2|{\bf 
q}|^2),\;\; [\bar q^\mu :=(|{\bf q}|, -{\bf q})], \nonumber\\
\sum_\pm e_\pm^{\;\;\mu_1}(\hat q) e_\pm^{\;\;\mu_2}(\hat q)e_\pm^{\;\;\nu_1 
*}(\hat q)e_\pm^{\;\;\nu_2 *}(\hat q) & = & \frac12\left \{ \Pi^{\mu_1\nu_1}(\hat q)\Pi^{\mu_2\nu_2}(\hat q) +\Pi^{\mu_1\nu_2}(\hat q)\Pi^{\mu_2\nu_1}(\hat 
q)-\Pi^{\mu_1\mu_2}(\hat q)\Pi^{\nu_1\nu_2}(\hat q)\right \} . \nonumber\eeqn
 (\ref{defe}) and (\ref{deff}) are respectively the definition of the electric charge $e$ and the gravitational mass $f$. Then, the S matrix for this process is 
\beqn & & \frac{\delta_{\sigma_a\sigma_a '}\delta_{\sigma_b\sigma_b 
'}}{4\pi^2E_aE_bt}[e_ae_b(p_a\cdot p_b) +8\pi Gf_af_b\{ (p_a\cdot p_b)^2-m_a^2m_b^2/2\} ] \nonumber \\
& = & \frac{\delta_{\sigma_a\sigma_a '}\delta_{\sigma_b\sigma_b '}}{\pi t}
\left [-\frac{e_ae_b}{4\pi} +Gf_a\left (2E_a-\frac{m_a^2}{E_a}\right )f_bm_b\right ] \qquad\mbox{(if the particle $b$ is stopped)} ,\label{eg}\eeqn
where $t:=-(p_a -p_a ')\to 0$ is the transferred momentum. Therefore, we can identify $e_a$ as the electric charge of $a$. In the same way, effective gravitational mass behaves like 
\beqn \tilde m_a & := & f_a\{ 2E_a-(m_a^2/E_a)\} \nonumber \\
& \simeq & \left \{ \begin{array}{l}f_am_a \mbox{(when the particle $a$ is nonrelativistic)}\;\; .\\
2f_aE_a \mbox{(when the particle $a$ is relativistic)}\end{array}\right. \eeqn
 From (\ref{eg}) we can learn that the relativistic effective potential for this process in the center of mass frame is 
\eq \left [-\frac{e_ae_b}{4\pi t}\left (1+\frac{{\bf p}^2}{\sqrt{(\mu^2+{\bf p}^2)M^2-2{\bf p}^2\mu M+{\bf p}^4}}\right )+\frac{Gf_af_b}{2t}\left ( 4{\bf p}^2+\frac{(\mu^2+2{\bf p}^2)M^2-4{\bf p}^2\mu M+4{\bf p}^4}{\sqrt{(\mu^2+{\bf p}^2)M^2-2{\bf p}^2\mu M+{\bf p}^4}}\right ) \right ] ,\label{emp} \en 
where ${\bf p}:=({\bf p_a} -{\bf p_b})/2$, $M:=(m_a+m_b)c$, and $\mu :=m_am_bc/(m_a+m_b)$. 
These interactions monotonically increase as the relative velocity of both bodies grows. The running effect of these coupling constants is $\to\times 2$  for electricity and $\displaystyle\to\times\frac{8{\bf p}^2}{\mu M}$ for gravity (${\bf p}={\bf 0}\to\infty$). Even a  massless particle can feel the electromagnecity and gravity of a massless particle. 

 A spin 2 massless boson always reduces to the Einstein equation and thus is identified with a graviton. 

\subsection{Electro and weak magnetic masses}
 Let a left handed particle $P$ with a radius $a$, charge $Q$, mass $M$, velocity $v_i:=(0,\; 0,\; v)$, and the third component of weak hypercharge $T_3$, moving slowly compared with the light velocity. From (\ref{LAG}), the electroweak vector potentials  generated by $P$ can be written as 
\beqn A_i:=\frac{\mu_0Qv_i}{4\pi r}\;\;\mbox{ and }\;\; 
Z_i:=\frac{\mu_0T_3v_i}{4\pi r}
e^{-mr}, 
\eeqn
where $\mu_0$ is the permeability of vacuum and $m$ is the mass of $Z^0$ boson.
Then, the electroweak magnetic mass for $P$ is calculated as follows\cite{EM}:
\beqn B_i^{EM}&:=&({\bf\nabla\times A})_i=\frac{\mu_0Q({\bf v\times r})_i}{4\pi r^3} \\ 
\mbox{ and }\;\; B_i^{W}&:=&({\bf\nabla\times Z})_i=\frac{\mu_0T_3({\bf v\times r})_i}{4\pi r^3}(1+mr)e^{-mr}.\eeqn 
Noting ${\bf v\times r}=v(xe_y-ye_x)$ in terms of the bases $e_x,\; e_y$, the electroweak magnetic energy generated by $P$ is 
\beqn 
\int\frac{B^{EM\; 2}+B^{W\; 2}}{2\mu_0}dxdydz&=&\frac{1}{2\mu_0}\left (\frac{\mu_0v}{4\pi}\right )^2\int\left \{Q^2+T_3^2(1+2mr+m^2r^2)e^{-2mr}\right \}\frac{x^2+y^2}{r^6}dxdydz\nonumber \\ 
&=&\frac{1}{2\mu_0}\left (\frac{\mu_0v}{4\pi}\right )^2\frac23\int\left \{
Q^2+T_3^2(1+2mr+m^2r^2)e^{-2mr}\right \}\frac{x^2+y^2+z^2}{r^6}r^2drd\Omega\nonumber \\ 
&=&\frac{\mu_0Q^2}{4\pi}\frac{v^2}{3}\int_a^\infty\frac{dr}{r^2}+
\frac{\mu_0T_3^2}{4\pi}\frac{v^2}{3}\int_a^\infty 
(1+2mr+m^2r^2)e^{-2mr}\frac{dr}{r^2}\nonumber \\ 
&=&\frac{\mu_0Q^2}{4\pi}\frac{v^2}{3a}+
\frac{\mu_0T_3^2}{4\pi}\frac{v^2}{3}\left (
\frac{1}{a}+\frac{m}{2}\right )e^{-2ma}. 
\eeqn
Thus, we can express the total kinematical energy for $P$ as 
\beqn W&:=&\frac12\left (M+M_{EM}+M_W\right )v^2,\nonumber \\
\mbox{where}&&M_{EM}:=\frac{\mu_0}{4\pi}\frac{2Q^2}{3a}
\;\;\mbox{ and }\;\; M_{W}:=\frac{\mu_0}{4\pi}
\frac{2T_3^2}{3}\left (\frac{1}{a}+\frac{m}{2}\right )e^{-2ma}
\eeqn
are respectively the electro and weak magnetic masses. Interestingly, 
$M_W$ includes the contribution not diverges in the limit $a\to 0$. 
This can be a radiative origin of the small neutrino masses.
 Notice that if electromagnetic and weak interactions are of the same 
origin, we should take the summation of $B^{EM}$ and $B^{W}$ {\it before} 
taking absolute square, which has the effect of adding the cross term mass 
\beqn M_{EMW} &:=& \frac{\mu_0QT_3}{4\pi}\frac{4}{3}\int_a^\infty 
(1+mr)e^{-mr}\frac{dr}{r^2}\nonumber \\ 
&=& \frac{\mu_0}{4\pi}\frac{4QT_3}{3a}e^{-ma}\eeqn
to $W$. If $Q=-T_{3}$, the total mass converges for $a\to 0$.
\subsection{Conclusion for this section}
We took account of spins to make the discussion in section \ref{AHU} more strict. The previous result also holds for the Dirac equation with $j=1/2$ (total angular momentum), that a neutrino never feels a dominant $1/r^2$-like long range force as the usual standard model does not contain gravity, 
and so it feels a dominant Yukawa type potential and thus is not $L^2$ 
for $E=Mc^2$. Then, it is easy to show that the kinetic term can not cancel the divergence of a mass term. 
If above discussion is valid, we can conclude that neutrinos are massless at 
the tree level. Then, neutrinos naturally break the chiral symmetry. This can explain why weak bosons couple only to left handed fermions (the parity violation of weak interaction). Further we can easily calculate the electromagnetic mass for the Yukawa-type potential\cite{EM}. This is the effect of radiative corrections not included in static potentials. This idea may explain the smallness of neutrino mass without assuming the see-saw mechanism\cite{TY}. 

\sect{Conclusion} 
In this thesis we first constructed a type of singularities of a potential for 
the spherical symmetric Klein-Gordon equation. Instead of introducing a cut-off, we studied the effect of assuming that a  
potential $V$ has at least one $C^2$ class eigen function. The result  
crucially depends on the analytic property of the eigen function near its 
0 point. 

 Then we discussed a natural possibility that gravity and weak coupling constants $g_G$ and $g_W$ are defined after $g_{EM}$. According to this unified description of gauge fields via $U(1)$, a photon that feels a dominant $1/r^2$-like force (of centrifugal nature or not) can not create a long range force. Thus the photon can naturally be identified as a $Z^0$ boson. We found also that the mass term of the free field Lagrangian for a scalar particle that feels a dominant Yukawa type short range force always diverges in the limit of a rest mass and a long distance. The kinematical term can not cancel the divergence and thus such a scalar must be massless. This is almost valid for fermions if the total angular momentum $j=1/2$. 

 Gravity is the only long range force that can not be screened. It also acts between any particles with energies. Therefore, gravity always contributes to the $1/r$ term of a potential which is an integral constant for a $3$ dimensional Laplacian. 
 We tried to associate every pairs of the standard model elementary particles to their corresponding asymptotic expansion. The fact that the iterative solution inevitably includes several infinite series of different order in one expansion may be the origin of the non-commutative gauge invariance. In this form, we can naturally understand QCD color confinement via the normalization of gluons. 

 Based on few assumptions almost evident and independent of above discussions, we next considered the meaning of internal and external degrees of freedom for a 2 body problem. How many degrees of freedom are inherent to the two frames $X$ and $Y$? -We found a phase which can not reduce to the inertial motion of one frame. The Poincar\'{e} group with spins includes in itself an $U(1)$ phase $\theta$, or the angle between two spins. 
 This $\theta$ can formally be written as imaginary rotations via asymmetric spinor representations. We tried to derive gauge fields from this phase $\theta$. Such an idea to derive electromagnetic fields via the nonintegrability of a phase dates back to Dirac\cite{Dirac}. We extended his idea to other fields in this thesis, making use of the phase $\theta$. 

Furthermore, {\it there are 4 internal degrees of freedom of the past that can not be changed by future performance}. A rotational frame is not an inertial frame. Therefore, the spin of a particle on the origin of an inertial frame $X$ is the intrinsic property of itself, independent of motions of the other frame for the 2 body system. This implies geometric origin of the Cabibbo-Kobayashi-Maskawa matrix\cite{KobaMas}: The first 2 are due to the direction of spin X viewed from the other frame Y. The second 2 are due to anisotropic nature of the particle X itself, i.e., `the fact that a rotation around the navel and a rotation around the neck are different'. Both are included in steps of the section \ref{Sup}. Thus, {\it a neutrino oscillation, if it really happened, would not only imply nonzero neutrino masses, but also imply an internal structure for leptons, so far not detected}.

 As a spin-off, supersymmetry is regarded as a kind of Mach's principle for spinning frames. A transformation of two fermions operators to one boson operator is physically natural. For example, it really appears to treat Cooper pairs in superconductivity. However, an elementary boson can not always be divided into two fermions. Supersymmetric transformations in elementary particle physics defined as the mutual replacement of a fermion and a boson of the same mass are by no means natural. We can even regard them as a primary mistake to treat the rotational frame as an inertial frame. Such kinds of Ptolemaic or geocentric theories seem incorrect.

 To treat gravity, we first reviewed famous results in general relativity. Then we discussed several paradoxes, and possible solutions for them. Replacement of the potential with that of S. Weinberg\cite{WeinG} will give the correct energy dependence of gravity. We took account of angular momentums to make previous discussions more precise. Finally, we can explain why gravity is always attractive, by naturally treating it as a subsidiary force derived by the $\bf L\cdot S$ interaction of a dominant Yukawa-type potential.
\section*{Acknowledgments} 
I am grateful to Prof. Izumi Tsutsui and Dr. Toyohiro Tsurumaru for  
useful discussions. This thesis is partially motivated by 
some implications given by Prof. Tsutomu Kambe and Prof. Kazuo Fujikawa. 
I wish to thank Dr. Yukinari Sumino and Prof. Stanley J. Brodsky for helpful 
discussions. Prof. Tsutomu Yanagida and Prof. Tohru Eguchi, Prof. Yutaka Matsuo, and Dr. Ken-Ichi Izawa kindly read manuscripts and gave me suggestions for improvement. I also appreciate Dr. Syu Kato and my family Yuko, Takeo, Megumi, and Masahide for spiritual support. 
This thesis is partly supported by the Japan Scholarship Foundation. 
Computer facilities and printers, maintained by the people of our laboratory were very helpful to perform this study. 
Finally, I thank Dr. Maya Yokoi for her respectable life. 
\appendix
\setcounter{footnote}{1}
\section{Some remarks for section 2}\label{L2}
\subsection{Comment on uniqueness-1} 
In our construction of a generalized Taylor expansion of the type (g) 
in section \ref{Sing}, all coefficients in exponents are assumed real. 
This is because the convergence is not guaranteed if the coefficients 
include imaginary parts. 
However in some special cases, the expansions with imaginary 
coefficients are also defined. 
 For example, 
a type (g) expansion $f(z)$ is already in an ascending order, and multiplication of all terms by a common complex phase like 
$f(z)\to f(z)e^{i\theta (z)} \quad (\theta\in$ {\bf R} $)$ 
does not lose the convergence, if $\theta (z)$ converges. 
Further, we consider possible extensions including imaginary coefficients. \\ 
\begin{description}
	\item[(1)] If we assume that $f(z)=\sum_nf_n(z)$ converges uniformly and absolutely on any compact subset in the annular domain $D=\{ z\in {\bf C}|\; 0<|z|<r\}$, multiplication of all terms by different complex phases like $f_n(z)\to f_n(z)e^{i\theta_n(z)} \quad (\theta_n(z)\in$ {\bf R} $)$ does not lose the convergence
. If we do not expand the complex phases and keep the absolute ascending order, the expansion is uniquely defined. 

	\item[(2)] Even if $f(z)$ converges uniformly on compact sets in the real domain $D=\{ z\in {\bf R}|\; 0<|z|<r\}$ but does not absolutely converge, multiplication of finite number of terms respectively by different complex phases does not lose the convergence. An example (not included in the type (g)) is $f(z)=\sum_{n=1}^\infty (-1)^{n-1}f_n(z),\quad f_n(z)=(z^{-1/n}-1)\quad (0<z<1)$, where every term $f_n(z)$ is positive and larger than $(1-z)/n$, ascending power of $z$ and the sum is bounded as $0<f(z)<z^{-1}-1$. If we regard each $f_n(z)$ as a single term, the expansion is uniquely arranged in an ascending order. 
	\item[(3)] Replacement of an original term of $f(z)$ with several complex terms is also valid. 
\end{description}

 However, in cases (2) and (3) a careful estimation of limit value is needed. For example, this enables us to replace
$f(z)=(z^{-1}-1)-(z^{-1/2}-1)+e^{-2/z}+\cdots$ (for (2)) or $f(z)=1+e^{2/z}+\cdots$ (for (3))
 by 
$f(z)=e^{-iz}+e^{\sqrt{2}iz}+e^{-2/z}+e^{-2/z-i/z}+e^{-2/z+2iz}+\cdots
 (z\in {\bf R})$, 
when several number of the same absolute order terms appear. 
In this case, the meaning of an `ascending order' becomes unclear and we 
should consider all terms of the same absolute order at the same time. 
Notice that we must not expand the complex phases to keep the 
absolute ascending order and then the expansion is not unique.
\subsection{Comment on uniqueness-2}
The solution of the following `2 dimensional weak exterior Dirichlet problem' is not unique:\\
{\bf `The function $u(x, y)$ is defined on and on the exterior of the circle 
$x^2+y^2=a^2$ (called C), satisfying the Laplace equation $\Delta u(x, y)=0$, and being $0$ on C. Determine $u(x, y)$.'}\\
The proof is as follows. Let $u(x, y)$ be {\bf Re }$(z-\Frac{a^2}{z})$, where $z=x+iy$. 
Because of the Cauchy-Riemann equation, the real part of an analytic function becomes 
automatically harmonic, which shows that $u(x, y)$ thus defined is a nontrivial solution. It clearly is $0$ on C. In fact there are infinite number of solutions to the problem, because the equation is linear and we can replace $z\to 
\Frac{z^{n+1}}{a^n}$. Furthermore, if we include multi-valued functions, another type of solutions can be found by the following procedure.
\begin{enumerate}
\item Let us take two multi-valued functions $f(z)$ and $g(z)$ analytic on and on the exterior of C. Both are assumed to 
have a single-valued branch defined outside the cut ${\bf Im }z=0, 0\leq$ {\bf Re }$z$.
\item Combine them so as not to have a gap at the cut on C. For example, define a new function 
$h(z):= \Delta_g f(z)-\Delta_f g(z)$, where $\Delta_f:= f(ae^{+i0})- f (ae^{i(2\pi -0)}), \Delta_g:= g(ae^{+i0})-g(ae^{i(2\pi -0)})$ are the gaps of $f(z)$ and $g(z)$.
\item As the branch {\bf Re }$h(z)\quad (z=re^{i\theta}, 0\leq\theta\leq 2\pi )$ is continuous 
on C and is $0$ at $z=a$, there is the unique sine Fourier expansion of it on C, that is, {\bf Re }$h(ae^{i\theta})=\Sum_{n=1}^\infty h_n\sin(\Frac{n\theta}{2})$.
\item There is the corresponding function $\hat h(z):=\Sum_{n=1}^\infty 
-h_n(\Frac{a}{z})^{\Frac{n}{2}}$ such that {\bf Im }$\hat h(z)=\Sum_{n=1}^\infty h_n\sin(\Frac{n\theta}{2})$ on C.
\item Define $\hat H(x, y):=$ {\bf Re }$h(z)-${\bf Im }$\hat h(z)$, where $z:=x+iy$. Then, the branch 
$\hat H(x, y)$ can be a nontrivial solution to the problem. For example, 
$f(z)=\Frac{1}{\log z},\; g(z)=\Frac{1}{z\log z}+1$. Notice that $\hat H(x, y)$ is generally multi-valued on the cut except for the point $(a, 0)$.
\end{enumerate}
In fact, there are innumerable many solutions, for a Laplacian is simply an operator to take the average of every neighbor point next to the center point.
\subsection{Comments on $L^2$ normalizability condition} 
If $R(r)$ is not a $L^2$ function, it does not always mean a 
contradiction. I think that $\delta$ function like sharpness of $R(r)$
is not realistic but for $r\to\infty$ the `generalized expectation value' 
of a physical operator $A$ can be defined for $R(r)$ as follows:
\beqn <A>:=\lim_{L\to\infty}\Frac{\int_0^L r^{N-1}dr R^*(r)AR(r)}
{\int_0^L r^{N-1}dr R^*(r)R(r)}. \label{ko32} \eeqn
 However, the $L^2$ condition for an eigen function is naturally required. We crucially make use of the $L^2$ normalizability of an eigen function as the consistency condition in section \ref{AHU}. Precise discussion of the necessity of this condition will make a book\cite{Comp}. However, we here review some examples related to this problem. 

 As the first example\cite{Isham}, let $\widehat{A}$ and $\widehat{B}$ be arbitrary self conjugate operators such that 
\beqn [\widehat{A},\;\widehat{B}]=i\hbar\label{AB},\eeqn 
where $|a>$ is an eigen vector of $\widehat{A}$ for the corresponding eigen value $a$. Let us take diagonal matrix elements of (\ref{AB}) and we obtain 
\beqn <a|[\widehat{A},\; \widehat{B}]|a>&=&<a|\widehat{A}\widehat{B}-\widehat{B}\widehat{A}|a>=(a-a)<a|\widehat{B}|a>=0.\label{[]}\eeqn
Substituting (\ref{AB}) to L.H.S. of (\ref{[]}) and dividing both sides by $i\hbar$, we can prove the famous result $1=0$! From this paradox we can learn that careless treatment of normalizability of an eigen function leads to physically wrong (and mathematically nonsense) results. Indeed, in the case where eigen spectra of both operators are continuous, the eigen vector is not normalizable and the scalar product in (\ref{[]}) does not make sense. For example, if $\widehat{A}=x$ and $\widehat{B}=\widehat{p}=-i\hbar \Frac{\part}{\part x}$, 
we can take $|a>=e^{i\widehat{a}x}$ which is not $L^2$({\bf R}). 

 The second example is that we can not take a variation around a nonexistent 
equilibrium of a functional. Indeed, a continuous function on a finite closed set does not always have a maximum value\cite{Zenk} shown as follows. 
Let us take $D=(0,\; 1)$ and consider the maximum value of the functional 
\eq \Phi (f):=\int_0^1xf(x)^2dx\en
defined on a finite closed set 
\eq B=\left \{f\in L^2(D):\;\; \int_0^1f(x)^2dx\leq 1\right \}.\en 
Clearly, 
\eq\Phi (f)=\int_0^1xf(x)^2dx<\int_0^1f(x)^2dx\leq 1\en 
and if we take 
\beqn f_n(x)&=&\left \{\begin{array}{ll}
0&(0<x<1-\Frac{1}{n})\\ \sqrt{n}&(1-\Frac{1}{n}<x<1)\end{array} 
\right. ,\\
\Phi (f_n)&=&\int_{1-\Frac{1}{n}}^1nxdx=1-\Frac{1}{2n}\to 1\quad (n\to\infty ).
\eeqn
Thus, $\Phi (f)$ does not have a maximum value. This is because 
the finiteness condition of eigen values $\alpha_n$ of the operator 
$A\psi_n(x)=\alpha_n\psi_n(x)$
\beqn \sum_{n=1}^\infty |\alpha_n|^2=\int_D\int_D|\sum_{n=1}^\infty A\psi_n(x)\psi_n(y)|^2dxdy<\infty\eeqn 
is not satisfied in this case $A=x$, where $\{\psi_n(x)\}$ are 
complete and orthonormal bases such that 
\eq ||\psi_n(x)||=1,\qquad 
(\psi_n,\;\; \psi_m)=0\quad (n\neq m),\qquad Af=\sum_{n=1}^\infty 
\alpha_n(f,\;\; \psi_n)\psi_n.\en 
A more realistic example is known that 
the area a needle sweeps when it comes back to the same position but 
with both sides reversed (that is, after turning around) does not have 
a minimum\cite{Ochiai}. Thus, in general a variation problem does not necessarily have an 
extreme point. The $L^2$ condition is necessary to avoid such ill-posed 
problems. \\

 Notice that historically A. Einstein said to W. Heisenberg\cite{Ein}, 
{\it `Only theory can decide what we can observe'}. Indeed, in quantum 
mechanics 
we can find many examples when an eigen function does not satisfy normalization or required analyticity conditions or selection rules and thus is not physical. 
Only theoretically consistent states can be observed.\\

 Notice also that recently S. Weinberg said in his preprint\cite{Weinberg:2002kg} 
that \\
{\it `We include neutrinos in the radiation, neglecting the anisotropic part of their energy-momentum tensor, which makes possible a purely analytic treatment.'} \\
after 10 days of our preprint\cite{Mu3}. Further G. 't Hooft said
 in his preprint\cite{'tHooft:2002wz} that \\
{\it `$\cdots$There appears to be no objection against a wider use of such a procedure. If our higher order amplitudes exhibit infrared divergences, besides the ultra-violet ones, we could absorb them in $\Delta{\cal L}$ as well. This time, $\Delta{\cal L}$ is not expected to affect the coupling
strengths and the field operators, but a quantity that is ideally suited to be renormalized by such terms is the effective Coulomb potential. Thus, after already having dealt with the ultraviolet divergences, we add to the ``lowest order Lagrangian'' ${\cal L}^{\mbox{renorm}}$, a further term $\Delta{\cal L}$ 
that affects the Coulomb potential. We just borrow this term from the higher order corrections. As soon as these are calculated, we will be obliged to return the loan.'}\\

 Our view in section \ref{AHU} is closely related to these preprints. Both 
show that normalizability of the total Lagrangian is physically important. 
As we shall show in section \ref{Mom}, eigen functions for a `stopped' neutrino with $E=Mc^2,\;\; j=\frac12$ (total angular momentum) is not $L^2$, 
which may explain the smallness of neutrino mass without assuming the 
see-saw mechanism\cite{TY}. 

\section{Calculation of asymmetric spinor representations}\label{sigma}
Let us calculate the transformation rule for $\sigma_\mu\to g\sigma_\mu 
g'^\dagger$, where $g:=e^{\alpha^i\sigma_i}, \;\; g':=e^{\beta^i\sigma_i}
\quad (\alpha^i, \beta^i\in{\bf C}^3)$ are Lorenz transformations operating on $\sigma_\mu$ respectively from the left and the right.
\beqn
\mbox{With the identities } & & \{\sigma_i, \sigma_j\} =2\delta_{ij},\qquad [\sigma_i, \sigma_j]=2i\epsilon_{ijk}\sigma_k,\qquad\sigma_i\sigma_j=\delta_{ij}+i\epsilon_{ijk}\sigma_k\\
\mbox{it follows that }\;\;\sigma_i\sigma_\mu &=&\delta_{i\mu}+\delta_{0\mu}
\sigma_i+\frac12[\sigma_i, \sigma_\mu ]=2\delta_{i\mu}+2\delta_{0\mu}
\sigma_i-\sigma_\mu\sigma_i=\delta_{i\mu}+\delta_{0\mu}\sigma_i
-i\epsilon_{\mu ik}\sigma_k,\\
e^{\alpha^i\sigma_i} =\sum_{n=0}^\infty\frac{(\alpha^i\sigma_i)^n}{n!}
&=&\sum_{n=0}^\infty\frac{1}{(2n)!}(\alpha_1^{\;\; 2}+\alpha_2^{\;\; 2}
+\alpha_3^{\;\; 2})^{2n}\left (\sigma_0+\frac{\alpha^i\sigma_i}{2n+1}\right )
=:\sigma_0\cosh\alpha +\sigma_i\sinh\alpha^i\\
\mbox{and then, }\;\;\quad g\sigma_\mu g'^\dagger 
&:=& e^{\alpha^i\sigma_i}\sigma_\mu e^{\beta^{*j}\sigma_j}
=(\sigma_0\cosh\alpha +\sigma_i\sinh\alpha^i)\sigma_\mu
(\sigma_0\cosh{\beta^*} +\sigma_j\sinh{\beta^*}^j)\nonumber \\ 
&&\hspace{-35mm}=\;\;\sigma_\mu\cosh\alpha\cosh{\beta^*}+\left\{\delta_{i\mu}
+\delta_{0\mu}\sigma_i-i\epsilon_{\mu ik}\sigma_k\right\}\sinh\alpha^i
(\sigma_0\cosh{\beta^*} +\sigma_j\sinh{\beta^*}^j)
+\sigma_\mu\sigma_j\cosh\alpha\sinh{\beta^*}^j\nonumber \\ 
&&\hspace{-35mm}=\;\;\sigma_\mu\cosh\alpha\cosh{\beta^*}+\left\{\delta_{i\mu}
+\delta_{0\mu}\sigma_i-i\epsilon_{\mu ik}\sigma_k\right\}\sinh\alpha^i
\cosh{\beta^*}+\delta_{i\mu}\sigma_j\sinh\alpha^i\sinh{\beta^*}^j\nonumber \\
&&\hspace{-35mm}\qquad +\delta_{0\mu}(\delta_{ij}+i\epsilon_{ijk}\sigma_k)\sinh
\alpha^i\sinh{\beta^*}^j-i\epsilon_{\mu ik}(\delta_{kj}+i\epsilon_{kjl}\sigma_l)\sinh\alpha^i\sinh{\beta^*}^j+\sigma_\mu\sigma_j\cosh\alpha\sinh{\beta^*}^j
\nonumber \\ 
&&\hspace{-35mm}=\;\;\left\{\begin{array}{ll}
\sigma_0\left (\cosh\alpha\cosh{\beta^*}+\sinh\alpha^i\sinh{\beta^*}^i\right )
&\\ \quad 
+\sigma_i\left (\sinh\alpha^i\cosh{\beta^*}+\cosh\alpha\sinh{\beta^*}^i
+i\epsilon_{ijk}\sinh\alpha^j\sinh{\beta^*}^k\right )&(\mu =0)\\
\sigma_0\left (\sinh\alpha^\mu\cosh{\beta^*}+\cosh\alpha\sinh{\beta^*}^\mu 
-i\epsilon_{\mu ij}\sinh\alpha^i\sinh{\beta^*}^j\right )
+\sigma_\mu\cosh\alpha\cosh{\beta^*}&\\
\quad +\sigma_l\left\{i\epsilon_{\mu il}
\left (\cosh\alpha\sinh{\beta^*}^i-\sinh\alpha^i\cosh{\beta^*}\right )
+\sinh\alpha^\mu\sinh{\beta^*}^l+
\epsilon_{\mu ik}\epsilon_{kjl}\sinh\alpha^i\sinh{\beta^*}^j\right\}
&(\mu\neq 0)\end{array}\right.
\nonumber \\ 
&&\hspace{-35mm}=\;\;\sum_{n=0}^\infty\frac{1}{(2n)!}(\alpha_1^{\;\; 2}+\alpha_2^{\;\; 2}+\alpha_3^{\;\; 2})^{2n}\sum_{m=0}^\infty\frac{1}{(2m)!}({\beta^*}_1^{\;\; 2}+{\beta^*}_2^{\;\; 2}+{\beta^*}_3^{\;\; 2})^{2m}\nonumber \\
&&\hspace{-7mm}\times\left\{\begin{array}{ll}
\sigma_0\left (1+\frac{\alpha^i{\beta^*}^i}{(2n+1)(2m+1)}\right )
+\sigma_i\left (\frac{\alpha^i}{2n+1}+\frac{{\beta^*}^i}{2m+1}+i\epsilon_{ijk}
\frac{\alpha^j}{2n+1}\frac{{\beta^*}^k}{2m+1}\right )&(\mu =0)\\
\sigma_0\left (\frac{\alpha^\mu}{2n+1}+\frac{{\beta^*}^\mu}{2n+1}
-i\epsilon_{\mu ij}\frac{\alpha^i}{2n+1}\frac{{\beta^*}^j}{2m+1}\right )
+\sigma_\mu &\\
\quad +\sigma_l\left\{i\epsilon_{\mu li}
\left (\frac{\alpha^i}{2n+1}-\frac{{\beta^*}^i}{2m+1}\right )
+\frac{\alpha^\mu}{2n+1}\frac{{\beta^*}^l}{2m+1}+
(\delta_{li}\delta_{j\mu}-\delta_{l\mu}\delta_{ji})\frac{\alpha^i}{2n+1}
\frac{{\beta^*}^j}{2m+1}\right\}&(\mu\neq 0)\end{array}\right.\nonumber \\ 
&&\hspace{-35mm}=\;\;\left\{\begin{array}{ll}
\sigma_0\cosh (\alpha +{\beta^*} )+\overrightarrow{\sigma}\cdot\left (
\overrightarrow{\sinh\alpha}\cosh{\beta^*}+\overrightarrow{\sinh{\beta^*}}
\cosh\alpha +\; i\;\overrightarrow{\sinh\alpha}
\times\overrightarrow{\sinh{\beta^*}}\right )&(\mu =0)\\
\sigma_0\left (\overrightarrow{\sinh\alpha}\cosh{\beta^*}
+\overrightarrow{\sinh{\beta^*}}\cosh\alpha 
-\; i\;\overrightarrow{\sinh\alpha}\times\overrightarrow{\sinh{\beta^*}}
\right )&\\
\quad +\overrightarrow{\sinh\alpha}\left (\overrightarrow{\sigma}\cdot
\overrightarrow{\sinh{\beta^*}}\right )+\overrightarrow{\sinh{\beta^*}}
\left (\overrightarrow{\sigma}\cdot\overrightarrow{\sinh\alpha}\right )
+\overrightarrow{\sigma}\left (\times\; i\;\overrightarrow
{\sinh ({\alpha -\beta^*})}+\cosh (\alpha -{\beta^*})\right )&(\mu\neq 0)
\end{array}\right.\nonumber \\ 
&&\hspace{-35mm}\to\left [\begin{array}{ll}\left\{\begin{array}{ll}
\sigma_0\cosh (2\alpha )+\overrightarrow{\sigma}\cdot
\overrightarrow{\sinh (2\alpha )}&(\mu =0)\\
\;\sigma_0\;\overrightarrow{\sinh (2\alpha )}
+\; 2\;\overrightarrow{\sinh\alpha}\left (\overrightarrow{\sigma}\cdot
\overrightarrow{\sinh\alpha}\right )+\overrightarrow{\sigma}
&(\mu\neq 0)\end{array}\right.
&(\overrightarrow{\alpha}\to\overrightarrow{\beta^*})\\
\left\{\begin{array}{ll}
\sigma_0&(\mu =0)\\
-\; 2\;\overrightarrow{\sinh\alpha}\left (\overrightarrow{\sigma}\cdot
\overrightarrow{\sinh\alpha}\right )
+\overrightarrow{\sigma}\left (\times\; i\;\overrightarrow
{\sinh (2\alpha )}+\cosh (2\alpha )\right )&(\mu\neq 0)
\end{array}\right.&(\overrightarrow{\alpha}\to -\overrightarrow{\beta^*})\\
\left\{\begin{array}{ll}
\sigma_0\cosh (2x)+\overrightarrow{\sigma}
\cdot\left (\overrightarrow{\sinh (2x)}+2\cosh x
\cos y\;\overrightarrow{\sinh x}\times\overrightarrow{\sin y}\right )
&(\mu =0)\\
\sigma_0\left [\overrightarrow{\sinh (2x)}-2\cosh x
\cos y\;\overrightarrow{\sinh x}\times\overrightarrow{\sin y}
\right ]&\\
\quad +2\cos^2y\;\overrightarrow{\sinh x}
\left (\overrightarrow{\sigma}\cdot\overrightarrow{\sinh x}\right )
+2\cosh^2x\;\overrightarrow{\sin y}\left (
\overrightarrow{\sigma}\cdot\overrightarrow{\sin y}\right )&\\
\quad +\left (\cosh (2y)+\overrightarrow{\sin (2y)}\;\times\right )
\overrightarrow{\sigma}&(\mu\neq 0)
\end{array}\right.&
\left (\begin{array}{l}\overrightarrow{\alpha}\to\overrightarrow{\beta}\\ 
=:\overrightarrow{x}+i\;\overrightarrow{y}, \\ 
\overrightarrow{x}, \overrightarrow{y}\in{\bf R}^3\end{array}\right ).
\end{array}\right.\label{asym}\\
\mbox{Therefore, }
\qquad\qquad&&e^{x^i\sigma_i}\sigma_\mu e^{x^j\sigma_j}
\simeq\mbox{\bf Re }\sigma_\mu e^{2x^{*j}\sigma_j}
\simeq\mbox{\bf Re }e^{2x^i\sigma_i}\sigma_\mu ,\\
&&e^{iy^i\sigma_i}\sigma_\mu e^{-iy^j\sigma_j}
\simeq\mbox{\bf Re }\sigma_\mu e^{-2iy^j\sigma_j}
\simeq\mbox{\bf Re }e^{2iy^j\sigma_j}\sigma_\mu ,\eeqn
are approximately valid up to the first order of the pure infinitesimal boost 
$x^i\sigma_i$ or rotation $iy^i\sigma_i\;\; 
(\overrightarrow{x}, \overrightarrow{y}\in{\bf R}^3)$, 
where {\bf Re} is an operator to eliminate imaginary coefficients of 
$\sigma_\mu$, for example, {\bf Re} $(\delta_{j\mu}+\delta_{0\mu}\sigma_j
+i\epsilon_{\mu jk}\sigma_k)=\delta_{j\mu}+\delta_{0\mu}\sigma_j$.
\section{Quantum gravity with minimal assumptions}\label{QGMA}
We review here past works of quantum gravity from particle point of view.
The purpose is to construct the quantum field 
theory including gravity, based on 
physical assumptions as few as possible.
This consists of 4 subjects.

The first subject, and probably suits this purpose the most is the work by Steven Weinberg, in which he derived the Einstein equation from the Lorenz invariance of a S-matrix. According to his old paper\cite{Wein}, gravity is derived without assuming a curved space-time. Therefore, general covariance and the geometric property of gravity are possibly subsidiary or mere approximations.

The second subject is that, according to an effective field theory, we
can make a prediction without knowing the underlying fundamental
theory. For example, John F. Donoghue\cite{Dono} calculated one loop quantum
corrections to the Newtonian potential explicitly, by assuming the 
Einstein-Hilbert action and fluctuations around the flat metric, and by
making use of the result of 't Hooft and Veltman\cite{tHVel}. The potential naturally 
contains classical corrections of general relativity\cite{Dono}. 

The third subject is, what will happen if we loosen the
assumption on coordinates in the standard model that all physical
coordinates are transformed to the Minkowski space-time by a Poincar\'{e} transformation. There are some troubles in treating gravitational fields under classical approximations assuming a curved space-time\cite{Wald}. 
It is known that for the standard model of elementary particles, the anomaly cancellation condition in a curved space-time with torsion is the same as in a flat space-time\cite{Doba}.

The fourth subject is to clarify the inevitable ambiguities of a
theory. For example, the vacuum state in a curved space-time is not
unique and there exist several theories those can not be distinguished
by finite times of measurements\cite{Wald}. This is a theorem on
the ambiguity related to the problem of divergence. For another example,
a higher-derivative theory includes non-physical solutions those can not
be Taylor expanded. This can be the origin of the gauge ambiguity. If we
exclude superfluous solutions by imposing the perturbative constraint
conditions, it means a gauge fixing and the theory is reduced to local and lower-derivative\cite{Simon}. This treatment is  known to be equivalent to the treatment of a constraint system by Dirac brackets\cite{Wein}. 
\section{Gravitational lens effects in special relativity}\label{Lens}
The fact $\delta_{NG}=0$ is clear if we take the limit $L'\to\infty$ at first, but we can learn something from detailed calculation process without taking this limit. The solution of (\ref{spepr}) is 
\vspace{-4.1mm}\beqn z&=&(1+\epsilon\cos\gamma\varphi )/l_0,\qquad\gamma^2 := 1-\left (\frac{GM_0}{L'c}\right )^2 \\
\mbox{or in the same notation as in (\ref{solens}), }\quad
&=&\frac{GM_0E'}{(L'\gamma )^2}+\left (z_0-\frac{GM_0E'}{(L'\gamma )^2}\right )\cos \gamma\varphi ,\quad
\mbox{which vanishes at }\nonumber\\ 
\cos\gamma\varphi &=&\frac{GM_0E'}{GM_0E'-(L'\gamma )^2z_0},\\
\mbox{where taking }\quad\varphi =0,\quad &&\mbox{(\ref{energ}) and (\ref{angl}) reduce to }\nonumber \\
-\frac{v^2}{1-(v/c)^2}=c^2\left \{1-\left (E'+\frac{a}{r}\right )^2\right \}&\Rightarrow& c^2\left \{1-\left (E'+az_0\right )^2\right \}
=-(L'z_0)^2\nonumber \\
&&\hspace{-70mm}\Rightarrow z_0=\frac{1}{a^2c^2-L'^2}\left (-ac^2E'\pm
\sqrt{(ac^2E')^2+c^2(1-E'^2)(a^2c^2-L'^2)}\right ),\\
\cos\gamma\varphi =\mp\frac{GM_0E'}{\sqrt{(GM_0)^2+(L'c)^2(E'^2-1)}}
&\to& \mp\frac{GM_0}{L'c}=\mp\frac{GM_0z_0}{c^2}\quad (E'\to\infty ),
\mbox{ for }ds\to 0.\\ \mbox{Therefore, the angle between the two }
&&\hspace{-10mm}\mbox{asymptotic lines (in Newtonian gravity) is}\\
\hspace{-56mm}\delta_{NG}&=&\frac{2GM_0z_0}{c^2}.
\eeqn 
\section{The Lagrangian for the standard model}\label{SM}
We can write the Lagrangian for the standard model in the following  concise form\cite{Doba}: 
 
\beqn {\cal L }_{SM}&:=&{\cal L }_m+{\cal L }_{Higgs}+{\cal L }_{YM}+{\cal L }_{Yukawa}, \qquad\mbox{where}\\ 
\mbox{\bf matter terms}\qquad\qquad\;\;\; {\cal L }_m&:=&i\bar{\cal Q }\gamma^\mu D_\mu^{\cal Q }{\cal Q }+i\bar{\cal L }\gamma^\mu D_\mu^{\cal L }{\cal L }, \label{lLm}\\ 
\mbox{\bf Higgs terms}\qquad\quad\;\;\; {\cal L }_{Higgs } & := & (D^\mu\phi )^\dagger (D_\mu\phi )-\lambda (\phi^\dagger\phi -\frac{\mu^2 }{2\lambda })^2, \label{LH} \\
\mbox{\bf Yang-Mills terms}\qquad {\cal L }_{YM } & := & -\frac14(\nabla_\mu G_\nu^{\;\; a }-\nabla_\nu G_\mu^{\;\; a } +g_Sf^{abc }G_\mu^{\;\; b }G_\nu^{\;\; c})^2\nonumber \\
&&-\frac14(\nabla_\mu W_\nu^{\;\; a }-\nabla_\nu W_\mu^{\;\; a } +gf^{abc }W_\mu^{\;\; b }W_\nu^{\;\; c})^2-\frac14(\nabla_\mu B_\nu-\nabla_\nu B_\mu )^2,\label{YM}\\
\mbox{\bf Yukawa terms}\qquad {\cal L }_{Yukawa} & := & 
-\overline{P_L{\cal Q }}\Phi H_{\cal Q}P_R{\cal Q }-\overline{P_R{\cal Q }}\;\overline{\Phi H_{\cal Q}}P_L{\cal Q }-\overline{P_L{\cal L }}\Phi H_{\cal L}P_R{\cal L}-\overline{P_R{\cal L}}\;\overline{\Phi H_{\cal L}}P_L{\cal L}, \label{LY} \\ 
\mbox{where }\; \mbox{\bf quark fields}\qquad {\cal Q } &:=& \left (\begin{array}{ccc} u^\alpha & c^\alpha & t^\alpha \\ d^\alpha  & s^\alpha & b^\alpha \end{array}\right )=:\left (\begin{array}{c}{\cal U} \\				{\cal D} \end{array}\right ), \label{Q } \\
\mbox{\bf lepton fields}\qquad {\cal L} &:=& \left (\begin{array}{ccc} \nu_e&\nu_\mu&\nu_\tau \\
e^-&\mu^- & \tau^- \end{array}\right )=:\left(\begin{array}{c} {\cal N} \\ {\cal E} \end{array}\right), \label{L}\\ 
\mbox{and }\;\mbox{\bf Higgs scalars}\quad\; 
\Phi &:=&\left (\begin{array}{cc}\phi^{0\dagger}& i\phi^+ \\ i\phi^-&\phi^0 \end{array}\right ) \quad\mbox{such that}\quad\langle 0| \phi (x)|0\rangle  =:\frac{1 }{\sqrt 2 }\left (\begin{array}{c}0 \\ v \end{array}\right ),\quad\mbox{where}\label{H}\\ 
\phi (x)&:=&\left (\begin{array}{c}i\phi^+ \\ \phi^0 \end{array}\right ):=\frac{1 }{\sqrt 2 }\left (\begin{array}{c}\varphi_2 +i\varphi_1 \\ v+\varphi -i\varphi_3 \end{array}\right ),\qquad \phi^-:=(\phi^+)^\dagger\label{VEV}\\
\mbox{are }SU(2)\mbox{ doublets, each }&\mbox{with}& \mbox{ covariant derivatives }\nonumber \\ 
D_\mu^{\cal Q } & := & \nabla_\mu -ig_SG_\mu -igW_\mu^{\;\; a }T^a P_L -ig'(Y_L^{\cal Q } P_L +Y_R^{\cal Q } P_R)B_\mu ,  \label{DQ} \\
 D_\mu^{\cal L } & := & \nabla_\mu -igW_\mu^{\;\; a }T^a P_L -ig'(Y_L^{\cal L } P_L +Y_R^{\cal L }P_R)B_\mu ,\label{DL} \\
D_\mu & := & \nabla_\mu -igW_\mu^{\;\; a }T^a-ig'B_\mu Y,\quad\mbox{where}
\label{DH} \\
Y:=\frac12\left (\begin{array}{cc}1&0 \\ 0&1 \end{array}\right ),\quad 
T^1&:=&\frac12\left (\begin{array}{cc}0&1 \\ 1&0 \end{array}\right ),\quad 
T^2:=\frac12\left (\begin{array}{cc}0&-i \\ i&0 \end{array}\right ),\quad 
T^3:=\frac12\left (\begin{array}{cc}1&0 \\ 0&-1 \end{array}\right ),\label{Ts} \\ 
&&\hspace{-56mm}\lambda_{\alpha\beta }^1:=\frac12\left (\begin{array}{ccc}0&1&0 \\ 1&0&0\\ 0&0&0 \end{array}\right ),\;\;\lambda_{\alpha\beta }^2:=\frac12\left (\begin{array}{ccc}0&-i&0 \\ i&0&0\\ 0&0&0 \end{array}\right ),\;\;\lambda_{\alpha\beta }^3:=\frac12\left (\begin{array}{ccc}1&0&0 \\ 0&-1&0\\ 0&0&0 \end{array}\right ),\;\;\lambda_{\alpha\beta }^4:=\frac12\left (\begin{array}{ccc} 0&0&1 \\ 0&0&0\\ 1&0&0 \end{array}\right ),\nonumber \\ 
&&\hspace{-56mm}\lambda_{\alpha\beta }^5:=\frac12\left (\begin{array}{ccc}0&0&-i \\ 0&0&0\\ i&0&0 \end{array}\right ),\;\;\lambda_{\alpha\beta }^6:=\frac12\left (\begin{array}{ccc} 0&0&0 \\ 0&0&1 \\ 0&1&0\\ \end{array}\right ),\;\;\lambda_{\alpha\beta }^7:=\frac12\left (\begin{array}{ccc}0&0&0 \\ 0&0&-i \\ 0&i&0\\ \end{array}\right ),\;\;\lambda_{\alpha\beta }^8:=\frac{1}{2\sqrt{3}}\left (\begin{array}{ccc}1&0&0 \\ 0&1&0 \\ 0&0&-2\\ \end{array}\right )\nonumber\\
\mbox{and {\bf Yukawa couplings}}\qquad H_{\cal Q }&:=&\left (\begin{array}{cc} H_{\cal U} & 0 \\ 0 &H_{\cal D} \end{array}\right )\quad ,\qquad 
H_{\cal L}=:\left(\begin{array}{cc} 0&0 \\ 0&H_{\cal E} \end{array}\right). \label{Y}  
\eeqn
Here $T$ stands for the transposition, the upper index $\alpha$ and $-,\; 0,\; +$ respectively for colors and electric charges, while $a$ for spinor indices. \footnote{The covariant derivative 
$\nabla_\mu$, where 
\beqn{t_{\mu_1\cdots\mu_q }^{\sigma_1\cdots\sigma_p } }_{;\nu }: & = & \nabla_\nu t_{\mu_1\cdots\mu_q }^{\sigma_1\cdots\sigma_p } \nonumber \\
 & := & \part_\nu t_{\mu_1\cdots\mu_q }^{\sigma_1\cdots\sigma_p } +\hat\Gamma_{\; \nu\rho }^{\sigma_1 } 
t_{\mu_1\cdots\mu_q }^{\rho\sigma_2\cdots\sigma_p } +\cdots + \hat\Gamma_{\; 
\nu\rho }^{\sigma_p }t_{\mu_1\cdots\mu_q }^{\sigma_1\sigma_2\cdots\rho } 
-\hat\Gamma_{\; \nu\mu_1 }^{\rho }t_{\rho\mu_2\cdots\mu_q }^{\sigma_1\cdots\sigma_p }-\cdots-\hat\Gamma_{\; \nu\mu_q }^{\rho } 
t_{\mu_1\mu_2\cdots\rho }^{\sigma_1\cdots\sigma_p },\\
\mbox{with the }&&\mbox{\bf Christoffel symbol}\qquad 
\Gamma_{\; \mu\nu }^\rho :=\frac12g^{\rho\sigma }(\part_\mu 
g_{\nu\sigma } +\part_\nu g_{\mu\sigma }
-\part_\sigma g_{\mu\nu }) \label{Chri } \eeqn
for a tensor $t_{\mu_1\cdots\mu_q }^{\sigma_1\cdots\sigma_p }$ with a contravariant rank $p$ and a covariant rank $q$, is equivalent to $\part_\mu$ in the usual standard model. Some can possibly generalize it\cite{Doba} to include the 
\beqn  \mbox{\bf torsion }\qquad T_{\; \mu\nu }^\sigma &:=&\hat\Gamma_{\; \mu\nu }^{\sigma }-\hat\Gamma_{\; \nu\mu }^\sigma=-T_{\; \nu\mu }^\sigma\quad 
\mbox{and} \label{tort }\\ 
\mbox{\bf curvature}\qquad R_{\sigma\mu\nu }^\rho &:=&\part_\mu\hat\Gamma_{\; \nu\sigma }^\rho -\part_\nu\hat\Gamma_{\; \mu\sigma }^\rho +\hat\Gamma_{\; \nu\sigma }^\eta \hat\Gamma_{\; \mu\eta }^\rho-\hat\Gamma_{\; \mu\sigma }^\eta\hat\Gamma_{\; \nu\eta }^\rho =-R_{\; \sigma\nu\mu }^\rho\; . \label{Riem }\eeqn 
However, throughout this thesis the metric is Minkowski $g_{\mu\nu}=\eta_{\mu\nu}$ and they are neglected. In the same way, the usual standard model assumes massless neutrinos. Then, total degrees of freedom for the standard model (in flat spacetime) are 18 (3 for coupling constants $g_S,\; g',\; g$; 2 for Higgs parameters $\mu^2,\; \lambda$; 9 for Yukawa couplings i.e. masses respectively for every quark or lepton except for massless neutrinos; 4 for $V_{\mbox{KM}}$ of which 1 is a complex phase and causes the $CP$-violation for the weak interaction). }
\beqn P_L:=\frac{1-\gamma_5 }{2 },\;\;P_R:=\frac{1 +\gamma_5 }{2 } \eeqn
respectively are projective operators to left and right chiralities of fermions. $G_\mu, W_\mu, B_\mu$ respectively are gauge fields and transform under the following infinitesimal transformations of the corresponding group $SU(3)_c, SU(2)_L, U(1)_Y$ with completely anti symmetric structure constants $f^{abc },\; \epsilon^{abc }$ respectively for $SU(3)_c, SU(2)_L$  
\beqn q^\alpha (x) & \to & q^\alpha (x) 
-i\theta_c^a(x)\frac{\lambda_{\alpha\beta }^a }{2 }q^\beta (x),\qquad (q\mbox{ is an arbitrary flavor of quarks }u,d,s\cdots ) \\
\phi (x) & \to & \phi (x) -i\theta^a(x)T^a\phi (x), \nonumber \\
{\cal L } & \to & {\cal L }-i\theta_L^a(x)T^aP_L{\cal L }, \nonumber \\ 
{\cal Q } & \to & {\cal Q }-i\theta_L^a(x)T^aP_L{\cal Q }, \\
\phi (x) & \to & \phi (x)-i\theta_Y(x)Y\phi (x), \nonumber \\
\psi & \to & \psi (x)-i\theta_Y(x)({\bf Y_L }P_L +{\bf Y_R }P_R)\psi (x)\quad (\psi\mbox{ is an arbitrary flavor of fermions }e,\nu ,u,d\cdots ) \label{hct } \eeqn
as follows: 
\beqn G_\mu^a & \rightarrow & G_\mu^a-\frac1{g_s }\partial_\mu\theta_c^a(x) + 
f^{abc }\theta_c^b(x)G_\mu^c,\; \; a=1\cdots 8 ,\label{45 } \\
W_\mu^a & \rightarrow & W_\mu^a-\frac1g\partial_\mu\theta_L^a(x) + 
\epsilon^{abc }\theta_L^b(x)W_\mu^c,\; \; a=1\cdots 3 ,\label{46 } \\
B_\mu^a & \rightarrow & B_\mu^a +\frac{1 }{g' }\partial_\mu\theta_Y^a(x). \label{47 } 
\eeqn
The hypercharge matrices $Y_L^{\cal Q}$, $Y_L^{\cal L}$, $Y_R^{\cal Q}$, $Y_R^{\cal L}$ concisely describe the eigenvalues of $2\times 2$ diagonal matrices ${\bf Y}$  (with indices $L$, $R$ because their dependence on chirality) which depends on $\psi$ belongs to which of $\cal U$, $\cal D$, $\cal E$, $\cal L$. They satisfy $Q=T^3 +Y$ for every field, where $T^3$ is the matrix defined in (\ref{Ts}) and $Q$ is the electric charge of $\psi$. The explicit definitions of them and $f^{abc}$ are as follows\cite{Doba}\cite{TY}: 
\begin{center}$\hspace{-14mm}\begin{array}{llcl}{\bf Table \hspace{1mm} \ref{SM}.1: } & \mbox{Relations of hypercharges and }& \qquad\qquad & \mbox{{\bf Table \ref{SM}.2\hspace{1mm}: }Explicit values of }f^{ABC}.\\ & \mbox{electric charges for fermions.}& \qquad &\end{array}$\\ 
$\begin{array}{ll}
\begin{tabular}{| c| c| c| c| c| c| c| c|}
\hline  & ${\cal U}_R$ & ${\cal U}_L$ & ${\cal D}_R$ & ${\cal D}_L$ & ${\cal E}_R$ & ${\cal E}_L$ & ${\cal N}_L$ \\ 
\hline & & & & & & & \\ 
$Y$ & $\displaystyle\frac23$ & $\displaystyle\frac16$ & $\displaystyle -\frac13$ & $\displaystyle\frac16$ & $\displaystyle -\frac12$ & $\displaystyle -1$ & $\displaystyle -\frac12$ \\ & & & & & & & \\ \hline & & & & & & & \\ $T_3$ & $0$ & $\displaystyle\frac12$ & 0 & $\displaystyle -\frac12$ & $\displaystyle\frac12$ & $0$ & $\displaystyle -\frac12$  \\ & & & & & & & \\ \hline & & & & & & & \\ $Q$ & $\displaystyle\frac23$ & $\displaystyle\frac23$ & $\displaystyle -\frac13$ & $\displaystyle -\frac13$ & $0$ & $-1$ & $-1$ \\ & & & & & & & \\ \hline 
\end{tabular}
&
\begin{tabular}{| c| c| c| c| c| c| c| c| c| c |}
\hline & & & & & & & & & \\ $abc$ & 123 & 147 & 156 & 246 & 257 & 345 & 367 & 458 & 678 \\ & & & & & & & & & \\ 
\hline & & & & & & & & & \\ $f^{abc}$ & 1 & $\displaystyle\frac12$ & $\displaystyle -\frac12$ & $\displaystyle\frac12$ & $\displaystyle\frac12$ & $\displaystyle\frac12$ & $\displaystyle -\frac12$ & $\displaystyle\frac{\sqrt{3}}{2}$ & $\displaystyle\frac{\sqrt{3}}{2}$ \\ & & & & & & & & & \\ \hline 
\end{tabular}
\end{array}$
\end{center}
Parameters $g_s$, $g$, $g'$ and $\theta_c^a(x)$, $\theta_L^a(x)$, $\theta_Y(x)$ respectively stand for the coupling constants of the corresponding groups and the amount of local gauge transformations. 

 Photons and $Z^0$ bosons are defined as 
\beqn A_\mu &:=&\sin\theta_WW_\mu^3+\cos\theta_WB_\mu ,\\
Z_\mu &:=& \cos\theta_WW_\mu^3-\sin\theta_WB_\mu ,\\
\mbox{where}\quad \theta_W&&\mbox{is the Weinberg angle defined by }
\qquad \tan\theta_W:=\frac{g'}{g}.\label{defW}
\eeqn 
 Finally, if we diagonalize the Yukawa terms by some unitary transformations between different chiral fermions generations
\footnote{Notice that in our notation of writing fermions by 
$2\times 3$ matrices, generation transformation matrices rather 
`acts as if from the right'. }
\beqn {\cal D}'_{L, R}&=&U^{\cal D}_{L, R}\;{\cal D}_{L, R},\qquad
{\cal U}'_{L, R}=U^{\cal U}_{L, R}\;{\cal U}_{L, R},\qquad
{\cal E}'_{L, R}=U^{\cal E}_{L, R}\;{\cal E}_{L, R}\\ 
\mbox{into the form }\qquad{\cal M}^i_{\mbox{diag}}&=&v\; U^i_LH_iU^{i\dagger}_R,\qquad\mbox{where }\qquad i={\cal U},\;\;{\cal D},\;\;{\cal E},
\eeqn
$W_\mu$ interactions change as 

\beqn i\bar{\cal Q}\gamma^\mu W_\mu P_L{\cal Q}&=&
\frac{g}{2}\left (\overline{{\cal U}_L}\gamma^\mu W^3_\mu{\cal U}_L
+\sqrt{2}\;\overline{{\cal U}_L}\gamma^\mu W^-_\mu{\cal D}_L
+\sqrt{2}\;\overline{{\cal D}_L}\gamma^\mu W^+_\mu{\cal U}_L
-\overline{{\cal D}_L}\gamma^\mu W^3_\mu{\cal D}_L\right ).\\
&&\hspace{-40mm}\mbox{Therefore }\quad\frac{g}{\sqrt{2}}\left (\overline{{\cal U}_L}\gamma^\mu 
W^-_\mu{\cal D}_L+\overline{{\cal D}_L}\gamma^\mu W^+_\mu{\cal U}_L\right )=
\frac{g}{\sqrt{2}}\left (\overline{{\cal U}'_L}\gamma^\mu 
W^-_\mu V_{\mbox{\small KM}}{\cal D}'_L+\overline{{\cal D}'_L}V_{\mbox{\small KM}}^\dagger\gamma^\mu W^+_\mu{\cal U}'_L\right ),\\
\mbox{where }&& V_{\mbox{\small KM}}:=U_L^{\cal U}U_L^{{\cal D}\dagger}\qquad
\mbox{is the Cabibbo-Kobayashi-Maskawa matrix.}\\
&&\hspace{-40mm}\mbox{From (\ref{LH}) and (\ref{LY}) we find}\qquad\phi^\dagger\phi =\frac{\mu^2}{2\lambda}=\frac{v^2}{2}\\
&&\hspace{-40mm}\mbox{and then, masses of gauge bosons are }\qquad 
M_W=\frac{gv}{2},\qquad M_Z=\frac{M_W}{\cos\theta_W},\qquad M_A=0.\label{ZW}\\ 
&&\hspace{-40mm}\mbox{Experimental values of them are\cite{TY}\cite{Part}}\nonumber\\ 
&& v=246\mbox{ GeV}/c^2,\quad \sin^2\theta_W=0.2276\pm 0.0075,\\ 
&& M_W=80.423\pm 0.39+i(2.118\pm 0.042)\mbox{ GeV}/c^2,\\ 
&& M_Z=91.1876\pm 0.0021+i(2.4952\pm 0.0023)\mbox{ GeV}/c^2,\label{EZW}\\
&&\hspace{-40mm}\mbox{where an imaginary mass has the meaning of decay width\cite{TY}}\qquad \Gamma =1/\tau ,\\
&&\hspace{-40mm}\mbox{where }\tau\mbox{ is the life time of the corresponding particle.} \nonumber\eeqn

 Then, the explicit form of the electroweak Lagrangian is as follows\cite{FJ}: 
\beqn {\cal L}&=&\frac12(\part_\mu\varphi )^2-\lambda v^2\varphi^2 
-\lambda v\varphi^3-\frac{\lambda}{4}\varphi^4+W_\mu^+W^{-\mu}
\left (M_W+\frac{g}{2}\varphi\right )^2-\frac12|\part_\mu W_\nu^+ 
-\part_\nu W_\mu^+|^2\nonumber \\ 
&&+\frac12(Z_\mu )^2\left (M_Z+\frac{\sqrt{g^2+g'^2}}{2}\varphi\right )^2-\frac14(\part_\mu Z_\nu -\part_\nu Z_\mu )^2 
-\frac14(\part_\mu A_\nu -\part_\nu A_\mu )^2 \nonumber \\ 
&&+\frac{ig}{2}(W^{+\mu}W^{-\nu}-W^{-\mu}W^{+\nu})
[\part_\mu (Z_\nu\cos\theta_W+A_\nu\sin\theta_W)-
\part_\nu (Z_\mu\cos\theta_W+A_\mu\sin\theta_W)] \nonumber \\ 
&&+\frac{ig}{2}(\part^\mu W^{+\nu}-\part^\nu W^{+\mu}) 
[W_\mu^-(Z_\nu\cos\theta_W+A_\nu\sin\theta_W) 
-W_\nu^-(Z_\mu\cos\theta_W+A_\mu\sin\theta_W)] \nonumber \\ 
&&-\frac{ig}{2}[(Z_\nu\cos\theta_W+A_\nu\sin\theta_W)W_\mu^+ 
-(Z_\mu\cos\theta_W+A_\mu\sin\theta_W)W_\nu^+] 
(\part^\mu W^{-\nu}-\part^\nu W^{-\mu}) \nonumber \\ 
&&+\frac{g^2}{4}(W_\mu^{+}W_\nu^{-}-W_\nu^{+}W_\mu^{-})^2 
-\frac{g^2}{2}|W_\mu^+ (Z_\nu\cos\theta_W+A_\nu\sin\theta_W)-
W_\nu^+(Z_\mu\cos\theta_W+A_\mu\sin\theta_W)|^2 \nonumber \\ 
&& +\bar{\cal E}(i\gamma^\alpha\part_\alpha -m_{\cal E}){\cal E} 
+\bar\nu_{\cal E} i\gamma^\alpha\part_\alpha\left (
\frac{1-\gamma_5}{2}\right )\nu_{\cal E} -eA_\alpha\left \{\bar{\cal E}
\gamma^\alpha{\cal E}-\frac23\bar{\cal U}
\gamma^\alpha{\cal U}+\frac13\bar{\cal D}\gamma^\alpha{\cal D}
\right \}+\nonumber \\ 
&&\frac{g}{2\sqrt 2}\left [W_\alpha^+\left \{\bar\nu_{\cal E}
\gamma^\alpha (1-\gamma_5){\cal E}+\bar{\cal U}\gamma^\alpha (1-\gamma_5)V_{\mbox{KM}}{\cal D}\right \}+W_\alpha^-\left \{\bar{\cal E}
\gamma^\alpha (1-\gamma_5)\nu_{\cal E}
+\bar{\cal D}\gamma^\alpha (1-\gamma_5)
V_{\mbox{KM}}^+{\cal U}\right \}\right ]\nonumber \\ 
&&-\frac{\sqrt{g^2+g'^2}}{4} Z_\alpha \left [\bar\nu_{\cal E}\gamma^\alpha (1-\gamma_5)\nu_{\cal E}-\bar{\cal E}\gamma^\alpha (1-\gamma_5){\cal E}+4\sin^2\theta_W\bar{\cal E}\gamma^\alpha{\cal E}\right ]\nonumber \\ 
&&-\frac{\sqrt{g^2+g'^2}}{4} Z_\alpha \left [ \bar{\cal U}
\gamma^\alpha (1-\gamma_5){\cal U}-\bar{\cal D}\gamma^\alpha 
(1-\gamma_5){\cal D}-4\sin^2\theta_W\left \{\frac23\bar{\cal U}
\gamma^\alpha{\cal U}-\frac13\bar{\cal D}\gamma^\alpha{\cal D}
\right \}\right ]\nonumber \\ 
&&-\frac{m_{\cal E}}{v}\varphi\bar{\cal E}{\cal E}
-\frac{m_{\cal Q}}{v}\varphi\bar{\cal Q}{\cal Q}
\left (+i\frac{m_{\cal E}}{v}\varphi_3\bar{\cal E}\gamma_5{\cal E}
-i\frac{\sqrt 2m_{\cal E}}{v}[\bar\nu_L{\cal E}_R\phi^+
-\bar{\cal E}_R\nu_L\phi^-]\mbox{etc.}\right ), \label{LAG}\\ 
&&\mbox{where terms in (+}\cdots\mbox{etc.) vanish in the Unitary gauge}\nonumber \\
&&\varphi_1(x)=\varphi_2(x)=\varphi_3(x)=0\quad\\ 
\mbox{and}&&W_\mu^\pm :=\frac{1}{\sqrt 2}(W_\mu^1\mp 
iW_\mu^2),\quad e:=\frac{gg'}{\sqrt{g^2+g'^2}}=g\sin\theta_W.\eeqn

\section{Is the specific charge a `particle' at all?}\label{Spe}
Some people try to explain the Higgs mechanism through the phase transition, 
in a similar way as in super conductivity. However, the definition of the vacuum in particle physics is different from as in condensed matter physics. In particle physics, it is defined as the poincar\'{e} invariant state where not any kind of a particle nor its anti particle is present. Of course there is no interaction, for the gauge particle to intermediate the force (i.e., a photon, a graviton, a $Z^0$ boson,$\cdots$) is absent. This is slightly different from that of general relativity, where the vacuum admits the presence of gravity. And it is a little more different from that of condensed matter physics, where the vacuum must not contain a plasmon nor a phonon nor a Cooper pair-of course they are not an elementary particle-, but admits the matter as medium of them (consists of an order of 1 mol $\sim 6\times 10^{23}$ of atoms). A phase transition defined in statistical mechanics needs a huge number of particles. Neither a temperature nor a specific heat can be defined for a particle. 

 Quantum field theory is a tool to deal with many body problems. Feynman diagrams for calculation of the Lamb shift involve many photons and electrons. In fact, this means many times of interactions of an electron with the electric field of the center proton. As a photon has no charge and its anti particle is also a photon, the number of photons is not conserved. However, an electron has a charge and the total charge is conserved for one atom. Therefore, however many virtual electron-positron pairs contribute to the diagram, the total number of electrons is always one. Exponentially many complex diagrams just correspond to exponentially rare possible reactions appearing in higher orders of perturbation. The Higgs scalar in the standard model works only as the specific charge of an electron. It is well defined also for the lonely electron and without any assumption on time evolution, thermal equilibrium nor many body problems. That is why naive extrapolations of a phase transition to the Higgs mechanism are physically misleading. We should be sincere in applying such a doubtful origin of Higgs mechanism to cosmology. Some people call the current theory of inflation as `the standard theory of cosmology', but this is very confusing. The original standard model for elementary particles never refers to the origin of the Higgs field. Rather, the idea that a specific charge is not a constant and obeys an equation of motion, rolling down the potential made by itself seems to me utter nonsense. Of course only experiments can judge and do justice to the theory. 

\end{document}